%Paper: hep-th/9403137
%From: preskill@theory.caltech.edu (John Preskill)
%Date: Tue, 22 Mar 94 15:13:40 PST

% 22 March 1994
%
% Plain TeX, uses the macro jnl.tex
%
% 8 postscript figures have been submitted as a uuencoded tar file,
% with instructions for unpacking.
%
% To include the figures, the macro epsf.tex is required
%
% If you do not have the figures, just enter a carriage return each
% time that TeX pauses.
%
%%                              JNL.TEX
%%
%%                This is JNL.TEX Version 0.3 as of 6/12/85.
%%
%%      This is a set of TeX 82 macros designed to produce scientific
%%      papers with a minimum of fuss and using as much of plain.tex as
%%      possible.  The user need only know what is in the TeXbook, and
%%      the macros under ``user definitions'' below.  Also, the user
%%      definitions are intended to be as simple as possible, so that
%%      the user may change them as desired.

%%
%%  Font definitions suitable for the IMAGEN (Written by Tony Kennedy)
%%

%  Define a whole menagerie of pseudo-12pt fonts

\font\twelverm=cmr10 scaled 1200    \font\twelvei=cmmi10 scaled 1200
\font\twelvesy=cmsy10 scaled 1200   \font\twelveex=cmex10 scaled 1200
\font\twelvebf=cmbx10 scaled 1200   \font\twelvesl=cmsl10 scaled 1200
\font\twelvett=cmtt10 scaled 1200   \font\twelveit=cmti10 scaled 1200

\skewchar\twelvei='177   \skewchar\twelvesy='60

%  Define \...point macros to change fonts and spacings consistently

\def\twelvepoint{\normalbaselineskip=12.4pt
  \abovedisplayskip 12.4pt plus 3pt minus 9pt
  \belowdisplayskip 12.4pt plus 3pt minus 9pt
  \abovedisplayshortskip 0pt plus 3pt
  \belowdisplayshortskip 7.2pt plus 3pt minus 4pt
  \smallskipamount=3.6pt plus1.2pt minus1.2pt
  \medskipamount=7.2pt plus2.4pt minus2.4pt
  \bigskipamount=14.4pt plus4.8pt minus4.8pt
  \def\rm{\fam0\twelverm}          \def\it{\fam\itfam\twelveit}%
  \def\sl{\fam\slfam\twelvesl}     \def\bf{\fam\bffam\twelvebf}%
  \def\mit{\fam 1}                 \def\cal{\fam 2}%
  \def\tt{\twelvett}
  \textfont0=\twelverm   \scriptfont0=\tenrm   \scriptscriptfont0=\sevenrm
  \textfont1=\twelvei    \scriptfont1=\teni    \scriptscriptfont1=\seveni
  \textfont2=\twelvesy   \scriptfont2=\tensy   \scriptscriptfont2=\sevensy
  \textfont3=\twelveex   \scriptfont3=\twelveex  \scriptscriptfont3=\twelveex
  \textfont\itfam=\twelveit
  \textfont\slfam=\twelvesl
  \textfont\bffam=\twelvebf \scriptfont\bffam=\tenbf
  \scriptscriptfont\bffam=\sevenbf
  \normalbaselines\rm}

%       tenpoint

\def\tenpoint{\normalbaselineskip=12pt
  \abovedisplayskip 12pt plus 3pt minus 9pt
  \belowdisplayskip 12pt plus 3pt minus 9pt
  \abovedisplayshortskip 0pt plus 3pt
  \belowdisplayshortskip 7pt plus 3pt minus 4pt
  \smallskipamount=3pt plus1pt minus1pt
  \medskipamount=6pt plus2pt minus2pt
  \bigskipamount=12pt plus4pt minus4pt
  \def\rm{\fam0\tenrm}          \def\it{\fam\itfam\tenit}%
  \def\sl{\fam\slfam\tensl}     \def\bf{\fam\bffam\tenbf}%
  \def\smc{\tensmc}             \def\mit{\fam 1}%
  \def\cal{\fam 2}%
  \textfont0=\tenrm   \scriptfont0=\sevenrm   \scriptscriptfont0=\fiverm
  \textfont1=\teni    \scriptfont1=\seveni    \scriptscriptfont1=\fivei
  \textfont2=\tensy   \scriptfont2=\sevensy   \scriptscriptfont2=\fivesy
  \textfont3=\tenex   \scriptfont3=\tenex     \scriptscriptfont3=\tenex
  \textfont\itfam=\tenit
  \textfont\slfam=\tensl
  \textfont\bffam=\tenbf \scriptfont\bffam=\sevenbf
  \scriptscriptfont\bffam=\fivebf
  \normalbaselines\rm}

%%
%%      Various internal macros
%%

\def\beginlinemode{\endmode
  \begingroup\parskip=0pt \obeylines\def\\{\par}\def\endmode{\par\endgroup}}
\def\beginparmode{\endmode
  \begingroup \def\endmode{\par\endgroup}}
\let\endmode=\par
{\obeylines\gdef\
{}}
\def\singlespace{\baselineskip=\normalbaselineskip}

\def\oneandahalfspace{\baselineskip=\normalbaselineskip
  \multiply\baselineskip by 3 \divide\baselineskip by 2}
\def\doublespace{\baselineskip=\normalbaselineskip \multiply\baselineskip by 2}

\newcount\firstpageno
\firstpageno=2
\footline={\ifnum\pageno<\firstpageno{\hfil}%
\else{\hfil\twelverm\folio\hfil}\fi}
\let\rawfootnote=\footnote              % We must set the footnote style
\def\footnote#1#2{{\rm\singlespace\parindent=0pt\rawfootnote{#1}{#2}}}
\def\raggedcenter{\leftskip=4em plus 12em \rightskip=\leftskip
  \parindent=0pt \parfillskip=0pt \spaceskip=.3333em \xspaceskip=.5em
  \pretolerance=9999 \tolerance=9999
  \hyphenpenalty=9999 \exhyphenpenalty=9999 }
\def\dateline{\rightline{\ifcase\month\or
  January\or February\or March\or April\or May\or June\or
  July\or August\or September\or October\or November\or December\fi
  \space\number\year}}
\def\received{\vskip 3pt plus 0.2fill
 \centerline{\sl (Received\space\ifcase\month\or
  January\or February\or March\or April\or May\or June\or
  July\or August\or September\or October\or November\or December\fi
  \qquad, \number\year)}}

%%
%%      Page layout, margins, font and spacing (feel free to change)
%%

\hsize=6.5truein
%\hoffset=1truein
\vsize=8.9truein
%\voffset=1truein
\parskip=\medskipamount
\twelvepoint            % selects twelvepoint fonts (cf. \tenpoint)
\doublespace            % selects double spacing for main part of paper (cf.
                        %       \singlespace, \oneandahalfspace)
\overfullrule=0pt       % delete the nasty little black boxes for overfull box

%%
%%      The user definitions for major parts of a paper (feel free to change)
%%

    % Preprint number at upper right of title page

\def\title                      %  Title on title page
  {\null\vskip 3pt plus 0.2fill
   \beginlinemode \doublespace \raggedcenter \bf}

\def\author                     %  Author(s) name(s)  on title page
  {\vskip 3pt plus 0.2fill \beginlinemode
   \singlespace \raggedcenter}

\def\affil                      % Affiliations (can intermix with \author)
  {\vskip 3pt plus 0.1fill \beginlinemode
   \oneandahalfspace \raggedcenter \sl}

\def\abstract                   % Begin abstract
  {\vskip 3pt plus 0.3fill \beginparmode
   \doublespace \narrower ABSTRACT: }

\def\endtitlepage               % End title page, begin body of paper
  {\endpage                     %       This subsumes \body
   \body}

\def\body                       % Begin text body;  can be used to end
  {\beginparmode}               % \title, \author, \affil, \abstract,
                                % \reference, or \figurecaption modes

\def\head#1{                    % Head;  NOTE enclose the text in {}
  \filbreak\vskip 0.5truein     %  e.g., \head{I. Introduction}
  {\immediate\write16{#1}
   \raggedcenter \uppercase{#1}\par}
   \nobreak\vskip 0.25truein\nobreak}

\def\subhead#1{                 % Subhead;  NOTE enclose the text in {}
  \vskip 0.25truein             % e.g., \subhead{A. History of the Problem}
  {\raggedcenter #1 \par}
   \nobreak\vskip 0.25truein\nobreak}

\def\refto#1{$|{#1}$}           % For references in text as superscript

\def\references                 % Begin references -- basic format is Phys Rev
  {\subhead{References}         % I.e., volume, page, year (space after commas)
   \beginparmode
   \frenchspacing \parindent=0pt \leftskip=1truecm
   \parskip=8pt plus 3pt \everypar{\hangindent=\parindent}}

\gdef\refis#1{\indent\hbox to 0pt{\hss#1.~}}    % Ref list numbers.

\gdef\journal#1, #2, #3, 1#4#5#6{               % Journal reference.  Comma set
    {\sl #1~}{\bf #2}, #3, (1#4#5#6)}           % off: name, vol, page, year

\def\refstylenp{                % Nucl Phys(or Phys Lett) ref style: V, Y, P
  \gdef\refto##1{ [##1]}                                % Reference in text []
  \gdef\refis##1{\indent\hbox to 0pt{\hss##1)~}}        % Ref list numbers)
  \gdef\journal##1, ##2, ##3, ##4 {                     % Journal reference
     {\sl ##1~}{\bf ##2~}(##3) ##4 }}

\def\refstyleprnp{              % Input like pr, output like np!!
  \gdef\refto##1{ [##1]}                                % Reference in text []
  \gdef\refis##1{\indent\hbox to 0pt{\hss##1)~}}        % Ref list numbers)
  \gdef\journal##1, ##2, ##3, 1##4##5##6{               % Journal reference
    {\sl ##1~}{\bf ##2~}(1##4##5##6) ##3}}

\def\endreferences{\body}

\def\figurecaptions             % Begin figure captions
  { \beginparmode
   \subhead{Figure Captions}
}

\def\endpage                    %  Eject a page
  {\vfill\eject}

\def\endpaper                   %  Ways to say goodbye
  {\endmode\vfill\supereject}

\def\endit
  {\endpaper\end}

%%
%%      Various little user definitions
%%

\def\ref#1{Ref. #1}                     %       for inline references
\def\Ref#1{Ref. #1}                     %       ditto

          % For citation of equation numbers
        %       ditto
                     %       ditto
                     %       ditto
                   %       ditto
                   %       ditto
\def\frac#1#2{{\textstyle{#1 \over #2}}}

\def\sla{\raise.15ex\hbox{$/$}\kern-.57em}
\def\leaderfill{\leaders\hbox to 1em{\hss.\hss}\hfill}
\def\twiddle{\lower.9ex\rlap{$\kern-.1em\scriptstyle\sim$}}
\def\bigtwiddle{\lower1.ex\rlap{$\sim$}}
\def\gtwid{\mathrel{\raise.3ex\hbox{$>$\kern-.75em\lower1ex\hbox{$\sim$}}}}
\def\ltwid{\mathrel{\raise.3ex\hbox{$<$\kern-.75em\lower1ex\hbox{$\sim$}}}}
\def\square{\kern1pt\vbox{\hrule height 1.2pt\hbox{\vrule width 1.2pt\hskip 3pt
   \vbox{\vskip 6pt}\hskip 3pt\vrule width 0.6pt}\hrule height 0.6pt}\kern1pt}

\def\m@th{\mathsurround=0pt }
\def\leftrightarrowfill{$\m@th \mathord\leftarrow \mkern-6mu
 \cleaders\hbox{$\mkern-2mu \mathord- \mkern-2mu$}\hfill
 \mkern-6mu \mathord\rightarrow$}
\def\overleftrightarrow#1{\vbox{\ialign{##\crcr
     \leftrightarrowfill\crcr\noalign{\kern-1pt\nointerlineskip}
     $\hfil\displaystyle{#1}\hfil$\crcr}}}

%% *********** New stuff follows *******************

\font\titlefont=cmr10 scaled\magstep3

\def\martinstyletitle                      %  Title on title page
  {\null\vskip 3pt plus 0.2fill
   \beginlinemode \doublespace \raggedcenter \titlefont}

\font\twelvesc=cmcsc10 scaled 1200

\def\author                     %  Author(s) name(s)  on title page
  {\vskip 3pt plus 0.2fill \beginlinemode
   \singlespace \raggedcenter\twelvesc}

%%
%%      AmSTeX compatability definitions
%%
%%      To run a TeX file originally intended for AmSTeX, only small changes
%%      should be necessary (I hope).  Use the line \input jnl at the start.
%%      Remove the lines \input amstex, \documentstyle{itpjnl} at the
%%      beginning;  also remove all the page layout stuff (\parindent=1cm,
%%      \hsize=5.28125in etc.)  The page layout is now done automatically.
%%      Also OMIT the qualifier \magnification=1200 when you IMPRINT the
%%      .dvi file.  (\TagsOnRight is harmless, you can take it out or leave
%%      it in.)  I believe most AmSTeX will work with no change.  One problem
%%      is \footnote, which is a little different in that it now needs to
%%      have an explicit asterisk *  (or whatever) included, like this:
%%              \footnote*{Text winds up at bottom of page.}
%%      This is discussed on p. 116 of the TeXbook.  IGNORE the AmSTeX
%%      documentation (if you can call it that);  refer to the TeXbook.
%%
%%      Note that many commands in AmSTeX have their equivalents in the
%%      TeXbook, perhaps with different names and slightly differing
%%      usage. E.g., the old \align in AmSTeX is replaced by \eqalign
%%      (p. 190) and \aligntag is replaced by \eqalignno (p. 192).
%%      \align and \aligntag still work, but I recommend that you use
%%      \eqalign and \eqalignno in documents run under jnl.
%%
%%      See me if you have any problems  -- Doug.
%%

\def\heading                            % Heading
  {\vskip 0.5truein plus 0.1truein      % e.g., \heading I. NOTES \endheading
   \beginparmode \def\\{\par} \parskip=0pt \singlespace \raggedcenter}

\def\subheading                         % Subheading
  {\vskip 0.25truein plus 0.1truein     % e.g., \subheading{A. The Problem}
   \beginlinemode \singlespace \parskip=0pt \def\\{\par}\raggedcenter}

\def\tag#1$${\eqno(#1)$$}

\def\align#1$${\eqalign{#1}$$}

\def\aligntag#1$${\gdef\tag##1\\{&(##1)\cr}\eqalignno{#1\\}$$
  \gdef\tag##1$${\eqno(##1)$$}}

\def\endaligntag{}

\def\overset #1\to#2{{\mathop{#2}\limits^{#1}}}
\def\underset#1\to#2{{\let\next=#1\mathpalette\undersetpalette#2}}
\def\undersetpalette#1#2{\vtop{\baselineskip0pt
\ialign{$\mathsurround=0pt #1\hfil##\hfil$\crcr#2\crcr\next\crcr}}}

%%
%%      Various little user definitions
%%

\def\ref#1{Ref.~#1}                     %       for inline references
\def\Ref#1{Ref.~#1}                     %       ditto
\def\[#1]{[\cite{#1}]}
\def\cite#1{{#1}}
\def\(#1){(\call{#1})}
\def\call#1{{#1}}
\def\taghead#1{}
\def\frac#1#2{{#1 \over #2}}

\def\12{{1\over2}}

\def\sla{\raise.15ex\hbox{$/$}\kern-.57em}
\def\leaderfill{\leaders\hbox to 1em{\hss.\hss}\hfill}
\def\twiddle{\lower.9ex\rlap{$\kern-.1em\scriptstyle\sim$}}
\def\bigtwiddle{\lower1.ex\rlap{$\sim$}}
\def\gtwid{\mathrel{\raise.3ex\hbox{$>$\kern-.75em\lower1ex\hbox{$\sim$}}}}
\def\ltwid{\mathrel{\raise.3ex\hbox{$<$\kern-.75em\lower1ex\hbox{$\sim$}}}}
\def\square{\kern1pt\vbox{\hrule height 1.2pt\hbox{\vrule width 1.2pt\hskip 3pt
   \vbox{\vskip 6pt}\hskip 3pt\vrule width 0.6pt}\hrule height 0.6pt}\kern1pt}
\def\tdot#1{\mathord{\mathop{#1}\limits^{\kern2pt\ldots}}}

\def\pmb#1{\setbox0=\hbox{#1}%
  \kern-.025em\copy0\kern-\wd0
  \kern  .05em\copy0\kern-\wd0
  \kern-.025em\raise.0433em\box0 }

\catcode`@=11
\newcount\tagnumber\tagnumber=0

\immediate\newwrite\eqnfile
\newif\if@qnfile\@qnfilefalse
\def\write@qn#1{}
\def\writenew@qn#1{}
\def\w@rnwrite#1{\write@qn{#1}\message{#1}}
\def\@rrwrite#1{\write@qn{#1}\errmessage{#1}}

\def\taghead#1{\gdef\t@ghead{#1}\global\tagnumber=0}
\def\t@ghead{}

\expandafter\def\csname @qnnum-3\endcsname
  {{\t@ghead\advance\tagnumber by -3\relax\number\tagnumber}}
\expandafter\def\csname @qnnum-2\endcsname
  {{\t@ghead\advance\tagnumber by -2\relax\number\tagnumber}}
\expandafter\def\csname @qnnum-1\endcsname
  {{\t@ghead\advance\tagnumber by -1\relax\number\tagnumber}}
\expandafter\def\csname @qnnum0\endcsname
  {\t@ghead\number\tagnumber}
\expandafter\def\csname @qnnum+1\endcsname
  {{\t@ghead\advance\tagnumber by 1\relax\number\tagnumber}}
\expandafter\def\csname @qnnum+2\endcsname
  {{\t@ghead\advance\tagnumber by 2\relax\number\tagnumber}}
\expandafter\def\csname @qnnum+3\endcsname
  {{\t@ghead\advance\tagnumber by 3\relax\number\tagnumber}}

\def\equationfile{%
  \@qnfiletrue\immediate\openout\eqnfile=\jobname.eqn%
  \def\write@qn##1{\if@qnfile\immediate\write\eqnfile{##1}\fi}
  \def\writenew@qn##1{\if@qnfile\immediate\write\eqnfile
    {\noexpand\tag{##1} = (\t@ghead\number\tagnumber)}\fi}
}

\def\callall#1{\xdef#1##1{#1{\noexpand\call{##1}}}}
\def\call#1{\each@rg\callr@nge{#1}}

\def\each@rg#1#2{{\let\thecsname=#1\expandafter\first@rg#2,\end,}}
\def\first@rg#1,{\thecsname{#1}\apply@rg}
\def\apply@rg#1,{\ifx\end#1\let\next=\relax%
\else,\thecsname{#1}\let\next=\apply@rg\fi\next}

\def\callr@nge#1{\calldor@nge#1-\end-}
\def\callr@ngeat#1\end-{#1}
\def\calldor@nge#1-#2-{\ifx\end#2\@qneatspace#1 %
  \else\calll@@p{#1}{#2}\callr@ngeat\fi}
\def\calll@@p#1#2{\ifnum#1>#2{\@rrwrite{Equation range #1-#2\space is bad.}
\errhelp{If you call a series of equations by the notation M-N, then M and
N must be integers, and N must be greater than or equal to M.}}\else %
{\count0=#1\count1=#2\advance\count1 by1\relax\expandafter\@qncall\the\count0,%
  \loop\advance\count0 by1\relax%
    \ifnum\count0<\count1,\expandafter\@qncall\the\count0,%
  \repeat}\fi}

\def\@qneatspace#1#2 {\@qncall#1#2,}
\def\@qncall#1,{\ifunc@lled{#1}{\def\next{#1}\ifx\next\empty\else
  \w@rnwrite{Equation number \noexpand\(>>#1<<) has not been defined yet.}
  >>#1<<\fi}\else\csname @qnnum#1\endcsname\fi}

\let\eqnono=\eqno
\def\eqno(#1){\tag#1}
\def\tag#1$${\eqnono(\displayt@g#1 )$$}

\def\aligntag#1\endaligntag
  $${\gdef\tag##1\\{&(##1 )\cr}\eqalignno{#1\\}$$
  \gdef\tag##1$${\eqnono(\displayt@g##1 )$$}}

\def\eqalignno#1{\displ@y \tabskip\centering
  \halign to\displaywidth{\hfil$\displaystyle{##}$\tabskip\z@skip
    &$\displaystyle{{}##}$\hfil\tabskip\centering
    &\llap{$\displayt@gpar##$}\tabskip\z@skip\crcr
    #1\crcr}}

\def\displayt@gpar(#1){(\displayt@g#1 )}

\def\displayt@g#1 {\rm\ifunc@lled{#1}\global\advance\tagnumber by1
        {\def\next{#1}\ifx\next\empty\else\expandafter
        \xdef\csname @qnnum#1\endcsname{\t@ghead\number\tagnumber}\fi}%
  \writenew@qn{#1}\t@ghead\number\tagnumber\else
        {\edef\next{\t@ghead\number\tagnumber}%
        \expandafter\ifx\csname @qnnum#1\endcsname\next\else
        \w@rnwrite{Equation \noexpand\tag{#1} is a duplicate number.}\fi}%
  \csname @qnnum#1\endcsname\fi}

\def\ifunc@lled#1{\expandafter\ifx\csname @qnnum#1\endcsname\relax}

\let\@qnend=\end\gdef\end{\if@qnfile
\immediate\write16{Equation numbers written on []\jobname.EQN.}\fi\@qnend}

\catcode`@=12

\catcode`@=11
\newcount\r@fcount \r@fcount=0
\newcount\r@fcurr
\immediate\newwrite\reffile
\newif\ifr@ffile\r@ffilefalse
\def\w@rnwrite#1{\ifr@ffile\immediate\write\reffile{#1}\fi\message{#1}}

\def\writer@f#1>>{}
\def\referencefile{%			  Stuff to write .REF file
  \r@ffiletrue\immediate\openout\reffile=\jobname.ref%
  \def\writer@f##1>>{\ifr@ffile\immediate\write\reffile%
    {\noexpand\refis{##1} = \csname r@fnum##1\endcsname = %
     \expandafter\expandafter\expandafter\strip@t\expandafter%
     \meaning\csname r@ftext\csname r@fnum##1\endcsname\endcsname}\fi}%
  \def\strip@t##1>>{}}

\def\citeall#1{\xdef#1##1{#1{\noexpand\cite{##1}}}}
\def\cite#1{\each@rg\citer@nge{#1}}	% Variable No. of args, separated by

\def\each@rg#1#2{{\let\thecsname=#1\expandafter\first@rg#2,\end,}}
\def\first@rg#1,{\thecsname{#1}\apply@rg}	% each@ag is a general purpose
\def\apply@rg#1,{\ifx\end#1\let\next=\relax%	  variable no. of arg. macro.
\else,\thecsname{#1}\let\next=\apply@rg\fi\next}% args separated by commas

\def\citer@nge#1{\citedor@nge#1-\end-}	% Check for M-N range (M and N numbers)
\def\citer@ngeat#1\end-{#1}
\def\citedor@nge#1-#2-{\ifx\end#2\r@featspace#1 % Single argument
  \else\citel@@p{#1}{#2}\citer@ngeat\fi}	% M-N range of arguments
\def\citel@@p#1#2{\ifnum#1>#2{\errmessage{Reference range #1-#2\space is bad.}%
    \errhelp{If you cite a series of references by the notation M-N, then M and
    N must be integers, and N must be greater than or equal to M.}}\else%
 {\count0=#1\count1=#2\advance\count1 by1\relax\expandafter\r@fcite\the\count0,
  \loop\advance\count0 by1\relax%	  Loop from M to N
    \ifnum\count0<\count1,\expandafter\r@fcite\the\count0,%
  \repeat}\fi}

\def\r@featspace#1#2 {\r@fcite#1#2,}	% Eat spaces at beginning or end of arg
\def\r@fcite#1,{\ifuncit@d{#1}%		  Cite individual reference
    \newr@f{#1}%
    \expandafter\gdef\csname r@ftext\number\r@fcount\endcsname%
                     {\message{Reference #1 to be supplied.}%
                      \writer@f#1>>#1 to be supplied.\par}%
 \fi%
 \csname r@fnum#1\endcsname}
\def\ifuncit@d#1{\expandafter\ifx\csname r@fnum#1\endcsname\relax}%
\def\newr@f#1{\global\advance\r@fcount by1%
    \expandafter\xdef\csname r@fnum#1\endcsname{\number\r@fcount}}

\let\r@fis=\refis			% Save old \refis, redefine
\def\refis#1#2#3\par{\ifuncit@d{#1}%      Use two params #2 #3 to strip blank
   \newr@f{#1}%
   \w@rnwrite{Reference #1=\number\r@fcount\space is not cited up to now.}\fi%
  \expandafter\gdef\csname r@ftext\csname r@fnum#1\endcsname\endcsname%
  {\writer@f#1>>#2#3\par}}

\def\ignoreuncited{%   redefine \refis if ignoring uncited references
   \def\refis##1##2##3\par{\ifuncit@d{##1}%
    \else\expandafter\gdef\csname r@ftext\csname r@fnum##1\endcsname\endcsname%
     {\writer@f##1>>##2##3\par}\fi}}

\def\r@ferr{\endreferences\errmessage{I was expecting to see
\noexpand\endreferences before now;  I have inserted it here.}}
\let\r@ferences=\references
\def\references{\r@ferences\def\endmode{\r@ferr\par\endgroup}}

\let\endr@ferences=\endreferences
\def\endreferences{\r@fcurr=0%		  Save old \endreferences, redefine
  {\loop\ifnum\r@fcurr<\r@fcount%	  Loop over refnum and produce text
    \advance\r@fcurr by 1\relax\expandafter\r@fis\expandafter{\number\r@fcurr}%
    \csname r@ftext\number\r@fcurr\endcsname%
  \repeat}\gdef\r@ferr{}\endr@ferences}

% Save old \endpaper, redefine it to write parting message.

\let\r@fend=\endpaper\gdef\endpaper{\ifr@ffile
\immediate\write16{Cross References written on []\jobname.REF.}\fi\r@fend}

\catcode`@=12

\citeall\refto		% These macros will generate citations
\citeall\ref		%
\citeall\Ref		%

\input epsf
\tolerance=10000
%\equationfile
%\referencefile

%\input pptxxx
\ignoreuncited
%\def\(#1){(\call{#1})}
%\def\call#1{{#1}}
%\def\Equation#1{Equation~\(#1)}         % For citation of equation
                                        % numbers
%\def\Equations#1{Equations~\(#1)}       %       ditto
%\def\Eq#1{Eq.~\(#1)}                    %       ditto
%\def\Eqs#1{Eqs.~\(#1)}                  %       ditto
%\def\eq#1{eq.~\(#1)}
%\def\eqs#1{eqs.~\(#1)}

\def\la{\lambda}

\def\Om{{\Omega}}

\def\bbR{{I\kern-0.3em R}}

\def\pmb#1{\setbox0=\hbox{$#1$}%
\kern-.025em\copy0\kern-\wd0
\kern.05em\copy0\kern-\wd0
\kern-.025em\raise.0433em\box0 }

\def\q2{{Q^2}}
\def\gtwid{\raise.3ex\hbox{$>$\kern-.75em\lower1ex\hbox{$\sim$}}}
\def\ltwid{\raise.3ex\hbox{$<$\kern-.75em\lower1ex\hbox{$\sim$}}}
\def\12{{1\over2}}

\def\part{\partial}

\def\topppageno1{\global\footline={\hfil}\global\headline
={\ifnum\pageno<\firstpageno{\hfil}\else{\hss\twelverm --\ \folio
\ --\hss}\fi}}

\def\toppageno2{\global\footline={\hfil}\global\headline
={\ifnum\pageno<\firstpageno{\hfil}\else{\rightline{\hfill\hfill
\twelverm \ \folio
\ \hss}}\fi}}

\def\boxit#1{\vbox{\hrule\hbox{\vrule\kern3pt
  \vbox{\kern3pt#1\kern3pt}\kern3pt\vrule}\hrule}}

\def\slS{\raise.15ex\hbox{$/$}\kern-.53em\hbox{$S$}}
\def\cI{{\cal I}}

{\hbox to\hsize{\tenpoint \baselineskip=12pt
        \hfil\vtop{
        \hbox{\strut CALT-68-1918}
%       \hbox{\strut **SB Preprint No?**}
        \hbox{\strut hep-th/9403137}}}}

\title BLACK HOLE THERMODYNAMICS AND INFORMATION LOSS IN TWO DIMENSIONS
\author Thomas M. Fiola,$^*$ ~John Preskill,$^\dagger$ ~Andrew Strominger,$^*$
{}~and
Sandip P. Trivedi$^\dagger$

\affil ${}^*$ Department of Physics
University of California at Santa Barbara, CA 93106-9530

\affil ${}^\dagger$ Lauritsen Laboratory of High Energy Physics
California Institute of Technology, Pasadena, CA 91125

\abstract{Black hole evaporation is investigated in a (1+1)-dimensional model
of quantum gravity.  Quantum corrections to the
black hole entropy are computed, and the fine-grained entropy of the Hawking
radiation is studied. A generalized second law of thermodynamics is formulated,
and shown to be valid under suitable conditions. It is also
shown that, in this model, a black hole can consume an
arbitrarily large amount of information.}
\endpage

\head{I. Introduction}

Hawking's discovery of black hole radiance [\cite{hawk}] established a deep and
satisfying link connecting gravitation, thermodynamics, and quantum theory.
But it also raised some disturbing puzzles.  Foremost among these are the
mystery of black hole entropy, and the paradox of information loss. These two
puzzles are closely related.  Together, they comprise a crisis in fundamental
physics.

Black hole thermodynamics has a compelling beauty.  Bekenstein's bold
conjecture [\cite{bek}]  that a generalized second law of thermodynamics
applies to processes involving black holes, combined with Hawking's explicit
calculation of the black hole temperature, led to the remarkable result that a
black hole has an intrinsic entropy given by ${1\over 4}$ the area of the event
horizon (in Planck units).  But previous efforts to verify the generalized
second law [\cite{zurekthorne,frolov}] have been limited to quasi-stationary
processes, and to the leading semiclassical approximation.  In this paper, we
will study black hole thermodynamics in two-dimensional spacetime.  For the
special case of two dimensions, we are able to go substantially
further than previous
analyses, by considering processes that are not quasi-stationary, and by taking
explicit account of quantum-mechanical back reaction effects. We will propose a
precise statement of the generalized second law, and will demonstrate that it
is valid in a particular two-dimensional model, under suitable conditions.

In Hawking's semiclassical theory of black hole evaporation [\cite{hawk}], the
radiation emitted by the black hole was found to be exactly thermal
[\cite{wald}].  Thus, in the leading semiclassical approximation, the radiation
carries no information about the initial quantum state of the object that
collapsed to form the black hole.  This property of the radiation led Hawking
to assert [\cite{hawk2}] that quantum-mechanical information can be destroyed
when a black hole forms and then subsequently evaporates completely.  Although
the semiclassical approximation is not exact, it is highly plausible that more
accurate calculations would still support the conclusion that the outgoing
radiation carries very little information; the key point is that, once it has
fallen past the global horizon, the collapsing body is out of causal contact
with the radiation emitted from the black hole.  Still, no complete analysis of
the microscopic state of the radiation has ever been carried out.  In this
paper, we study black hole evaporation in a two-dimensional model, taking into
account quantum-mechanical gravitational back reaction effects.  We find that
the microscopic state of the emitted radiation carries essentially no
information, as in the leading semiclassical calculations.  Thus, loss of
information really seems to occur in this model.  (Or, perhaps, the information
about the initial quantum state is retained inside a stable black hole remnant
[\cite{remnant}].)

It was emphasized in Ref.~[\cite{CGHS}] that two-dimensional models of quantum
gravity can serve as a theoretical laboratory for investigating the fundamental
issue of information loss.  A further motivation for studying the CGHS model
introduced in Ref.~[\cite{CGHS}] is that it can be viewed as the low-energy
effective field theory that governs the S-wave modes propagating on the
background of a magnetically charged dilaton black hole in four dimensions.
The (four-dimensional) dilaton black hole is of particular interest because it
is a classical solution to a field theory that arises as a low energy
approximation to string theory [\cite{gibbons}].

Though the CGHS model is far simpler than four-dimensional gravity, the full
quantum theory of the model is still difficult to analyze.  Therefore, CGHS
studied a particular limit in which the model simplifies further.  In this
limit, the number $N$ of matter field species tends to infinity, with $N\hbar$
held fixed.  Then, to leading order in an expansion in $1/N$, but all orders in
$N\hbar$, the quantum fluctuations of the dilaton and metric may be ignored,
and only the fluctuations of the matter fields need be retained.  Later, Russo,
Susskind, and Thorlacius (RST) [\cite{rst}] showed (expanding on ideas
introduced in Ref.~[\cite{emod}]) that the model can be simplified still
further by introducing a suitably chosen finite local counterterm.  Our
calculations in this paper will be carried out in the RST model, to leading
order in $1/N$.  We will review the RST model in Section II.

The generalized second law of thermodynamics states that the {\it total}
entropy is nondecreasing, where the total entropy is the sum of the intrinsic
entropy of the black hole and the thermodynamic entropy of the matter outside
the black hole.  To investigate the validity of the second law, we will carry
out a three-step program.  First, we must define precisely what is meant by the
entropy due to the matter ``outside'' the black hole, and we must calculate
this entropy.  Second, we must find the correct expression for the black hole
entropy in the RST model, including corrections to all orders in $N\hbar$ (but
to leading order in $1/N$).  Third, we must consider how the total entropy
evolves, for a variety of initial conditions satisfied by the ``collapsing''
matter.

To obtain an expression for the entropy outside the black hole, we erect a
sharp boundary at the apparent horizon, and then trace over the matter field
degrees of freedom behind the horizon to obtain a density matrix $\rho_{\rm
out}$ for the matter fields outside.  We then calculate the ``fine-grained''
entropy
$S_{\rm FG}=-{\rm tr}\left(\rho_{\rm out}\ln \rho_{\rm out}\right)$ of this
density matrix.  The fine-grained entropy quantifies the degree of entanglement
of the quantum fields outside the horizon with those inside.  We will see that
this quantity can also be interpreted as the thermodynamic entropy of the
matter outside the black hole.  (Actually this is not quite the whole story.
For a black hole formed from collapse, we will need to add to the fine-grained
entropy another term, the ``Boltzman entropy'' of the infalling matter.  This
will be explained in Section VI.)

Our calculations of the fine-grained entropy are performed in Section III. The
method that we use is a generalization of the technique introduced by Unruh
[\cite{unruh}] in his analysis of the thermal bath seen by a uniformly
accelerated observer, later extended to other cases by Holzhey
[\cite{holzhey}].  These calculations are of some intrinsic interest apart from
the relevance of the results to black hole physics, and we therefore discuss
them in detail.  As we will see, the
fine-grained entropy has an ultraviolet divergence that arises from the
entanglement of very-short-wavelength field fluctuations just inside and just
outside the boundary.  We regulate the divergence by introducing a
short-distance cutoff (or, equivalently, by smoothing the boundary). One way to
introduce this cutoff is to foliate the spacetime with spacelike slices; then
on each slice, we assign to the boundary at the apparent horizon a
``thickness'' of proper length $\delta$.   The resulting expression for the
fine-grained entropy depends on this length $\delta$, but it does not depend on
the choice of the foliation, or on the coordinates used on each slice.  In
particular, two slices that cross the apparent horizon at the same point, but
with a relative boost, yield the same value of the fine-grained entropy.  As
the
black hole evolves, the proper length $\delta$ is held fixed.

In two-dimensional spacetime, the ultraviolet divergence is logarithmic, and
the cutoff-dependent term in the entropy is merely a numerical constant.  (At
least, it is a constant from the time of formation of the black hole until its
ultimate disappearance.)
Thus, the divergence does not prevent us from making
statements about the {\it change} in the entropy that are free from cutoff
dependence.\footnote{*}{However, we will see that the change in the
entropy (as we define it)
at the moment of black hole formation, as well as the total
entropy produced by the entire formation/evaporation process, do
depend significantly on the cutoff.}  The situation seems to be quite different
in four
dimensions.
Then the divergence is quadratic, and proportional to the area of the horizon
[\cite{hooft2}].  Thus, the generalization of our analysis to four dimensions
is not straightforward.

The second step in our program, to find the corrected expression for the black
hole entropy in the RST model, is carried out in Section V.  We find a finite
correction to the entropy computed in the leading semiclassical theory;  the
correction arises from the back reaction on the geometry when the black hole
accretes or emits a small amount of radiation.  We regard the black hole
entropy as finite, and attribute the ultraviolet divergence in the total
entropy to the matter fields surrounding the black hole.  This is really a
matter of convention, as our calculations fix the black hole entropy only up to
an additive constant.  We have chosen to fix the constant by demanding that the
intrinsic entropy of the black hole vanishes as its mass goes to zero.

We assemble our expression for the total entropy in the RST model in Section
VI,
and analyze the evolution of the entropy in Sections VI and VII.  Section VII
contains our analysis of the generalized second law of thermodynamics.  To
prove the second law, we need to make some additional assumptions.  Most
notably, we assume that the state of the matter that collapses to form the
black hole is of a particular type---it is a coherent state built on the
asymptotic inertial vacuum.  Some such assumption seems to be necessary.  It is
possible to construct strange quantum states that pack a lot of entropy into a
region at a very low cost in energy
[\cite{holzhey,wilczek}], or states with negative energy density (though
this is not possible for coherent states).  By
preparing one of these strange states and dropping it into a black hole, the
generalized second law that we have formulated {\it can} be violated, at least
for a while. It would certainly be of interest to find a modified formulation
of the generalized second law
with more general validity and/or a concise characterization of how and when
our formulation breaks down.

Our expression for the fine-grained entropy also enables us to address the
question of information loss.  We can imagine sustaining a black hole for an
arbitrarily long time by feeding it mass to compensate for the Hawking
radiation that it emits.  It was emphasized in Ref.~[\cite{stromtriv}] that, if
we draw a suitable spacelike slice through the geometry of this black hole, the
amount of information stored in the portion of the slice that is behind the
global horizon can be arbitrarily large.  Thus one may argue that the number of
internal quantum states of a black hole is not limited by its intrinsic
Bekenstein-Hawking entropy.  In Section IV, we analyze this sustained black
hole
(in the RST model) from the viewpoint of an observer who remains outside the
horizon.  We show that the fine-grained entropy outside the horizon can
increase by an arbitrarily large amount.  In accord with the conclusion of
Ref.~[\cite{stromtriv}], then, we find that there is no consistent way to
regard the density matrix $\rho_{\rm out}$ as arising from the entanglement of
the degrees of freedom outside the horizon with a {\it finite} number of
internal degrees of freedom of the black hole.  Unless there are stable black
hole remnants with an infinite number of internal degrees of freedom
[\cite{remnant}], information is inevitably lost in the RST model.

In fact, the amount of lost information is even larger than one might have
naively expected.  The evaporation of a warm black hole into cold empty space
is a thermodynamically irreversible process---the increase in the thermodynamic
entropy of the emitted radiation is larger than the decrease in the entropy of
the black hole [\cite{zurek,page}].  (In one spatial dimension, it is larger by
a factor of two.)  We find in Section VI that the {\it fine-grained} entropy
outside the
horizon behaves like the thermodynamic entropy.  This means that the number of
bits of lost information exceeds the number of bits needed to describe the
initial quantum state of the collapsing matter, by a factor of
(approximately) two.  Thus, the
Bekenstein-Hawking entropy of the black hole formed in the initial collapse
does not correctly quantify the amount of information that is ultimately lost.

The fine-grained entropy can increase indefinitely because the field modes
localized close to the horizon are subjected to a red shift that increases
exponentially as the black hole evolves.  We introduced a short-distance cutoff
of fixed proper length at the apparent horizon.  But it follows that this
cutoff, when expressed in terms of the
asymptotically inertial coordinates used to
define the quantum vacuum (or, equivalently, in terms of the wavelength
measured at past null infinity), decreases exponentially
along the horizon.  As shorter and shorter wavelengths
come into play, the degree of entanglement between the fields inside and
outside the horizon increases correspondingly.  It is this feature of the
quantum state outside the horizon that is responsible for both the thermal
character of the outgoing radiation and for the loss of an indefinite amount of
information in the RST model.

It is evident that the conclusion that information is lost is predicated on
assumptions about how {\it extreme} Lorentz boosts act on the matter degrees of
freedom.  (This point has been especially emphasized by 't Hooft
[\cite{hooft}], Jacobson [\cite{jacobson}], Susskind [\cite{susskind}], and the
Verlindes [\cite{verl}].)  While loss of information apparently occurs in the
RST model, it might be avoided in a different model with different physics at
{\it very} short distances.  In such a model, it may be possible to
attribute the fine-grained entropy to entanglement with a finite number of
microscopic internal degrees of freedom of the black hole, and to interpret the
Bekenstein-Hawking entropy of the black hole in terms of these internal degrees
of freedom.  The explicit contruction of a model with these properties would be
of great interest.

The content of this paper overlaps with that of several other references that
have appeared while our work was being completed.  In particular, Keski-Vakkuri
and Mathur [\cite{mathur}] have also analyzed the fine-grained entropy outside
the horizon of an evaporating black hole.  Where there is overlap, our
conclusions are in agreement with theirs.  Calculations of the fine-grained
entropy for moving-mirror spacetimes (which closely resemble black hole
spacetimes) have been discussed by Holzhey, Larsen, and Wilczek
[\cite{holzhey_wilczek}].  Quantum corrections to the black
hole entropy have been considered recently by Susskind and Uglum
[\cite{uglum}], Callan and Wilczek [\cite{callan}], Kabat and Strassler
[\cite{strassler}], and Dowker [\cite{dowker}].

\head{II. Review of the RST Model}

An elegant model for two-dimensional black hole evaporation was
introduced by Russo, Susskind and Thorlacius [\cite{rst}], expanding on ideas
introduced in [\cite{emod}]. The RST model differs from the original
CGHS model [\cite{CGHS}]
by a finite counterterm that is fine-tuned to preserve a global symmetry. The
counterterm makes it possible to solve the model exactly
in the large-$N$ limit, where $N$ is the number of scalar matter fields.
Numerical analyses [\cite{lowe,pira}] of the
CGHS model indicate that it is qualitatively similar to the RST model, despite
the fine-tuning.

The original CGHS model [\cite{CGHS}] of two-dimensional dilaton gravity has
the classical action
$$
S_{\rm classical}={1\over 2\pi}\int
d^2x\sqrt{-g}\left[e^{-2\phi}\left(R+4(\nabla
\phi)^2 + 4\lambda^2\right) -{1\over 2}\sum_{i=1}^N(\nabla f_i)^2\right]~,
\eqno(CGHS)
$$
where $g$ is the metric, $R$ is the curvature scalar, $\phi$ is the dilaton
field, and the $f_i$ are the $N$ scalar matter fields.  This model can be
regarded as the low-energy effective action that governs the radial modes
propagating on the near-extreme magnetically charged black hole of
four-dimensional dilaton gravity.  The length scale $\lambda^{-1}$ is
proportional to the magnetic charge of the four-dimensional black hole.

Two-dimensional dilaton gravity has classical black hole solutions.  The mass
of a black hole can be expressed in terms of the value $\phi_H$ of the dilaton
field at the event horizon as
$$
M_{\rm BH}={\lambda\over \pi}e^{-2\phi_H}~.
\eqno(dilBHmass)
$$
We may also interpret Eq.~\(dilBHmass) as the deviation from the extremal limit
of the mass of a four-dimensional black hole.  Semiclassically, the
two-dimensional black hole has a nonzero Hawking temperature.  This can be
computed from the periodicity of the black hole solution in Euclidean time
[\cite{CGHS}], or from the Bogolubov transformation that relates the asymptotic
incoming modes of the matter fields to the asymptotic outgoing modes
[\cite{gine}].  The temperature is
$$
T_{\rm BH}={\lambda\over 2\pi}~,
\eqno(BHtemp)
$$
which is independent of the black hole mass.  Thus the two-dimensional black
hole has an infinite specific heat. The four-dimensional magnetically charged
dilaton black hole also has this property [\cite{gibbons}].  We obtain an
expression for the black hole entropy $S_{\rm BH}$ by integrating the
thermodynamic identity $dS=dM/T$; it is
$$
S_{\rm BH}={M_{\rm BH}\over T_{\rm BH}}=2e^{-2\phi_H}~,
\eqno(BHentropy)
$$
where we have fixed the constant of integration by demanding that $S_{\rm
BH}\to 0$ as $M_{\rm BH}\to 0$.  We may interpret Eq.~\(BHentropy) as ${1\over
4}$ the area of the event horizon of the classical four-dimensional dilaton
black hole.

CGHS considered the semiclassical corrections to this classical theory,
including the back reaction of the Hawking radiation on the geometry.  To make
the analysis tractable, they assumed that the number $N$ of scalar matter
fields is very large, and calculated the back reaction to leading order in an
expansion in $1/N$.  In leading order, the quantum fluctuations of the dilaton
and metric can be ignored, and we need only include the one-loop correction to
the energy momentum tensor of the scalars. This correction can be computed from
the conformal anomaly.  Equivalently, we add to the classical action
Eq.~\(CGHS)
the Polyakov-Liouville term [\cite{polyakov}]
$$
S_{\rm Liouville}=-{N\over 96\pi}\int d^2x\sqrt{-g(x)}\int d^2x'\sqrt{-g(x')}
R(x) G(x,x')R(x')~,
\eqno(liouville)
$$
where $G$ is a Green function of the operator $\nabla^2$.  This term expresses
the dependence on the background geometry of the functional measure for the
scalar fields.  The field equations derived from the action $S_{\rm
classical}+S_{\rm Liouville}$ have been studied numerically
[\cite{numerical,lowe,pira}], but analytic solutions have not been obtained.
However, RST (following Ref.~[\cite{emod}]) found that the model can be solved
exactly if a local counterterm
$$
S_{\rm c.t.} = -{N\over 48\pi}\int d^2x \sqrt{-g}\phi R
\eqno(RSTct)
$$
is added to the action.

To solve the model including \(RSTct), we introduce null coordinates
$x^{\pm}=x^0\pm x^1$ and
invoke the conformal gauge condition
$$
g_{+-}=g_{-+}=-{1\over 2} e^{2\rho}~,\quad g_{--}=g_{++}=0~.
\eqno(conformal_gauge)
$$
We then have
$$
\eqalign{S_{\rm classical}&={1\over\pi} \int d^2x \Bigg[2e^{-2\phi}
\part_+\part_-\rho
+e^{-2\phi}(\la^2e^{2\rho}-4\part_+\phi\part_-\phi)
+{1\over2}\sum^N_{i=1}\part_+f_i\part_-f_i\Bigg]~,\cr
&S_{\rm c.t.}= -{N\over 12\pi}\int
d^2x~\phi\partial_+\partial_-\rho~,\quad\quad
S_{\rm Liouville}= -{N\over 12\pi}\int d^2x~\partial_+\rho\partial_-\rho~.\cr}
\eqno(conformal_gauge_actions)
$$
We now perform the field redefinition\footnote{*}{Our conventions differ
slightly from
[\cite{rst}] and agree with [\cite{cpt}]. They are chosen so that  $\chi$ and
$\Omega$ are
held fixed as $N$ is taken to infinity.}
$$
\eqalignno{
\Om &={12\over N}e^{-2\phi}+{\phi\over2}+{1\over4}\ln{N\over 48}~, &(odef)\cr
\chi &={12\over N}e^{-2\phi}+\rho-{\phi\over2}-{1\over4}\ln{N\over 3}~. &(cdef)
\cr}
$$
In the large-$N$ limit, with $\chi$ and $\Om$ held fixed, the quantum effective
action is then
$$
S_{\rm eff} ={1\over\pi}\int d^2 x\Bigg[{N\over12}(-\part_-\chi
\part_+\chi+\part_+\Om\part_-\Om
+\la^2e^{2\chi-2\Om})+{1\over2}\sum^N_{i=1}\part_+f_i\part_-f_i\Bigg]~.
\eqno(nact)
$$
(The effects of ghosts may be ignored in the large-$N$
limit.)

There is a residual conformal gauge invariance in \(nact). We fix
this by the ``Kruskal gauge'' choice
$$
\chi=\Om~, \eqno(gchc)
$$
which implies
$$
\rho=\phi+{1\over2}\ln{N\over12}~. \eqno(rphi)
$$
In Kruskal gauge the equations of motion are simply
$$
\part_+\part_-\Om=-\la^2~; \eqno(oeom)
$$
the constraints can be expressed as
$$
\part^2_\pm\Om=-T^f_{\pm\pm}-t_{\pm}~. \eqno(cstr)
$$
Appearing on the right-hand side of Eq.~\(cstr) is the expectation field of the
scalar field energy-momentum tensor, which we have separated into two terms.
The first term $T^f$ is the ``classical'' piece that can be obtained by varying
the  matter action with respect to the metric, except that,
in order to simplify Eq.~\(cstr), we have chosen an unconventional
normalization, namely
$$
T^f_{\pm\pm}={12\pi\over N} \left(T^f_{\pm\pm}\right)_{\rm conv} ={6\over
N}\sum^N_{i=1}\part_\pm f_i\part_\pm f_i~.
\eqno(tdef)
$$
In particular, since ``Newton's constant'' is of order $1/N$, we have scaled
$T^f_{\pm\pm}$ by a factor of $1/N$, so that $T^f_{\pm\pm}$ of order one
produces a
back reaction of order one.  Fluctuations of the energy-momentum tensor about
its expectation value are suppressed by $1/N$, so the energy-momentum may be
treated as a classical quantity to leading order.

The functions $t_\pm(x^\pm)$ in Eq.~\(cstr) arise because the constraints in
Kruskal gauge
are governed by the energy-momentum tensor normal ordered with respect to the
``Kruskal vacuum'' state---the state that contains no quanta that are positive
frequency with respect to Kruskal time.  The quantum state of the scalar fields
can be expressed in terms of $f$ creation operators acting on the $f$-vacuum
state.  If this $f$-vacuum differs from the Kruskal vacuum, there is a finite
normal ordering correction to the energy momentum tensor, in addition to the
``classical'' term $T^f$.  In effect, this term arises because we must subtract
a $\rho$-dependent piece of the vacuum energy from both sides of the constraint
equation in order to express the left-hand side of Eq.~\(cstr) in terms of
$\Omega$.  It is important to recognize that Eq.~\(cstr) holds only in the
Kruskal gauge.  On the right-hand side of this equation, $T^f_{\pm\pm}$
transforms as a tensor, but $t_\pm$ does not.

In our analysis of black hole formation and evaporation, we will typically be
interested in incoming quantum states that are coherent states built on the
``$\sigma$ vacuum''.  The $\sigma^{\pm}$ coordinates are related to the Kruskal
coordinates $x^{\pm}$ by
$$
\lambda x^+=e^{\lambda \sigma^+}~,\quad \lambda x^-=-e^{-\lambda \sigma^-}~.
\eqno(sigma_define)
$$
These coincide with the inertial coordinates on ${\cal I}^-$;
thus, the $\sigma$ vacuum state $|0,\sigma\rangle$ is the state that appears to
contain no quanta according to inertial asymptotic observers in the past.  A
left-moving
coherent state can be built on this vacuum at ${\cal I}^-$, of the form
$$
|f^c,\sigma\rangle=A:e^{{i \over \pi}\sum^N_{i=1}\int
d\sigma^+\part_+f_i^c(\sigma^+)
 \hat f_i(\sigma^+) }:_\sigma|0,\sigma\rangle~ ,
\eqno(chst)
$$
where the normal ordering is with respect to the $\sigma$ vacuum, and $A$ is a
normalization constant.  In Eq.~\(chst), $\hat f$ denotes the quantum field,
and
$f^c$ is its expectation value,
$$
\langle f^c,\sigma|\hat f_i(\sigma^+)|f^c,\sigma\rangle = f_i^c(\sigma^+)~.
\eqno(expec_value)
$$
For the energy-momentum tensor $:\hat T_{++}(x^+):_K$ normal ordered with
respect to the
Kruskal vacuum $|0,K\rangle$, we then have
$$
\langle f^c,\sigma|:\hat T_{++}:_K|f^c,\sigma\rangle= T^{f^c}_{++}~+ \langle
0,\sigma|: \hat T_{++}:_K|0,\sigma\rangle ~;
\eqno(normal_order_correction)
$$
thus $t_+$ in Eq.~\(cstr) can be expressed as
$$
t_+=\langle 0,\sigma|: \hat T_{++}(x^+):_K|0,\sigma\rangle ~,
\eqno(calc_tplus)
$$
where it is understood that $\hat T_{++}$ has the unusual normalization in
Eq.~\(tdef), and that $\langle \hat T_{++}(x^+)\rangle$ is to be evaluated in
the Kruskal gauge.

In flat space with metric
$$
ds^2=-d\sigma^+d\sigma^-=-{dx^+ dx^-\over \lambda^2 x^+ x^-}~,
\eqno(flat_metric)
$$
we may use standard methods [\cite{fulling}] to compute
$$
t^0_\pm(x^\pm)=\langle 0, \sigma|:T_{\pm\pm}:_K|0, \sigma\rangle=-{12\pi\over
N}\cdot
{N\over 48\pi (x^\pm)^2}=- {1\over 4(x^\pm)^2}~.
\eqno(t_naught)
$$
The solution to Eq.~\(oeom) and Eq.~\(cstr) then becomes
$$
\Om=-\la^2x^+x^--{1\over4}\ln[-4\la^2x^+x^-]~, \eqno(oldv)
$$
or
$$
\phi=-{1\over2}\ln\Bigg[{-\la^2 Nx^+x^-\over12}\Bigg] =-\lambda\sigma^1
-{1\over 2}\ln\left({N\over 12}\right)~;\eqno(pldv)
$$
this is the ``linear dilaton vacuum'' solution, so called because $\phi$ is a
linear function of $\sigma^1={1\over 2}(\sigma^+-\sigma^-)$.  The solution
corresponding to general incoming matter from
$\cI^-$ is (in Kruskal gauge)
$$
\eqalign{
\chi(x^+,x^-)=\Om(x^+,x^-)
&=-{\la^2}x^+\left(x^-+{1\over\la^2}P_+(x^+)\right)+{1\over\la}M(x^+)\cr
&-{1\over4}\ln[-4\la^2x^+x^-]~, \cr} \eqno(gsol)
$$
where
$$
\eqalignno{
M(x^+) &= \la\int^{x^+} d \tilde x^+ \tilde x^+T^f_{++}(\tilde x^+)~,
&(mdef)\cr
P_+(x^+) &=\int^{x^+} d \tilde x^+T^f_{++}(\tilde x^+)~. &(pdef) \cr}
$$
(We have chosen the origin of the Kruskal coordinate system so as to remove
possible terms linear in  $x^+$ and $x^-$.)  Here $P_+(x_+)$ is the total
``Kruskal momentum'' that has flowed in from ${\cal I}^-$ up to retarded time
$x^+$.  If we express $M$ in terms of the energy-momentum in the $\sigma$ gauge
$$
{\cal E}(\sigma^+)=T_{++}^f(\sigma^+)
\eqno(sigma_gauge_energy)
$$
and recall that the $\sigma$ coordinates coincide with inertial coordinates on
${\cal I}^-$, we see that
$$
M(x^+)=\int^{\sigma^+}d\tilde\sigma{\cal E}(\tilde\sigma^+)
\eqno(sigma_mass)
$$
is the total ``energy-at-infinity'' that has flowed in from ${\cal I}^-$ up to
retarded time $x^+$.

If the incoming energy flux ${\cal E}(\sigma^+)$ satisfies suitable conditions
(described
below), this solution describes a black hole that forms and evaporates.  To
make sense of this statement, we must explain what is meant by a ``black hole''
in this two-dimensional model.  Since, in four--dimensional dilaton gravity,
$\Omega$ plays the role of the area of a two-sphere (as defined by the
canonical metric), we refer to the points
with $\partial_+\Omega<0$ and $\partial_-\Omega<0$ as ``trapped points''; the
``area'' necessarily decreases in the forward light cone of these points.  The
boundary of the region of trapped points, where $\partial _+\Omega=0$, is the
apparent horizon of a black hole.  From a two-dimensional viewpoint, the
significance of the apparent horizon is that $\Omega^{-1}$ is a coupling
constant that controls the higher--order quantum corrections in the model.
Thus, observers inside the apparent horizon are ineluctably drawn more deeply
into the strong-coupling region of the spacetime (at least for a while).

Viewed as a function of $\phi$, $\Om$ has a minimum at
$$
\eqalign{
\phi_{\rm cr} &=-{1\over2}\ln{N\over 48}~,\cr
\Om_{\rm cr} &={1\over 4}~. \cr}
\eqno(boundary)
$$
There is no real value of $\phi$ corresponding to $\Om<\Om_{\rm cr}$.  This
singular behavior occurs deep inside the strong-coupling region, where a
semiclassical analysis is no longer trustworthy.  Nevertheless, RST suggested
that a simple ``phenomenological'' description of this strong-coupling physics
might be possible.
They advocated that $\Om=\Om_{\rm cr}$
should be regarded as the analog of the origin of radial coordinates; it is a
boundary of spacetime, and one should not continue to negative ``radius.''
Instead, as long as the boundary is timelike, reflecting boundary conditions
(consistent with energy conservation) can
be imposed. Thus, RST propose
$$
f_i\biggr|_{\Om=\Om_{\rm cr}} =0~.
%\part_\pm\Om\biggr|_{\Om=\Om_{\rm cr}}=0~.
\eqno(rbc)
$$
RST also imposed boundary conditions on $\Omega$. Using these boundary
conditions, one can determine the dynamical motion of the line
$\Omega=\Omega_{\rm cr}$ in the $(x^+,x^-)$ plane.  However, it turns out to be
a delicate matter to impose quantum-mechanically consistent boundary
conditions. Fully consistent boundary
conditions will be discussed in Ref.~[\cite{andyandlarus}], but we
need not be concerned with such subtleties in this paper.

If the energy flux ${\cal E}$ of the incoming matter is at all times less than
the critical flux ${\cal E}_{\rm cr}={1\over 4}\lambda^2$, then the boundary
remains timelike, and the incoming matter is benignly reflected to future null
infinity ${\cal I}^+$ without any ``loss of information.''  However, when
${\cal E}$ exceeds ${\cal E}_{\rm cr}$, an apparent horizon appears and a black
hole forms.  Furthermore, behind the apparent horizon, the boundary becomes
spacelike, and the scalar curvature $R$ diverges on the spacelike portion of
the boundary.  It is no longer sensible to impose boundary conditions on the
fields when the boundary becomes spacelike.  Fig. 1 depicts the spacetime of a
black hole that forms from an initial incoming pulse of matter. After it forms,
the black hole emits Hawking radiation, and the apparent horizon recedes along
a timelike trajectory.  The global event horizon is the boundary of the region
in which all forward-directed timelike and null trajectories eventually meet
the spacelike singularity.  Of course, this singularity occurs deep within the
strongly-coupled region, and so might be absent in the full quantum theory.
But observers inside the global event horizon are inevitably drawn to the
strongly-coupled region where semiclassical methods are inapplicable.

\midinsert
\epsfysize=5in
\centerline{ \epsfbox{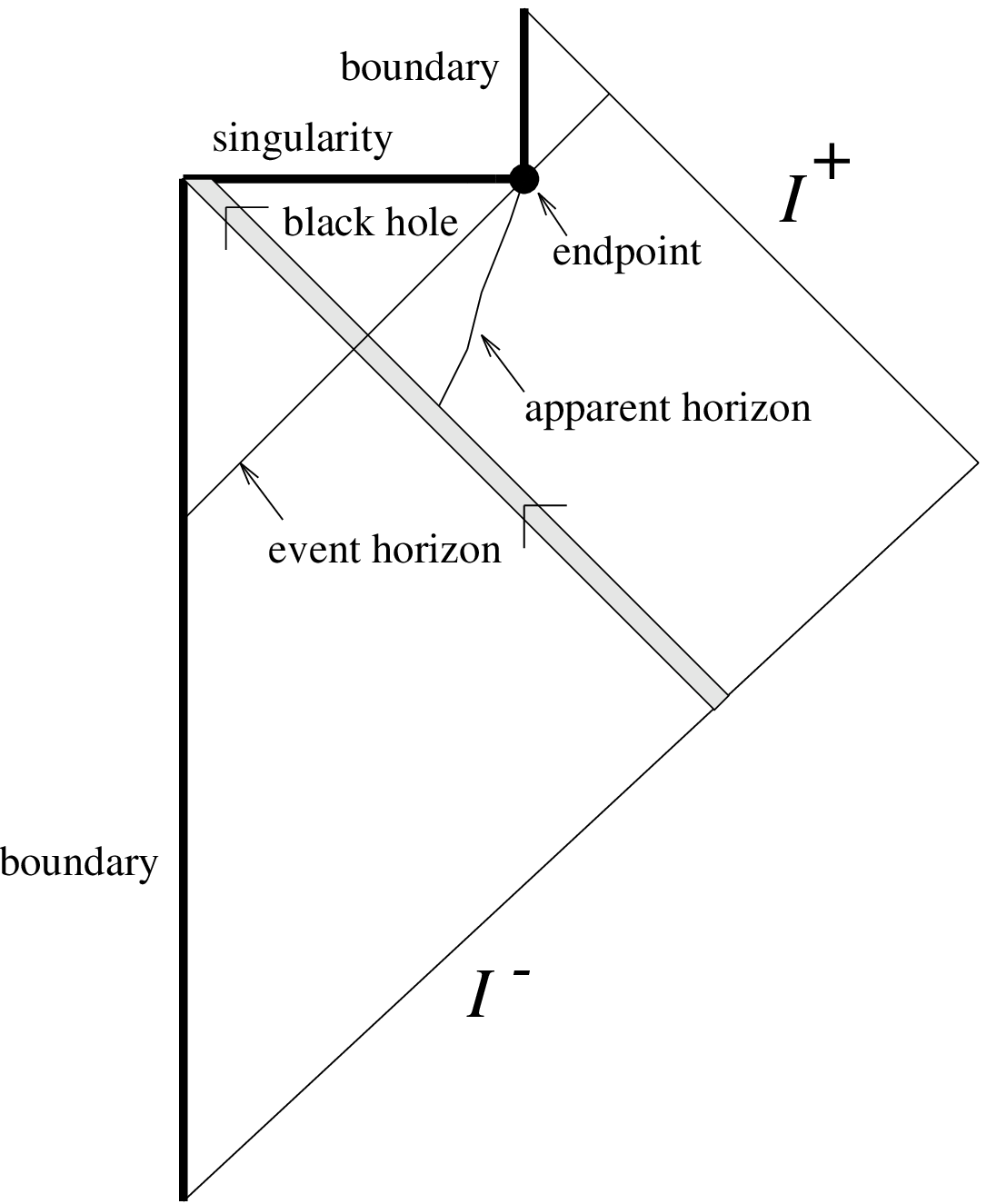}}
\bigskip
\centerline{FIGURE 1 (a).}
\medskip
{\centerline{\vbox{\hsize 5in \singlespace\tenrm \noindent
The two-dimensional spacetime of a black hole that
forms due to the collapse of a shock wave, and then evaporates completely.
After the black hole forms, the apparent horizon recedes along a timelike
trajectory, eventually meeting the singularity at the ``endpoint.''  The
timelike boundary and the spacelike singularity are in the strongly-coupled
region.  RST boundary conditions are imposed where the boundary is timelike.
 }}}

\endinsert

If the value $\Omega$ at the global horizon is large when the black hole first
forms, then semiclassical methods can be reliably used to analyze the evolution
of the geometry and of the quantum matter fields {\it outside} the global
horizon.  This remains true until just before the apparent horizon meets the
singularity at the ``endpoint'' shown in Fig. 1a.  The behavior of the
spacetime in the future of this endpoint cannot be unambiguously predicted
using semiclassical methods.  RST argued that, after the endpoint, the boundary
of the spacetime is again timelike, the matter fields again obey the boundary
condition Eq.~\(rbc), and the quantum state of the matter fields returns to the
vacuum state.  In their scenario, information about the quantum-mechanical
state of the original incoming matter is forever lost to asymptotic observers.
For most of our analysis of the evolving black hole,
we need not enter into speculation about what happens
beyond the endpoint.  It will suffice to analyze the quantum state of the
matter fields outside the horizon, without leaving the domain of validity of
semiclassical methods.

It will sometimes be convenient to consider an incoming quantum state that is a
coherent state built on the Kruskal vacuum state.  Then $t_{\pm}$ in
Eq.~\(cstr) vanish, and the general solution, in Kruskal gauge, is
$$
\chi(x^+,x^-)=\Om(x^+,x^-)
=-{\la^2}x^+\left(x^-+{1\over\la^2}P_+(x^+)\right)+{1\over\la}M(x^+)~,
\eqno(gsol_kruskal)
$$
with $M$ and $P$ again given by Eq,~\(mdef), \(pdef).  The (static) vacuum
solution with $P=0$ and constant $M$ describes a black hole in equilibrium with
a thermal
radiation bath.  Calculating the energy-momentum tensor of the matter fields in
the asymptotic region, we find that the incoming and outgoing energy flux are
both given by ${\cal E}_{\rm cr}$.  From the normalization condition
Eq.~\(tdef), we see that this corresponds to the conventionally normalized flux
$N\lambda^2/48\pi$, which is the thermal flux for N scalar fields at
temperature $T=\lambda/2\pi$.  Thus, we see that back reaction effects do not
modify the black hole temperature, to leading order in $1/N$.

The semiclassical field equations enable us to determine the evolution of the
expectation values of $\Omega$, $\chi$, and the $f_i$'s from specified initial
conditions (though of course we must fix the gauge to determine $\chi$).
However, in our analysis of black hole thermodynamics, we will need to keep
track of the entropy of the matter fields outside the apparent horizon of the
black hole.  For this purpose, it is not sufficient to know expectation values;
we must know the quantum states themselves.

Fluctuations of the energy-momentum tensor about its mean value will induce
correlations between the quantum state of the matter and the quantum state of
the dilaton field and of the geometry.  Fortunately, this entanglement of the
state of the matter with the state of the geometry is subdominant in the
large-$N$ limit and can be neglected to leading order.  Thus, the large-$N$
limit drastically simplifies the evolution of the quantum states.  To leading
order in $1/N$, we may regard the geometry and the dilaton field as a classical
background, dynamically determined by the expectation value of the
energy-momentum tensor, as prescribed by the semiclassical equations.  Evolving
the coherent state  of a free massless scalar field on this background is easy;
we need only choose the mean value $f^c_i$ in Eq.~\(chst) to be a solution to
the classical field equation.

The quantum states also depend on the position of the boundary through the
boundary condition Eq.~\(rbc).  If the incoming energy flux never exceeds
${\cal E}_{\rm cr}$, then the boundary remains timelike, and the incoming
matter is reflected off the boundary to ${\cal I}^+$.  Knowing the geometry and
the dynamically determined trajectory of the boundary, we can perform a
Bogolubov transformation and express the reflected state in terms of Fock space
states built on the inertial vacuum at ${\cal I}^+$.  (The state
$|f^c,\sigma\rangle$ will not, in general, be a simple coherent state in this
natural asymptotic Fock basis on ${\cal I}^+$.)  Thus, we can compute a unitary
$S$-matrix that relates the incoming and outgoing quantum states.

If the incoming energy flux ever exceeds ${\cal E}_{\rm cr}$, then a black hole
forms, and the boundary becomes spacelike.  Nevertheless, we can determine the
quantum state on a slice (like slice III in Fig.~1b) that penetrates inside the
black hole but avoids the spacelike singularity.  To do so we must again know
the dynamically determined trajectory of the boundary.  But in our calculations
in this paper, we will make the simplifying assumption that no incoming matter
meets the boundary before the global event horizon.  The trajectory
$x_B^-(x_B^+)$ of the boundary outside the global horizon is then determined by
setting $\Omega=\Omega_{\rm cr}={1\over 4}$ in the vacuum solution Eq.~\(oldv);
we find (in Kruskal coordinates)
$$
x_B^+ x_B^-=-{1\over 4\lambda^2}~.
\eqno(ldv_boundary)
$$
{}From this boundary trajectory and the semiclassically determined geometry,
the quantum state outside the global horizon can be completely determined to
leading order in $1/N$.  Our assumption that no matter meets the boundary
before the global horizon not only simplifies our calculations; it also enables
us to obtain results that are insensitive to any ambiguities concerning the
proper choice of the boundary conditions satisfied by $\Omega$.

\midinsert
\epsfysize=5in
\centerline{ \epsfbox{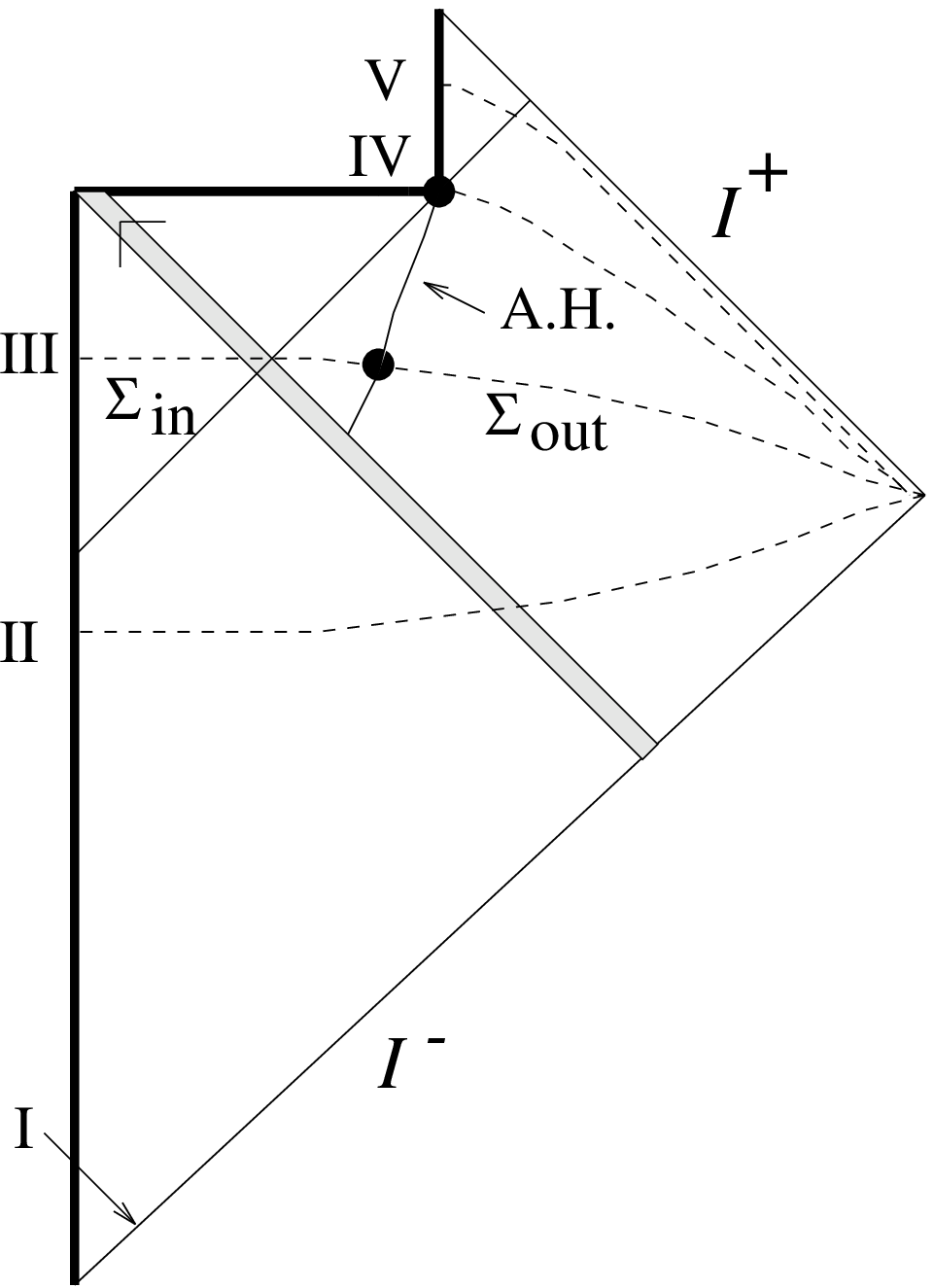}}
\bigskip
\centerline{FIGURE 1 (b).}
\medskip
{\centerline{\vbox{\hsize 5in \singlespace\tenrm \noindent
 Five spacelike slices through the spacetime, referred to in the
text.
 }}}

\endinsert

In principle, we could carry out the Bogolubov transformation and express the
outgoing quantum state in terms of the natural outgoing Fock basis.  We will
see in Section III, however, that the detailed form of this Bogolubov
transformation will not be needed in our calculation of the entropy of the
quantum state outside the apparent horizon of the black hole.

\head{III. Fine-grained entropy}

In our analysis of the formation and evaporation of a black hole in the RST
model, we will need to study the density matrix for the quantized matter fields
outside the apparent horizon of the black hole.  For a specified quantum state
of the matter fields, this density matrix $\rho$ is obtained by tracing over
the field degrees of freedom behind the horizon.  In this Section, we will
derive a formula for the ``fine-grained entropy'' $S_{\rm FG}=-{\rm tr}
\rho\ln\rho$ of this density matrix.  We will assume that the matter fields are
free massless scalar fields.

Our derivation will proceed in several steps.  First, we will consider a flat
two-dimensional spacetime, and suppose that the quantum state is the Minkowski
vacuum.  We imagine that a finite spatial region $R$ is inaccessible to an
observer.  The information accessible to this observer can therefore be encoded
in a density matrix $\rho$ that is obtained by tracing over the field degrees
of freedom inside region $R$.  We will calculate the entropy of this density
matrix.  (Our analytic formula for the entropy agrees with a numerical
calculation by Srednicki [\cite{srednicki}]. This formula was obtained earlier
by Holzhey [\cite{holzhey}], whose methods we follow closely.) We then proceed
to generalize the entropy formula to more general ``vacuum'' states, and to
curved spacetime.

In the RST model, scalar field modes are reflected by the boundary of the
spacetime; this reflection induces correlations between left-moving and
right-moving modes, which must be taken into account in the computation of the
entropy.  Thus, we consider a spacetime with a moving mirror, and derive a
formula for the entropy of the density matrix that is obtained by tracing over
a region that contains the mirror, when the quantum fields are in a ``vacuum''
state.  The curved-spacetime generalization of this formula can be directly
applied to the RST model.

Finally, in Appendix A, we consider more general quantum states, namely,
coherent states built upon a specified ``vacuum.''  We show (somewhat
surprisingly) that the fine-grained entropy for any such coherent state takes
the same value as for the corresponding ``vacuum.''  Thus, the quantum fields
inside and outside of region $R$ are no more entangled in an arbitrary coherent
state than in the vacuum.

\subhead{A. Minkowski vacuum}
We begin with the case of the Minkowski vacuum in flat two-dimensional
spacetime.  Let us imagine that the only observables that we can measure have
support outside of a finite spatial region $R$.  In the vacuum state, the
fields inside $R$ are correlated with the fields outside $R$.  Thus, even
though the state of the whole system is pure, the density matrix $\rho$
obtained by tracing over the inaccessible degrees of freedom inside $R$ is
mixed.  We wish to calculate the entropy
$$
S_{\rm FG}=-{\rm tr} \rho\ln\rho\eqno(finegrained)
$$
of this density matrix, which we will refer to as the fine-grained entropy of
the state outside $R$.  Note that we could just as well imagine that we are
able to measure only observables {\it inside} $R$.  The two density matrices
obtained by tracing over degrees of freedom inside or outside the region have
the same nonzero eigenvalues, and hence the same entropy.

For massless free fields in two dimensions, the right-moving and left-moving
modes are uncoupled, so it is sufficient to consider, say, the right-movers
alone.  It is convenient to use the null coordinates
$$
U=t-x~,\quad V=t+x~;\eqno(nullcoord)
$$
for the right-movers, we may specify the region $R$ as the interval $[U_1,U_2]$
in null coordinates.
To proceed with the entropy calculation, we must contruct a complete set of
(right-moving) modes localized inside this interval, and a complete set of
modes localized outside.  Then we must decompose the Minkowski vacuum state in
a basis consisting of states that are tensor products of states localized
inside with states localized outside.  Finally, we trace over the degrees of
freedom outside $R$ to obtain $\rho_{\rm inside}$, and compute $S_{\rm FG}$.

This seems a daunting task at first but upon reflection we recognize that we
already know how to do the calculation when the region $R$ is the half line.
The right-moving modes with $U<0$ are those that are accessible to a (Rindler)
observer who accelerates uniformly to the right. (See Fig.~2.) The density
matrix seen by the Rindler observer was computed long ago by
Unruh [\cite{unruh}].  We need only generalize Unruh's calculation to the case
where the region $R$ is the finite interval $U_1\le U\le U_2$ rather than the
half line $U<0$.

\midinsert
\epsfysize=5in
\centerline{ \epsfbox{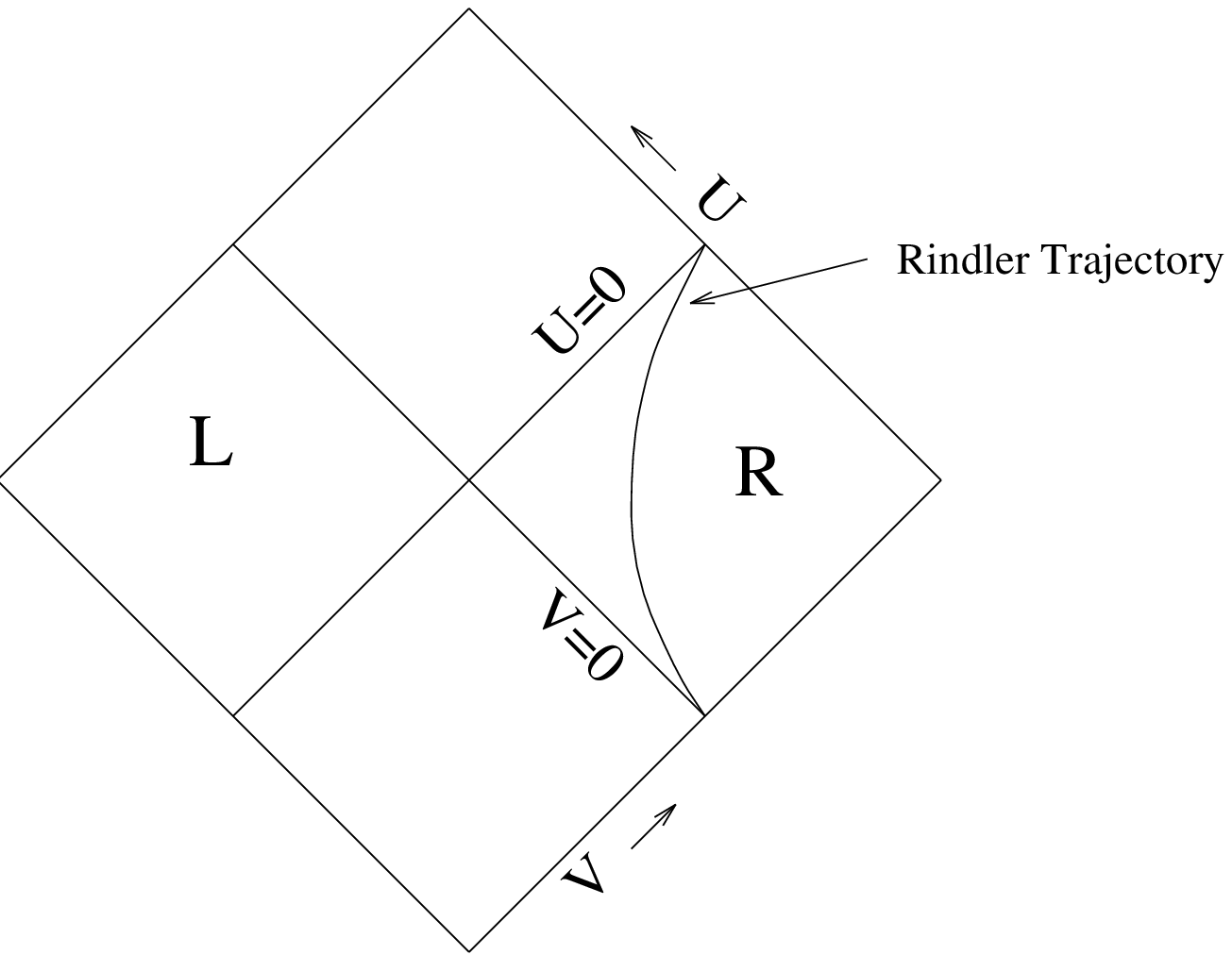}}
\bigskip
\centerline{FIGURE 2.}
\medskip
{\centerline{\vbox{\hsize 5in \singlespace\tenrm \noindent
Rindler spacetime.  The ``right wedge,'' with $U<0$ and
$V>0$, is accessible to a ``Rindler observer'' that accelerates uniformly to
the right.
The ``left wedge,'' with $U>0$ and $V<0$, is accessible to an observer that
accelerates uniformly to the left.
 }}}

\endinsert

First we briefly recall Unruh's reasoning.  The entropy does not depend on the
bases that we use for the modes that are localized in $U<0$ and $U>0$, so we
are free to choose these bases in any convenient way that simplifies the
calculation.  Unruh introduces Rindler coordinates $u_R$ and $u_L$ in the right
and left Rindler wedges that are related to the Minkowski coordinates by
$$
\eqalign{
u_R&=-\ln(-U)~,\quad U<0~,\cr
u_L&=-\ln(~U~)~,\quad U>0~.\cr}\eqno(rindler)
$$
The Rindler time defined by this transformation actually runs backwards in the
left wedge.  Therefore, the modes
$$
\eqalign{
\phi_{R,\omega}&= \theta(-U)~e^{-i\omega u_R}~,\cr
\phi_{L,\omega}&=\theta(U)~e^{i\omega u_L}~,\cr}\eqno(landrmodes)
$$
($\omega>0$) are positive frequency modes with respect to Rindler time in the
right and left wedges, respectively.
Since the coordinate $u_R$ covers the right wedge $U<0$ as $u_R$ varies from
$-\infty$ to $\infty$, arbitrary wave packets constructed from the modes
$\phi_{R,\omega}$ are localized in the right wedge; similarly, wave packets
constructed from the modes $\phi_{L,\omega}$ are localized in the left wedge.

If we choose as our basis these modes that have definite frequency with respect
to Rindler time, then, as Unruh noted [\cite{unruh}], it is easy to derive the
Bogolubov coefficients that relate these modes to the modes that have positive
frequency with respect to Minkowski time.  We need only recall that a
superposition of modes that are positive frequency with respect to Minkowski
time will be an analytic function of the Minkowski null coordinate $U$ in the
lower $U$ half plane.  Thus, by analytically continuing the mode
$\phi_{R\omega}$ to the left wedge, through the lower $U$ half plane, we obtain
$$
\phi_{1,\omega}=N_\omega\left(\phi_{R,\omega}+
e^{-\pi\omega}\phi^*_{L,\omega}\right)~.\eqno(minkmodeone)
$$
This combination of a positive frequency mode (with respect to Rindler time) in
the right wedge and a negative frequency mode in the left wedge is a
superposition of modes that have strictly positive frequency with respect to
Minkowski time; $N_\omega$ is a normalization factor.  Similarly, the
combination
$$
\phi_{2,\omega}=N_\omega\left(\phi_{L,\omega}+
e^{-\pi\omega}\phi^*_{R,\omega}\right) \eqno(minkmodetwo)
$$
is also positive frequency with respect to Minkowski time.

Using the Bogolubov coefficients Eq.~\(minkmodeone) and \(minkmodetwo), it is
straightforward to express the Minowski vacuum state $\vert 0_M\rangle$ in
terms of Rindler Fock space states.  (See Appendix A.)  One finds
$$
\eqalign{
\vert 0_M\rangle =& \prod_j (1-e^{-2 \pi \omega_j})^{1 \over 2}
                    \exp\left( e^{-\pi \omega_j}
                    a_{R,j}^\dagger \ a_{L,j}^\dagger\right) \vert 0_R\rangle
\otimes\vert 0_L\rangle \cr
			=&\prod_j (1-e^{-2 \pi \omega_j})^{1 \over 2}
			\sum_{n_j=0}^\infty
                    e^{-\pi \omega_j n_j} \vert n_j, R\rangle\otimes \vert n_j,
L \rangle~;\cr} \eqno(minvac)
$$
Here $\vert 0_{R}\rangle$ and $\vert 0_L\rangle$ denote the Rindler vacuum
states in the right and left wedges, and $\vert n_j,R\rangle$, $\vert n_j,
L\rangle$ are the states containing $n_j$ quanta with Rindler frequency
$\omega_j$.

We can now trace over the degrees of freedom in the left wedge to obtain the
density matrix for the state in the right wedge; it is
$$
\rho_R ={\rm tr_L} \vert 0_M\rangle\langle 0_M\vert
=\prod_j \left(\ \left(1-e^{-2 \pi \omega_j}\right) \sum_{n_j}
                       e^{- 2 \pi \omega_j n_j}
                       \vert n_{j, R}\rangle \langle n_{j,R} \vert \ \right)~.
                      \eqno(rhor)
$$
This is evidently a thermal density matrix with temperature
$$
T={1\over 2\pi}~.\eqno(rindlertemp)
$$
The temperature is dimensionless because we have chosen to express the
frequencies in terms of dimensionless Rindler time.  If we re-express the
frequency in terms of the proper time measured by the uniformly accelerated
Rindler observers, we find that $T=a/2\pi$, where $a$ is the proper
acceleration.  Thus we obtain Unruh's result [\cite{unruh}]:  a uniformly
accelerated observer in the Minkowski vacuum sees a thermal bath with
temperature $a/2\pi$.

In one spatial dimension, the energy density of a (right-moving) ideal gas is
$$
{\cal E}=\int_0^\infty {d\omega\over 2\pi}{\omega\over e^{\omega/T} -1}
={\pi\over 12} T^2~,\eqno(enerdens)
$$
and the entropy density is obtained from the thermodynamic relation
$$
{\cal S}=\int_0^T{d{\cal E}\over T}= {\pi\over 6} T~.\eqno(entropydens)
$$
Integrating this entropy density over the half line gives an infinite result.
We can obtain a finite answer by introducing ultraviolet and infrared cutoffs;
then we find the fine-grained entropy
$$
S_{\rm FG}\equiv -{\rm tr} \rho_R\ln\rho_R
= {\pi\over 6} T \left(u_{R,\rm max}-u_{R,\rm min}\right)= {1\over 12}
\ln\left({U_{\rm max}\over U_{\rm min}}\right)~. \eqno(halflineentropy)
$$
Of course, including the left-moving modes would result in the additional term
${1\over 12}\ln \left( V_{\rm max}/V_{\rm min}\right)$.

The logarithmic behavior of the fine-grained entropy is a consequence of the
scale invariance of the vacuum fluctuations of a massless scalar field.  Field
modes of all wavelengths contribute to the entanglement of the quantum state in
the right wedge with the quantum state in the left wedge.  To exploit the scale
invariance, it is convenient to construct a basis for the modes as follows:
{}From the modes with wavenumber between $k_0$ and $2k_0$, we construct a basis
of nonoverlapping wavepackets, each with width of order $k_0^{-1}$.  Among
these modes, only the one wavepacket that overlaps the boundary between the two
regions contributes to the entanglement.  Now complete the basis by replacing
$k_0$ by $2^j k_0$, for all integer $j$.  For each value of $j$, a single
wavepacket contributes to the entropy;  on dimensional grounds, the
contribution is a pure number of order one, and because of the scale
invariance, the
contribution is independent of $j$.  Summing over all modes, we thus obtain an
expression for the fine-grained entropy that diverges logarithmically in both
the ultraviolet and the infrared.  The divergent behavior of
Eq.~\(halflineentropy) as $U_{\rm min}$ approaches zero arises because
field modes that are localized just to the right of $U=0$ are entangled with
the modes that are localized just to the left of $U=0$, in the Minkowski vacuum
state.  In three spatial dimensions, because of the enhanced density of states,
the ultraviolet divergence becomes quadratic; the entropy is proportional to
the transverse area [\cite{hooft2,srednicki,bombelli}], and is infrared finite.

We now want to generalize Unruh's procedure to the case where the inaccessible
region is a finite interval $[U_1,U_2]$ rather than the half line.  (This
generalization was pioneered by Holzhey [\cite{holzhey}].)  Again, the key idea
is that, since the entropy is basis-independent, we are free to introduce bases
for the modes inside and outside the interval that make the computation of the
entropy easy.  Following Unruh, we seek handy coordinate systems that cover the
inside and outside regions, which are related to one another by analytic
continuation.  We will also impose an infrared cutoff by restricting the null
coordinate $U$ to the range $[-L,L]$.  Thus, we introduce the coordinate
$$
u(U)=\ln\left|{\sin\left({(U-U_1)\pi\over 2L}\right)\over
\sin\left({(U_2-U)\pi\over 2L }\right)}\right|~.\eqno(intervalcoord)
$$
Here the vertical bars denote absolute value.  Eq.~\(intervalcoord) really
describes two distinct corrdinate systems; one coordinate, which we call
$u_{\rm in}$, varies from $-\infty$ to $\infty$ as $U$ varies from $U_1$ to
$U_2$.  The other coordinate, $u_{\rm out}$, covers the region $[-L,L]$, {\it
excluding} the interval $[U_1,U_2]$.  This coordinate $u_{\rm out}$ approaches
$\infty$ as $U$ approaches $U_2$ (from above), and it approaches $-\infty$ as
$U$ approaches $U_1$ (from below).  It also satisfies
$$
u_{\rm out}(U=L)=u_{\rm out}(U=-L)~;\eqno(periodicity)
$$
Thus any wavepacket constructed as a function of $u_{\rm out}$ automatically
satisfies periodic boundary conditions as a function of $U$ on the interval
$[-L,L]$.  The time coordinate defined by the transformation
Eq.~\(intervalcoord) runs {\it backwards} in the region outside the interval
$[U_1,U_2]$

Now the modes of definite frequency with respect to $u$,
$$
\eqalign{
\phi_{{\rm in}, \omega}&=\theta (U-U_1) \theta (U_2-U) e^{-i \omega u_{\rm
in}}~,\cr
\phi_{{\rm out}, \omega}&=\left(\theta (U_1-U) +\theta (U-U_2) \right) e^{i
\omega u_{\rm out}}~,\cr}\eqno(inandoutmodes)
$$
are analogous to the Rindler modes Eq.~\(landrmodes).  Following Unruh, we can
calculate Bogolubov coefficients by analytically continuing these modes in the
lower $U$ half plane.  We thus construct the mode
$$
\phi_{1, \omega}=N_{\omega} \left( \phi_{{\rm in}, \omega} + e^{- \pi \omega}
                             \phi_{{\rm
out},\omega}^*\right)~;\eqno(minkinandoutone)
$$
this is a superposition of a positive frequency inside mode and a negative
frequency outside mode that is positive frequency with respect to Minkowski
time.  Similarly, the superposition
$$
\phi_{2, \omega}=N_{\omega} \left(\phi_{{\rm out},\omega} +e^{- \pi \omega}
                            \phi_{{\rm in}, \omega}^* \right)
\eqno(minkinandouttwo)
$$
is also positive frequency with respect to Minkowski time.

With our choice of coordinates, the Bogolubov coefficients
Eq.~\(minkinandoutone) and \(minkinandouttwo) are of just the same form as the
Bogolubov coefficients Eq.~\(minkmodeone) and \(minkmodetwo) for the Rindler
case.  Thus, the calculation of the density matrix obtained by tracing over the
degrees of freedom inside the interval $[U_1,U_2]$ proceeds exactly as
before---we obtain a thermal density matrix with temperature $T=1/2\pi$.  We
compute the entropy by integrating the thermal entropy density over the
interval.  As expected, the expression for the entropy has a logarithmic
ultraviolet divergence at each endpoint of the interval, arising from the
entanglement of the short-wavelength field fluctuations on either side of the
endpoint.  We can regulate the calculation by excluding the contribution due to
the radiation bath within (affine) distance $\delta_2$ of the upper endpoint
and distance $\delta_1$ of the lower endpoint.  Then the result becomes
$$
\eqalign{
S_{\rm FG}&\equiv -{\rm tr} \rho_{\rm in}\ln\rho_{\rm in}
={1\over 12}\left(u_{\rm in}(U_2-\delta_2)-u_{\rm in}(U_1+\delta_1)\right)\cr
&={1\over12}\ln\left({\sin \left({(U_2-U_1-\delta_2 ) \pi \over
2 \ L}\right)
                     \sin \left({(U_2-U_1-\delta_1)\pi \over2 \ L}\right)
                     \over \sin \left({\delta_1 \pi \over 2 \ L}\right)
                            \sin \left({\delta_2 \pi \over 2 \
L}\right)}\right)~,\cr}\eqno(FGinterval)
$$
This is our expression for the fine-grained entropy (due to right-movers only)
of the density matrix that is obtained by tracing over the field degrees of
freedom outside the interval $[U_1,U_2]$, in the Minkowski vacuum.  Note that
this expression is invariant if $U_2-U_1$ is replaced by $2L-(U_2-U_1)$; in
other words, we get the same entropy if we trace over the region outside the
interval as if we trace over the region inside.

If we choose $U_1=-L$ and $U_2=L$, then our interval is the whole (periodically
identified) box.  Thus the density matrix $\rho_{\rm in}$ becomes pure, and the
entropy should be zero.  We readily see that Eq.~\(FGinterval) has this
property.  We also note that $S_{\rm FG}$ has a finite limit as the size of the
box gets large; the entropy is infrared finite.  (But see below.)  If we take
the limit $L\to\infty$ with the size of the interval held fixed, we obtain
$$
S_{\rm FG}={1 \over 12}  \ln \left({(U_2-U_1)^2 \over \delta_1 \ \delta_2
}\right)~.  \eqno(entropylimit)
$$
Eq.~\(entropylimit) was first derived by
Holzhey [\cite{holzhey}]. Its curved space generalization will be used
repeatedly in this paper.

Eq.~\(FGinterval) has a simple interpretation.  It is just the sum of two
expressions of the form Eq.~\(halflineentropy), one associated with each
endpoint of the interval, and with the finite length of the interval acting as
an infrared cutoff.  However, there is an additional contribution to the
fine-grained entropy that we have not yet included---the contribution due to
the
$\omega=0$ mode, the mode that is constant in $[U_1,U_2]$.  This contribution
to the entropy is formally infinite, because the zero-frequency mode has an
infinite number of accessible quantum states.

If we were doing thermodynamics on the full line, rather than a finite
interval, we could argue that different values of the constant mode of the
field correspond to different superselection sectors of the quantum theory.
Then it would be appropriate to project out a particular value of the zero
mode, if we want to restrict our attention to one particular superselection
sector.  (Alternatively, we could impose boundary conditions, such as fixed end
or antiperiodic boundary conditions, that remove the zero mode.)  The infinite
zero-mode entropy is associated with the existence of an infinite number of
different superselection sectors, rather than an infinite contribution to the
entropy in any particular sector.

However, if we are considering the fine-grained entropy on a finite interval,
we do not have the option of projecting out the zero mode, or of removing it by
a particular choice of boundary conditions.  There are normalizable modes that
are constant in the interval $[U_1,U_2]$, and decay outside the interval.
These modes make a non-negligible contribution to the entanglement of the
fields inside and outside the interval.

It turns out that this additional term in the entropy will not be relevant to
our discussion of black hole thermodynamics.  But it is worthwhile to note that
this term can be easily estimated.  Suppose that we imagine using the
nonoverlapping wavepacket basis described following Eq.~\(halflineentropy).  In
Eq.~\(entropylimit), we have included the contributions to the entropy due to
wavepackets that are narrow compared to $U_2-U_1$, and that straddle either
the boundary at $U_1$ or the boundary at $U_2$.  What we are missing is the
contribution due to the wavepackets that are wide compared to $U_2-U_1$, and
that straddle the whole interval.

The essential insight is that these broad wavepackets produce a perfect
correlation between the value of the constant mode of the scalar field in the
interval $[U_1,U_2]$ with the entangled state of the long wavelength modes to
the left and right of the interval.  Thus, our calculation of the Rindler
entropy can be used to find the degree of entanglement of the constant mode in
the interval with the fields outside the interval.  The Minkowski vacuum state
has the form
$$
|0_M\rangle=|{\rm short}\rangle \otimes |{\rm long}\rangle \otimes |{\rm
uncorrelated}\rangle~,
\eqno(mink_mode_breakdown)
$$
where $|{\rm short}\rangle$ represents the product over entangled modes with
wavelength less than $U_2-U_1$, and $|{\rm long}\rangle$ is the product over
the
entangled modes with wavelength greater than $U_2-U_1$;  $|{\rm
uncorrelated}\rangle$ denotes the product over the modes that are well
localized either entirely inside the interval or entirely outside, and so do
not contribute to the entanglement.  Crudely speaking, the long-wavelength
entangled state has the form (up to normalization)
$$
|{\rm long}\rangle\sim\ \prod_{j=0}^{j_{\rm max}}\sum_{n_j}|n_j,R\rangle\otimes
|n_j,L\rangle\otimes |n_j, {\rm inside}\rangle~.
\eqno(long_entangled)
$$
Here the $j$th factor is the contribution due to a wavepacket mode of width
$2^j(U_2-U_1)$, centered at the interval, and $n_j$ labels the quantum state of
that mode.  The field fluctuations in this mode generate correlations between
the quantum state $|n_j,R\rangle$ of the portion of the wavepacket localized to
the right of the interval and the quantum state $|n_j,L\rangle$ of the portion
of the wavepacket that is localized to the left of the interval.  Furthermore,
these fluctuations are perfectly correlated with the quantum state $|n_j,{\rm
inside}\rangle$ of the constant mode inside the interval.
We see that tracing over the state of the constant mode inside the interval, to
obtain a
density matrix for the state outside, produces just the same density matrix as
if we traced over the left region to obtain a density matrix for the right
region. Thus, we can use the Rindler entropy formula Eq.~\(halflineentropy) to
estimate the long-wavelength contribution to the fine-grained entropy for a
finite interval, with the size of the interval playing the role of the
ultraviolet cutoff.  This contribution is
$$
S_{\rm FG,long}={1\over 12}\ln\left({U_{\rm max}\over U_2-U_1}\right)~.
\eqno(long_entropy)
$$
Now, we can find the total fine-grained entropy outside of an interval
of length $L$ on a slice of fixed time.  Combining the contributions of the
right-movers and left-movers, we obtain
$$
S_{\rm FG}={1\over 3}\ln\left({L\over \delta}\right)+ {1\over
6}\ln\left({L_{\rm max}\over L}\right)~,
\eqno(srednicki)
$$
where $\delta$ is the short-distance cutoff at both ends of the
interval,\footnote{*}{The distance $\delta$ is actually
$\left(\delta_R\delta_L\right)^{1/2}$, where $\delta_R$ and $\delta_L$ are
cutoffs for the right-movers and left-movers respectively.  It can be
interpreted as the invariant proper length over which the ends of the interval
are smoothed out on the time slice.} and $L_{\rm max}$ is an infrared cutoff.
The error
in Eq.~\(long_entropy) should be a (nonuniversal) constant of order one that
can be absorbed into $\delta$ in Eq.~\(srednicki). The result Eq.~\(srednicki)
agrees with a numerical calculation (for antiperiodic boundary conditions) that
was carried out by Srednicki [\cite{srednicki}].

\subhead{B. Curved spacetime}
So far, we have assumed that the state of the quantum field is the Minkowski
vacuum.  It is easy to extend the result to the case of a more general ``vacuum
state''  in flat spacetime.  Suppose that we introduce a new null coordinate
$\hat U (U)$, and define a vacuum relative to this new coordinate; that is, we
consider the state that contains no (right-moving) quanta that are positive
frequency with respect to the coordinate $\hat U$.  The same reasoning that we
used above for the Minkowski vacuum applies just as well to this case.  Thus,
if the size of the interval $[\hat U_1,\hat U_2]$ is small compared to the
infrared cutoff, the fine-grained entropy is again given by\footnote{*}{We are
again neglecting the (infrared sensitive) contribution due to the mode that is
constant in the interval.  The contribution of this mode to the entropy must be
considered separately.}
$$
S_{\rm FG}={1 \over 12}  \ln \left({(\hat U_2- \hat U_1)^2 \over \hat\delta_1 \
\hat\delta_2 }\right)~.  \eqno(hatentropy)
$$
The only new subtlety is that the short-distance cutoffs $\hat\delta_{1,2}$ are
here expressed in terms of the new $\hat U$ coordinate.  We can reexpress these
cutoffs in terms of the Minkowski (affine) distances $\delta_{1,2}$ using the
identities
$$
\hat\delta_1={\hat U}'_1\ \delta_1~,\quad \hat\delta_2={\hat U}'_2\ \delta_2~,
\eqno(cutoffscale)
$$
where the prime denotes a derivative with respect to $U$.  When the cutoff is
expressed in terms of the inertial coordinates, the entropy becomes
[\cite{holzhey}]
$$
S_{\rm FG}={1 \over 12}  \ln \left({(\hat U_2- \hat U_1)^2 \over {\hat U}'_1\
{\hat U}'_2\ \delta_1 \ \delta_2 }\right)~.  \eqno(hatentropyscaled)
$$

At this stage let us combine together the contributions to the entropy due to
the right-moving and left-moving modes.  Suppose that the left moving
``vacuum'' state is defined relative to the coordinate $\hat V(V)$.  We
consider a space-like slice $\Sigma$, and a region on this slice bounded on the
left by the point $({\hat U}_2,{\hat V}_2)$ and on the right by the point
$({\hat U}_1,{\hat V}_1)$, as shown in Fig.~3.  Tracing over the degrees of
freedom inside this region yields a total fine-grained entropy
$$
S_{\rm FG}={1 \over 12}  \ln \left({(\hat U_2- \hat U_1)^2 \over {\hat U}'_1\
{\hat U}'_2\ \delta_{1,R} \ \delta_{2,R}}\right) +
{1 \over 12}  \ln \left({(\hat V_2- \hat V_1)^2 \over {\hat V}'_1\ {\hat V}'_2\
\delta_{1,L} \ \delta_{2,L} }\right)~,  \eqno(leftandrightentropy)
$$
where, {\it e.g.}, $\delta_{1,R}$ denotes the short-distance cutoff, in
inertial coordinates, on the wavelength of the right-moving modes at endpoint
1.  By combining together the contributions of the right-movers and the
left-movers, we thus obtain an expression that is invariant under Lorentz
boosts, for the product $\delta_{R}\delta_L$ of the cutoffs on the right-moving
and left-moving modes is boost-invariant.  This quantity is just (the square
of) a proper length measured on the slice $\Sigma$.

\midinsert
\epsfysize=5in
\centerline{ \epsfbox{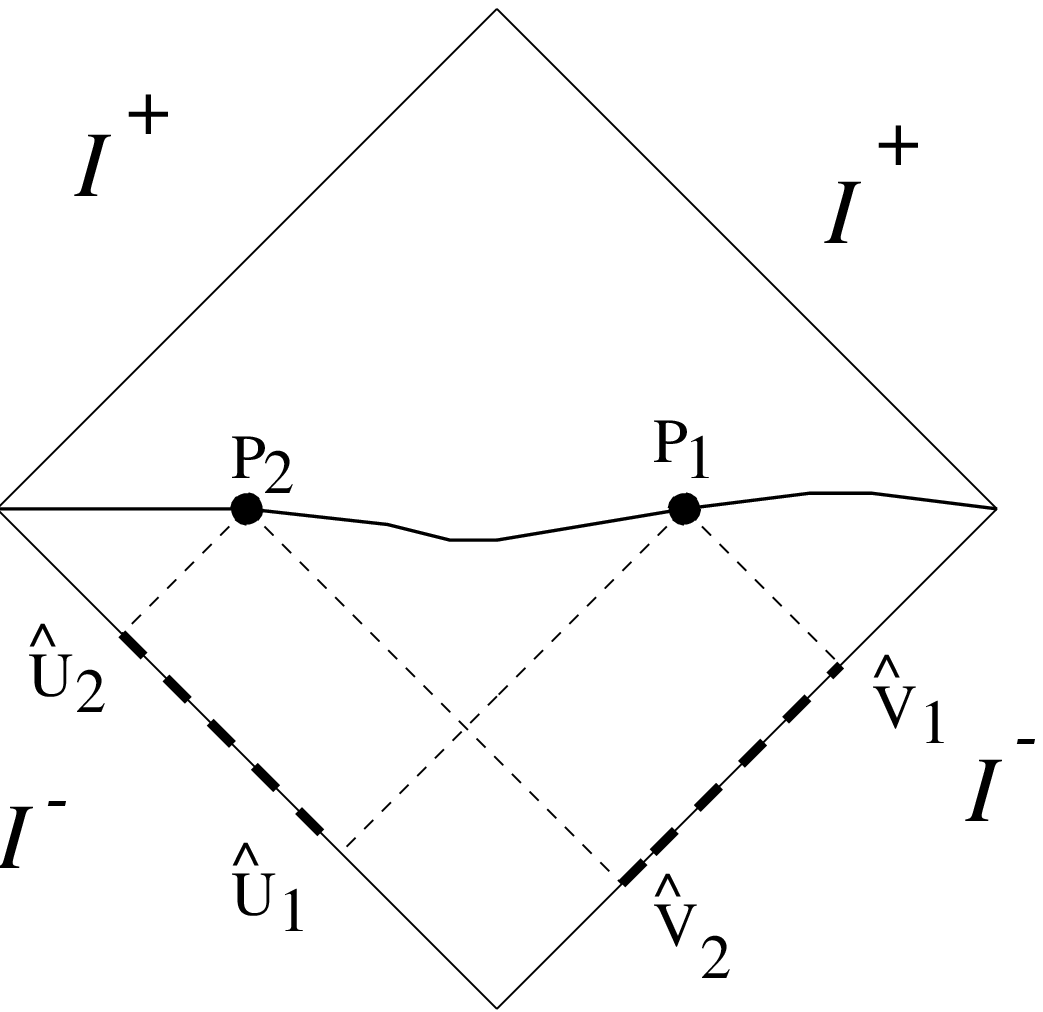}}
\bigskip
\centerline{FIGURE 3.}
\medskip
{\centerline{\vbox{\hsize 5in \singlespace\tenrm \noindent
 A spacelike slice through flat spacetime.  By tracing over
the field degrees of freedom on the portion of the slice in the region between
the points $P_1=(\hat U_1, \hat V_1)$ and $P_2=(\hat U_2,\hat V_2)$, we obtain
a density matrix $\rho_{\rm out}$ for the fields on the portion of the slice
outside that region.
 }}}

\endinsert

When expressed in terms of the new $(\hat U, \hat V)$ coordinates, the
Minkowski spacetime metric $ds^2=-dU\ dV$ becomes
$$
ds^2=-e^{2 \rho} \ d{\hat U} \ d{\hat V}~, \eqno(hatmetric)
$$
where
$$
e^{-2\rho}={\hat U}'\ {\hat V}'
\eqno(rhoUV)
$$
In terms of this metric, the expression Eq.~\(leftandrightentropy) for the
entropy becomes
$$
S_{\rm FG}={1\over 6}\left(\rho_1 +\rho_2\right) +
{1 \over 12}  \ln \left({(\hat U_2- \hat U_1)^2 \over  \delta_{1,R} \
\delta_{2,R}}\right) +
{1 \over 12}  \ln \left({(\hat V_2- \hat V_1)^2 \over  \delta_{1,L} \
\delta_{2,L} }\right)~.
\eqno(rhoentropy)
$$

This formula has the advantage that it can be applied to curved spacetime as
well.
In curved spacetime, there is no global inertial frame.  But we are free to
introduce coordinates $(\hat U,\hat V)$, and to consider the ``vacuum'' state
defined by these coordinates---the state that contains no quanta that are
positive frequency with respect to $\hat U$ and $\hat V$.  If the spacetime
metric has the form Eq.~\(hatmetric) in terms of these coordinates, then
Eq.~\(rhoentropy) gives the fine-grained entropy that results if we trace over
the field degrees of freedom contained in a finite interval of a spacelike
slice.  The cutoffs in Eq.~\(rhoentropy) are expressed in terms of the
locally flat coordinates $(U,V)$ at the endpoints of the interval, for
which the metric takes the form $ds^2=-dU\ dV$.  As noted above, the entropy is
unchanged by the local Lorentz transformations that preserve this metric.
Since our cutoff is in effect smeared over a region with width of order
$\delta$, it is implicit in Eq.~\(rhoentropy) that $\rho$ does not vary
appreciably over this region.

We should also remark that, for a given ``vacuum'' state, the coordinates
$(\hat U,\hat V)$ are not uniquely defined.  We have the freedom to perform an
$SL(2,{\bf C})$ transformation on the coordinates without changing the vacuum.
It is easy to check that Eq.~\(rhoentropy) is $SL(2,{\bf C})$-invariant.  As
expected, then, the conformal transformations that preserve the quantum state
of the fields also preserve our expression for the fine-grained entropy.

Finally, we note that our expression Eq.~\(halflineentropy) for the entropy on
the half line can also be easily generalized to curved spacetime.  Combining
the contributions of the right-movers and the left-movers, and expressing the
short-distance cutoffs $\delta_R,\delta_L$ in terms of locally inertial
coordinates at the boundary, we obtain
$$
S_{\rm FG}={1\over 6}\rho_P+{1\over 12}\ln\left(- \hat U_{\rm max}\hat V_{\rm
max}\over \delta_R\delta_L\right)~.
\eqno(curved_halfline_entropy)
$$
Here, again, the vacuum is defined with respect to the $(\hat U,\hat V)$
coordinates, and $\rho_P$ is the conformal factor in these coordinates at the
point $P$ that divides the space in half; $\hat U_{\rm max}$ and $\hat V_{\rm
max}$ are the infrared cutoffs.

\subhead{C. Moving mirror}

In a space without a boundary, the right-moving and left-moving modes of a free
massless scalar field are completely uncoupled, and the quantum states of the
right-movers and left-movers can be regarded as independent.  But if spacetime
has a reflecting boundary (as in the RST model) then correlations between the
right-moving and left-moving quantum states are induced.  These correlations
must be taken into account in the computation of the fine-grained entropy.

Suppose, then, that space is bounded on the left by a perfectly reflecting
mirror, as shown in Fig.~4.  We suppose that the mirror moves on some timelike
trajectory.  Then we can express the quantum state of the field as a
left-moving state at ${\cal I}^-$ (since there are no right-movers at ${\cal
I}^-$).  In particular, we can introduce a null coordinate $\hat V$, and
consider the ``vacuum'' state defined on ${\cal I}^-$ in terms of the $\hat V$
coordinate.  Then we may define a $\hat U$ coordinate by demanding $\hat U=\hat
V$ at the boundary, the position of the mirror.

\midinsert
\epsfysize=5in
\centerline{ \epsfbox{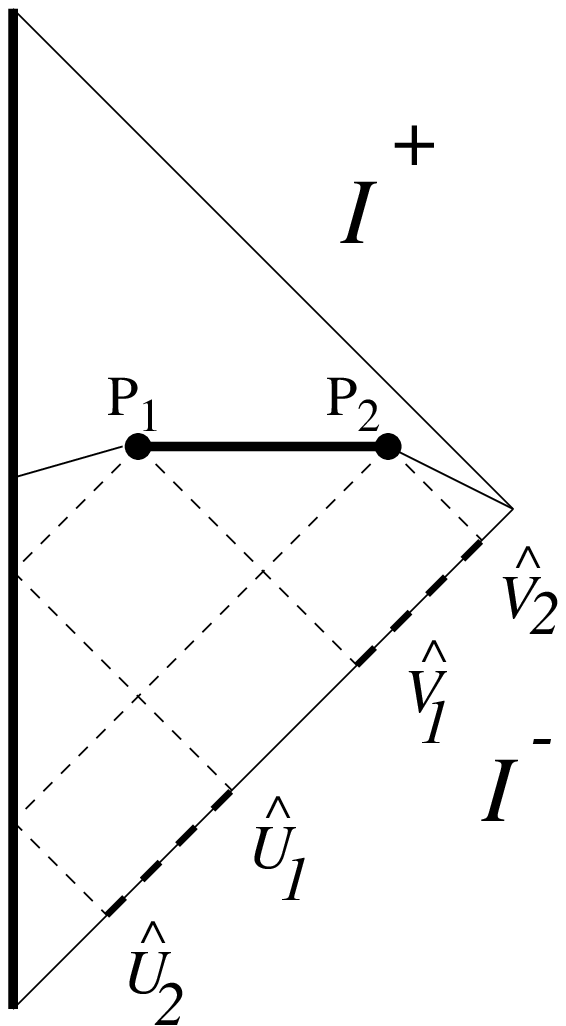}}
\bigskip
\centerline{FIGURE 4.}
\medskip
{\centerline{\vbox{\hsize 5in \singlespace\tenrm \noindent
A spacelike slice through the moving mirror spacetime.
Coordinates have been chosen so that the trajectory of the mirror is $\hat
V(\hat U)=\hat U$. By tracing over the field degrees of freedom on the portion
of the slice in the region between the points $P_1=(\hat U_1, \hat V_1)$ and
$P_2=(\hat U_2,\hat V_2)$, we obtain a density matrix $\rho_{\rm out}$ for the
fields on the portion of the slice outside that region.
 }}}

\endinsert

Now consider a spacelike slice $\Sigma$, and an interval on the slice bounded
by a point $P_1$ with coordinates
$({\hat U}_1,{\hat V}_1)$ and a point $P_2$
with coordinates $({\hat U}_2,{\hat V}_2)$.  As a warm-up for our analysis of
black holes (where the interval will correspond to the black hole interior), we
would like to trace over the
field degrees of freedom inside this interval, and obtain a density matrix for
the state on the slice outside the interval.  The right-moving and left-moving
modes in the interval are correlated.  In fact, as Fig.~4 shows, the
right-moving modes in the interval are the same as the left-moving modes on
${\cal I}^-$, in an interval ${\hat U}_2<\hat V< {\hat U}_1$.  Thus, tracing
over the left-movers and right-movers inside the interval bounded by $P_1$ and
$P_2$
on the slice $\Sigma$ is (almost) equivalent to tracing over the left-movers
only on ${\cal I}^-$, in the {\it union} of the two intervals ${\hat U}_2<\hat
V< {\hat U}_1$ and ${\hat V}_1<\hat V< {\hat V}_2$.  (But see the caveat
below.)

\midinsert
\epsfysize=5in
\centerline{ \epsfbox{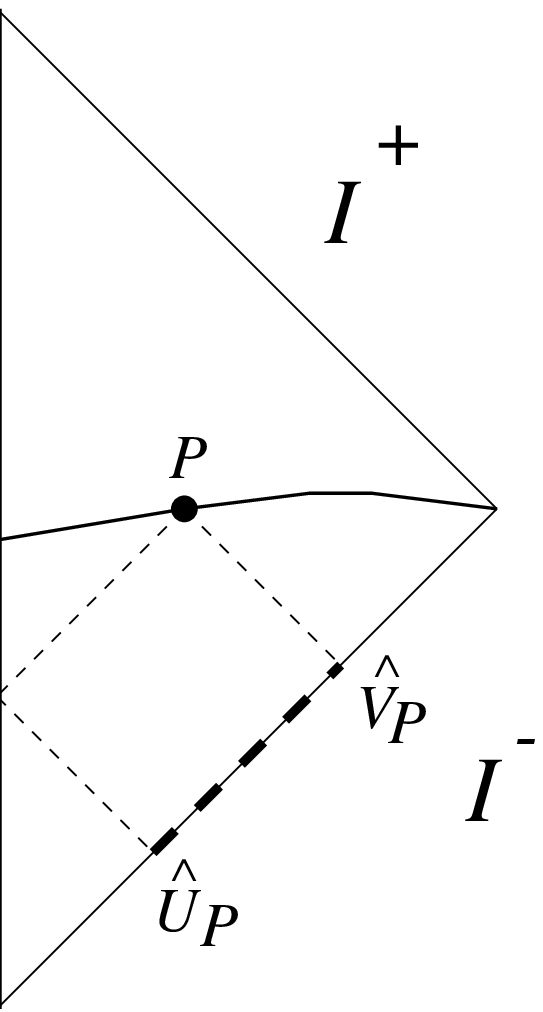}}
\bigskip
\centerline{FIGURE 5 (a).}
\medskip
{\centerline{\vbox{\hsize 5in \singlespace\tenrm \noindent
A spacelike slice through the moving mirror
spacetime.  Coordinates have been chosen so that the trajectory of the mirror
is $\hat V(\hat U)=\hat U$.  By tracing over the field degrees of freedom on
the portion of the slice between the point $P=(\hat U_P,\hat V_P)$ and the
mirror, we obtain a density matrix $\rho_{\rm out}$ for the fields on the
portion of the slice to the right of the point $P$.
 }}}

\endinsert

Tracing over the field degrees of freedom in a union of two disjoint intervals
is a bit complicated, but a simpler problem turns out to be adequate for our
purposes.  We consider a point $P$ with coordinates $({\hat U}_P, {\hat V}_P)$
on the slice $\Sigma$, and we trace over the field degrees of freedom on
$\Sigma$ between $P$ and the mirror.  As shown in Fig.~5a, this is (almost)
equivalent to tracing over the interval ${\hat U}_P<\hat V< {\hat V}_P$ on
${\cal I}^-$ (recalling that the $\hat U$ coordinate is defined by the
condition that $\hat U=\hat V$ at the boundary).  We may now appeal to
Eq.~\(entropylimit) to conclude that
$$
S_{\rm FG}={1\over 12}\ln\left({\left({\hat V}_P-{\hat U}_P\right)^2\over
\hat\delta_R\ \hat\delta_L}\right)~.
\eqno(mirrorentropy)
$$
Here $\hat\delta_R$ is the short-distance cutoff on the right-moving modes at
the point $P$, expressed in $\hat U$ coordinates; because of the way the $\hat
U$ coordinate has been defined, this is the same as the cutoff at $\hat V=\hat
U_P$ on ${\cal I}^-$, expressed in terms of $\hat V$ coordinates. In terms of
cutoffs $\delta_{R,L}$ expressed in terms of the vacuum coordinates
at the point $P$, Eq.~\(mirrorentropy) becomes
$$
S_{\rm FG}={1\over 6} \rho_P +{1\over 12}\ln\left({\left({\hat V}_P-{\hat
U}_P\right)^2\over \delta_R\ \delta_L}\right)~,
\eqno(mirrorentropyrho_first)
$$
where $\rho_P$ is the conformal factor of the metric Eq.~\(hatmetric) at the
point $P$.  The same derivation will of course apply if we choose the $\hat U$
coordinate so that $\hat V-\hat U$ is a nonzero constant, except that we will
now have
$$
S_{\rm FG}={1\over 6} \rho_P +{1\over 12}\ln\left({\left({\hat V}_P-{\hat
V}_B\right)^2\over \delta_R\ \delta_L}\right)~;
\eqno(mirrorentropyrho)
$$
here $\hat V_B$ is defined as the value of $\hat V$ at the point on the
boundary that is contained in a null line through $P$, as shown in Fig.~5b.

\midinsert
\epsfysize=5in
\centerline{ \epsfbox{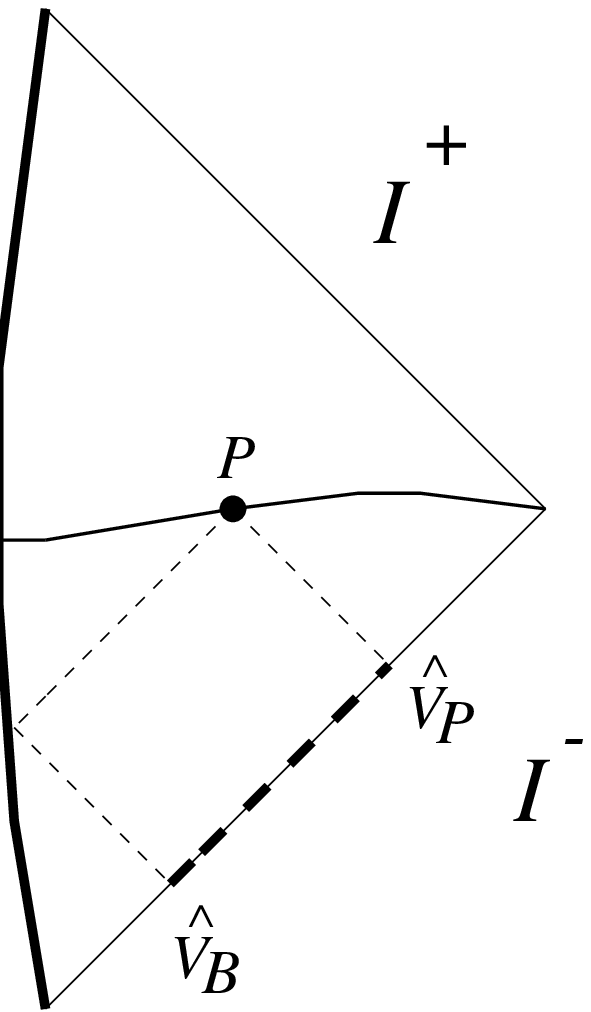}}
\bigskip
\centerline{FIGURE 5 (b).}
\medskip
{\centerline{\vbox{\hsize 5in \singlespace\tenrm \noindent
 If coordinates
are not chosen so that $\hat V=\hat U$ at the mirror, we define $V_B$ as the
retarded time of an incoming null ray that reflects off the mirror and then
passes through $P$.
 }}}

\endinsert

If we had imposed Neumann boundary conditions at the mirror, the model would be
equivalent to a model with left-movers only and no boundary.  Then
Eq.~\(mirrorentropy) would be the exact expression for the entropy due to the
modes that are not constant on the interval between the point $P$ and the
mirror.  In addition, there would be an infrared divergent contribution to the
entropy of the form Eq.~\(long_entropy), arising from modes that are constant
between $P$ and the mirror, and decay outside of $P$.  The situation with
Dirichlet boundary conditions is a bit different.  The condition that the
fields vanish at the mirror removes the mode that is constant behind $P$, and
as a result the entropy is infrared finite.  To understand why the entropy of
the left-movers defined at ${\cal I}^-$ is not exactly the same as the entropy
of the left-movers and right-movers on the spacelike slice, consider two
nonoverlapping wavepacket modes at ${\cal I}^-$, both localized
inside the interval $[\hat V_B,\hat V_P]$, and both entangled with modes
outside.  Suppose that one
of these wavepackets reflects from the mirror prior to the slice $\Sigma$, and
that the two wavepackets then interfere destructively on $\Sigma$.  Thus,
although the two modes are entangled with the fields outside the interval at
${\cal I}^-$, their coherent sum (namely zero) is not entangled with the fields
on $\Sigma$ outside of the point $P$.

While this error is quite small for the modes with wavelength much less than
the
width of the interval, it is significant for modes of long
wavelength.  However, on dimensional grounds, the total error in our estimate
of the entropy is a
constant of order one.  (The error is dimensionless, and does not depend on the
ultraviolet or infrared cutoffs.)  This constant can be absorbed into the
ultraviolet cutoff in Eq.~\(mirrorentropy), \(mirrorentropyrho_first), and
\(mirrorentropyrho).

Eq.~\(mirrorentropyrho) is our main result for the fine-grained entropy in the
moving mirror spacetime.  To summarize, the quantum state is the ``vacuum''
defined with respect to the $\hat V$ coordinate on ${\cal I}^-$, and $S_{\rm
FG}$ is the entropy of the density matrix that is obtained by tracing over the
field degrees of freedom on a spacelike interval between the point $P$ and the
mirror.  The $\delta_{R,L}$ are the cutoff wavelengths for left and right
movers at the point $P$, expressed in terms of the locally inertial coordinates
$U,V$ (such that the metric at $P$ has the form $ds^2=-dU\ dV$);  $\rho_P$ is
the value of the conformal factor at the point $P$ for the metric
$ds^2=-e^{2\rho}d\hat U\ d\hat V$, where the $\hat U$ coordinate is defined by
the condition $\hat V-\hat U={\rm constant}$ at the mirror.  (It is also
assumed that $\rho$ can be regarded as constant over a region with width
comparable to the cutoff length scale.) We recall that the product $\delta_R\
\delta_L$ (a proper length squared on the spacelike slice) is invariant under
local Lorentz boosts, and that $S_{\rm FG}$ is
unchanged by the $SL(2,{\bf C})$ transformations that modify the $\hat U$
coordinate without altering the vacuum state.  We also emphasize again that
this formula for $S_{\rm FG}$ applies in curved two-dimensional spacetime, as
well as in flat spacetime.

\subhead{D. Black hole}

The application of Eq.~\(mirrorentropyrho) to the RST model is immediate.  In
the spacetime of a black hole that forms due to infalling matter, there is a
timelike boundary, up until the formation of the spacelike singularity.  We
consider a spacelike slice $\Sigma$ (as in Fig.~6) that passes through the
apparent horizon at
the point $P$, and meets the timelike boundary behind the horizon.  Let the
quantum state be the vacuum defined by the coordinate $\sigma^+$---this is the
state that appears to contain no quanta to the inertial observers at ${\cal
I}^-$.  Construct a density matrix $\rho_{\rm out}$ outside the apparent
horizon by tracing over the field degrees of freedom behind the apparent
horizon.  Recalling that the RST model contains $N$ species of free massless
scalar field, Eq.~\(mirrorentropyrho) becomes
$$
S_{\rm FG}\equiv -{\rm tr}\left(\rho_{\rm out}\ \ln\rho_{\rm out}\right)
={N\over 6}\left(\rho_{H,\sigma}+\ln\left({\sigma^+_H-\sigma^+_B\over
\delta}\right)\right)~,
\eqno(RSTFG)
$$
where $\rho_{H,\sigma}=\rho(\sigma^-_H,\sigma^+_H)$ is the conformal factor of
the metric (in $\sigma$ coordinates) at the point $P$ where the slice crosses
the apparent horizon.  Here $\sigma^+_B$ is the value of $\sigma^+$ at the
point where the null line through $P$ meets the boundary, as indicated in
Fig.~6.  The cutoff $\delta$ is the proper length $(\delta_R\delta_L)^{1/2}$;
alternatively, we
may choose the local Lorentz frame at $P$ so that
$\delta_R=\delta_L\equiv\delta$.

\midinsert
\epsfysize=5in
\centerline{ \epsfbox{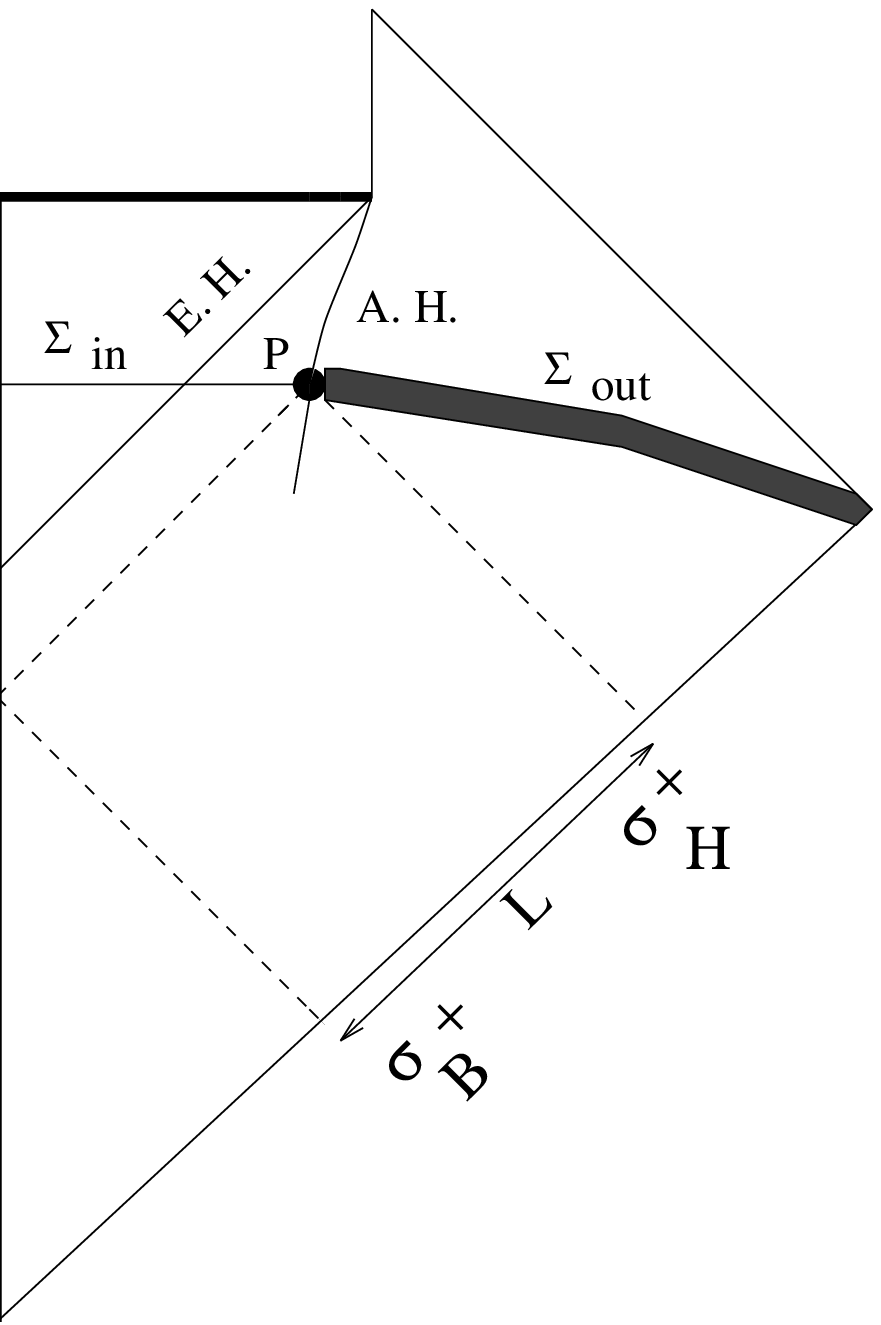}}
\bigskip
\centerline{FIGURE 6.}
\medskip
{\centerline{\vbox{\hsize 5in \singlespace\tenrm \noindent
 A spacelike slice $\Sigma$ through the black hole spacetime.
The slice crosses the apparent horizon at the point
$P=(\sigma_H^-,\sigma_H^+)$.  We define $\sigma_B^+$ as the retarded time of an
incoming null ray that reflects off the boundary and then passes through $P$.
Incoming null rays with retarded time between $\sigma_B^+$ and $\sigma_H^+$
cross $\Sigma$ inside the apparent horizon.
 }}}

\endinsert

Note that, in defining the conformal factor $\rho$ in Eq.~\(RSTFG), we have
implicitly used a $\sigma^-$ coordinate that satisfies
$$
\sigma^- = \sigma^+ + {\rm constant}\eqno(coordboundary)
$$
on the timelike boundary.  This $\sigma^-$ coordinate does not necessarily
coincide with the $\sigma^-$ coordinate that is defined in terms of  the
Kruskal coordinate $x^-$ by Eq.~\(sigma_define).  However, we saw in
Eq.~\(ldv_boundary) that, in the linear dilaton vacuum, the $\sigma$
coordinates defined by Eq.~\(sigma_define) satisfy
$$
\lambda\left(\sigma_B^+ -\sigma_B^-\right) = -2\ln 2
\eqno(ldv_sigma_boundary)
$$
at the boundary;  thus, Eq.~\(coordboundary) is satisfied, and the two
definitions of $\sigma^-$ {\it do} agree.  The same holds true if no infalling
matter has reached the boundary before the retarded time $\sigma^+=\sigma^+_B$.
 In our analysis of the thermodynamics of a black hole formed from collapse, we
will find it convenient to assume that this condition holds, so that the
Eq.~\(coordboundary) and Eq.~\(sigma_define) are both valid.

Under this assumption, we can re-express Eq.~\(RSTFG) in terms of the value
$\phi_H$ of the dilaton field at the apparent horizon.  First, we see from
Eq.~\(sigma_define) that the conformal factor $\rho_\sigma$ in $\sigma$ gauge
is related to the conformal factor $\rho_K$ in Kruskal gauge by
$$
ds^2=-e^{\left(2\rho_{\sigma}\right)}d\sigma^+
d\sigma^-=-e^{\left(2\rho_K\right)}dx^+dx^-
=-e^{\left(2\rho_K\right)}e^{\lambda(\sigma^+-\sigma^-)}d\sigma^+d\sigma^-~,
\eqno(Kandsigma)
$$
or
$$
\rho_\sigma=\rho_K + {\lambda\over 2}\left(\sigma^+-\sigma^-\right)~.
\eqno(rho_sigma_and_kruskal)
$$
For the point on the boundary with the same advanced time as the apparent
horizon (as in Fig.~6), we have $\sigma_B^-=\sigma_H^-$; thus, combining
Eq.~\(ldv_sigma_boundary), \(rphi), and \(boundary), we find that the value
$\rho_{H,\sigma}$ of the conformal factor at the apparent horizon, in $\sigma$
gauge, is
$$
\rho_{H,\sigma}=\phi_H -\phi_{\rm cr} + {1\over 2}\lambda\left(\sigma_H^+ -
\sigma_B^+\right)~.
\eqno(horizon_rho_from_phi)
$$
Our expression for the fine-grained entropy outside the apparent horizon then
becomes
$$
S_{\rm FG}={N\over 6}\left(\phi_H-\phi_{\rm cr}+{1\over 2}\lambda L + \ln
{L\over\delta}\right)~,
\eqno(entropy_phi)
$$
where we have defined
$$
L=\sigma^+_H -\sigma_B^+~.
\eqno(first_L_define)
$$
Roughly speaking, L is the affine volume (in $\sigma$ coordinates) behind the
horizon at retarded time
$\sigma_H^+$ (as shown in Fig.~6).

We derived Eq.~\(RSTFG) and \(entropy_phi) under the assumption that the
quantum state at ${\cal I}^-$ is the inertial vacuum.  However, we will show in
Appendix A that Eq.~\(RSTFG) and \(entropy_phi) still hold if the incoming
state is a coherent state built on this vacuum.  (Coherent states are
a  natural basis to use in the present context
because, in the large-$N$ limit, they are orthogonal
and have a simple evolution law.)
{\it If} we assume that the infalling matter is in a coherent state of this
type, and that no incoming matter reaches the boundary prior to the global
horizon, then Eq.~\(entropy_phi) is the correct expression  for the
fine-grained entropy of the matter fields outside the apparent horizon of the
black hole.

\head{IV. Evaporation and information}

When a black hole forms from collapsing matter, some of the information about
the initial quantum state of the matter becomes encoded in the correlations
of the quantum fields outside the horizon with the fields inside the horizon.
This information remains inaccessible to an observer who remains outside the
horizon at all times.  Our expression Eq.~\(RSTFG) for the fine-grained entropy
quantifies the amount of this missing information.  Thus, by studying the
behavior of $S_{\rm FG}$ as the black hole evolves, we can track the
information content of the Hawking radiation that is emitted.

The simplest case to consider is that in which the black hole remains
``critically illuminated'' for a long time.  That is, we imagine that the
incoming energy flux ${\cal E}(\sigma^+)$ matches the outgoing thermal flux
${\cal E}_{\rm cr}={1\over 4}\lambda^2$ due to the Hawking radiation.  During
the period of critical illumination, the black hole mass, and the value
$\phi_H$ of the dilaton field at the horizon, remain unchanged.
If the quantum state of the infalling matter is a coherent state built on the
asymptotic inertial vacuum, we may then use Eq.~\(entropy_phi)
to find the change in the fine-grained entropy of the matter fields outside the
horizon during the process; it is
$$
\Delta  S_{\rm FG}={N\over 6}\left({\lambda\over 2}\left(L_f-L_i\right) +\ln
{L_f\over L_i}\right)~,
\eqno(critical_entropy_change)
$$
where $L_i$ and $L_f$ denote the values of $L$ at the beginning and end of the
critical illumination.  (Note that, though our expression for the fine-grained
entropy depends on a short-distance cutoff, this entropy {\it change} is cutoff
independent.)  During critical illumination, the horizon is null and
$\sigma^+_B$ is fixed, so that
$dL/d\sigma_H^+=1$.  It is clear then, that if the critical illumination lasts
long enough, the increase in the fine-grained entropy may be as large as
desired.
We conclude that there is no limit to the amount of information
that can be destroyed by the black hole, or in other words, no limit to the
degree of entanglement of the fields outside the global horizon with those
inside.  It was argued in Ref.~[\cite{stromtriv}] that an arbitrary amount of
information can be stored on a slice inside the horizon of a black hole.
Eq.~\(critical_entropy_change) is the other side of the coin---there is no
limit to the
amount of information that can be {\it missing} from the region outside the
horizon.

Because the fine-grained entropy can increase without bound, while the black
hole mass remains fixed, it is not possible to attribute the fine-grained
entropy to the entanglement of the degrees of freedom outside the black hole
with a {\it finite} number of internal degrees of freedom of the black hole.
Unless there is a stable black hole remnant with an {\it infinite} number of
degrees of freedom [\cite{remnant}], information is unavoidably lost.

It is useful to recall the origin of the two terms in
Eq.~\(critical_entropy_change)
by referring to Eq.~\(RSTFG).  The value of $\rho$ at the apparent horizon in
sigma gauge is related to the dilaton field $\phi$ by
Eq.~\(horizon_rho_from_phi), or
$$
\rho_{H,\sigma}=\phi_H-\phi_{\rm cr}+{1\over 2}\lambda L~;
\eqno(rho_phi_L)
$$
the first term in Eq.~\(critical_entropy_change) is just
$(N/6)(\rho_{H,f}-\rho_{H,i})$.  Eq.~\(rho_phi_L) expresses the
familiar property that the field modes that cling near to the horizon for a
long while undergo an exponential redshift.  We recall that the cutoff $\delta$
is a fixed proper length at the apparent horizon.  This means that the cutoff
in $\sigma$ coordinates at the apparent horizon is shrinking
exponentially, according to
$$
\delta_\sigma^2 = e^{-2\rho_\sigma}\delta^2\sim e^{-\lambda L}~.
\eqno(cutoff_shrink)
$$
(In the second equality we have neglected the correction in Eq.~\(rho_phi_L)
due to the evolution of $\phi$.)  Since the $\sigma$ coordinates are the
inertial coordinates on ${\cal I}^-$, Eq.~\(cutoff_shrink) says that, as the
black hole evolves, shorter and shorter wavelength incoming
modes, as measured on ${\cal I}^-$, are
being included in the calculation of the fine-grained entropy.  It is the {\it
very-short-distance} correlations between these modes just inside and just
outside the horizon that are responsible for the dominant contribution to the
entropy in Eq.~\(critical_entropy_change).  The subdominant second term in
Eq.~\(critical_entropy_change) arises from the {\it long-distance} correlations
between
field modes inside and outside the horizon.

It may be appropriate to be somewhat more explicit about the connection between
the cutoff expressed in $\sigma$ coordinates and wavelengths measured at ${\cal
I}^-$.  In our analysis of the fine-grained entropy in Section III, we really
imposed two cutoffs, one on left-moving modes and one on right-moving modes.
In the case of the black hole background, these can both be expressed in terms
of the $\sigma^+$ coordinate---we see from Fig.~6 that there is a cutoff on
left-movers in the vicinity of $\sigma_B^+$, and another cutoff on left-movers
at $\sigma_H^+$.  Let us denote these two cutoffs by $\delta\sigma_H^-$ and
$\delta\sigma_H^+$.  (Recall that $\sigma_H^-=\sigma_B^+$  + constant.)
Individually, the two cutoffs have no invariant
significance; it is only their product
$\delta\sigma_H^-\delta\sigma_H^+=\delta_\sigma^2$ that is determined by
Eq.~\(cutoff_shrink).

The individual cutoffs $\delta\sigma_H^-$ and $\delta\sigma_H^+$ depend on how
we choose our time slices.  However, there is a natural way to foliate the
spacetime with spacelike slices.  We fix a position far from the black hole, by
specifying a value of the dilaton field $\phi$.  A family of observers, with
their clocks initially synchronized, fall freely toward the black hole from
this fixed position at regular intervals.  The natural time slices are those on
which all observers record the same proper time.\footnote{*}{This foliation
might not be globally defined on a general spacetime.  However, for our
purposes it is sufficient to define time slices locally in the vicinity of a
particular point on the apparent horizon.  Also, we note that the time slices
defined by the family of freely falling observers are {\it not} the same as the
slices of constant ``$\sigma$-time''
$\sigma^0={1\over 2}(\sigma^+ +\sigma^-)$. On the $\sigma^0$ time slices, we
have $\delta\sigma_H^-=\delta\sigma_H^+\propto e^{-\rho_{H,\sigma}}$.}  With
this choice, the cutoff
$\delta\sigma_H^+$ remains essentially constant along the apparent horizon, so
that the other cutoff shrinks according to
$$
\delta\sigma_H^-\propto e^{-2\rho_{H,\sigma}}~.
\eqno(sigma_shrink)
$$
Thus, as the black hole evolves, shorter and shorter wavelengths modes, as
measured on ${\cal I}^-$ near $\sigma^+=\sigma^+_B$, are being included in the
calculation of the fine-grained entropy . It is the very-short-distance
correlations between the modes localized just inside and just outside the
horizon that are responsible for the increase in the entropy.

Though there is a sense in which the dominant contribution to the entropy can
be
attributed to very-short-distance correlations, it is not correct to
say that the entropy can be very well localized near the horizon. Since the
cutoff is a fixed proper length at the horizon, an observer in the vicinity of
the horizon would conclude that ultra-short-distance modes (with wavelength
much less than $\delta$) make no contribution to the entropy.  It is only when
these modes are followed backwards to ${\cal I}^-$, where they are enormously
blueshifted, that ultra-short-distances need be considered.  In fact, on a
spacelike slice, most of the fine-grained entropy is due to the entanglement of
 fields far outside the horizon with fields that are far inside.

How secure is our conclusion that information is lost in the RST model?  One
potential worry is that it is a subtle task to control the fine-grained entropy
in a semiclassical calculation [\cite{page2}].  We have attempted to do so by
appealing to the $1/N$ expansion, so that we can neglect the quantum
fluctuations about the background geometry.  However, expanding the entropy in
powers of $\hbar$ (as we are attempting to do here\footnote{*}{Because $N\hbar$
is of order one, corrections higher order in $1/N$ are equivalent to
corrections higher order in $\hbar$.}) can be a tricky business, since the
$\hbar\to 0$ limit of the entropy may be highly singular.  For
example, knowing each matrix element of the density matrix $\rho_{\rm
out}$ to leading order in $1/N$ may not be sufficient to determine $S_{\rm
FG}$ to leading order, since the {\it size} of the matrix grows as
$N\to\infty$.  We believe, though, that this criticism does not apply to our
calculations.  We have derived an expression for the $S_{\rm FG}$ itself,
rather than the matrix elements of $\rho$, that is valid to leading order in
$1/N$.

A second worry [\cite{hooft,verl}] arises due to the extreme redshifting of the
field modes that are responsible for the emitted Hawking radiation.  If
information is {\it not} lost, then the fine-grained entropy of the Hawking
radiation can be attributed to entanglement with the internal degrees of
freedom of the black hole.  The number of internal degrees of freedom would
presumably be given by the Bekenstein number $e^{S_{\rm BH}}$.  Therefore, to
argue persuasively that information is lost, we must follow the evaporation of
the black hole long enough so that the increase of $S_{\rm FG}$ exceeds
$$
S_{\rm BH}={M_{\rm BH}\over T_{\rm BH}}= {2\pi M_{\rm BH}\over \lambda}~.
\eqno(bh_entropy_lambda)
$$
We thus require
$$
\lambda\left(L_f-L_i\right)\sim {24\pi M_{\rm BH}\over N\lambda}~.
\eqno(delta_lambda_info)
$$
(neglecting the logarithmic term in Eq.~\(critical_entropy_change)).
It follows that the quanta that are emitted during the late stages of the
critical illumination process are in modes that have been redshifted (relative
to their frequency at ${\cal I}^-$) by the factor
$$
e^{2(\rho_{H,f}-\rho_{H,i})}=\exp \left({24\pi M_{\rm BH}\over
N\lambda}\right)~.
\eqno(critical_redshift)
$$
In the RST model, it is understood that the incoming and outgoing energy
fluxes, and the mass of the black hole, are all quantities of order $N$.  Thus,
the argument of the exponential in Eq.~\(critical_redshift) is formally of
order one in the large-$N$ limit.  Still, $24\pi M_{\rm BH}/ N\lambda$ should
be large in
a well-controlled semiclassical calculation, so that this redshift factor is
truly enormous.
Because the Hawking radiation is being emitted in modes that have {\it very}
large energy as measured at ${\cal I}^-$, one may wonder whether there are
correspondingly large {\it fluctuations} in energy-momentum.  If so, the
response of the geometry to these fluctuations should be included when the
evolution of quantum states is studied.

In fact, we are not aware of any calculation that convincingly demonstrates
that these large fluctuations occur in the RST model, or that they precipitate
a breakdown of semiclassical methods.  At any rate, even if they do occur,
their effects are systematically suppressed in the $1/N$ expansion.  We can
always justify neglecting the response of the geometry to the fluctuations of a
mode that is blueshifted by the factor Eq.~\(critical_redshift), by allowing
$N$ to be sufficiently large.  For example, suppose we want the energy measured
at ${\cal I}^-$ of a typical mode to be less than some small fraction
$\epsilon$ of the mass of the black hole.  The typical quantum emitted in the
Hawking radiation has an energy of order $\lambda$, so that the energy measured
at ${\cal I}^-$ is of order $\lambda$ times the blueshift factor.  This
blueshifted energy is less than $\epsilon M_{\rm BH}$ provided that
$$
N>{1\over \epsilon}\left({N\lambda\over M_{\rm BH}}\right)\exp\left({24\pi
M_{\rm BH}\over N\lambda}\right)~.
\eqno(required_N)
$$
Since $M_{\rm BH}/N\lambda$ is a quantity of order one, this condition is
satisfied for $N$ sufficiently large (although the required value of $N$ grows
exponentially with the mass of the black hole).

While we believe that the above technical objections can be answered, our
discussion of ``loss of information'' in black hole evaporation should still
include some important caveats.  We can follow the evolution of a black
hole\footnote{*}{The case of (nearly) complete evaporation, as opposed to
critical illumination, will be further discussed in Section VI.C.} far enough
to exclude the scenario described by Page [\cite{page2}], in which the
fine-grained entropy begins to decrease sharply after about half of the mass
has been radiated away.  But we cannot follow the evolution all the way up to
the endpoint of the evaporation process (without additional assumptions about
the behavior of Planckian black holes).  It remains a logical possibility,
therefore, that the ``lost'' information is finally recovered in the very late
stages of the process, when the large-$N$ approximation breaks down.  (General
arguments [\cite{carlitz,jp}] indicate that, in this event, the final stage
would have to take an exceedingly long time.)

We also note that implicit assumptions have been made
about how physics in our toy model behaves under {\it extreme} boosts, and
these
assumptions might not be appropriate in the real world.
We remark again that, since the redshift factor $\exp(24\pi M_{\rm
BH}/N\lambda)$ is very large, the fine-grained entropy that we have computed is
dominated by the contributions due to field modes that are of extraordinarily
short wavelength on ${\cal I}^-$.  As has been emphasized by 't Hooft
[\cite{hooft}], Jacobson [\cite{jacobson}], Susskind [\cite{susskind}], and the
Verlindes [\cite{verl}], loss of information could conceivably be avoided if
ordinary relativistic field theory ceases to apply at sufficiently short
distances, so that our calculation of the fine-grained entropy is invalidated.
While loss of information appears to occur in the RST model (for sufficiently
large $N$), it might not occur
in a different model with different short-distance physics.

A related point is that we have made an assumption
about the nature of the cutoff that arises in the definition of the entropy.
This cutoff can be regarded as the proper length over which we have smeared the
boundary between the region inside the black hole and the region outside.  Our
procedure has been to keep this proper length fixed as the black hole evolves.
This procedure is the only reasonable one we could think of, but if some
justification could be found for varying the cutoff along the
horizon, our conclusions would be altered.

\head{V. Black hole entropy}

In Section III, we derived  an expression for the (fine-grained) entropy of the
matter fields
outside the apparent horizon of a black hole.  To do black hole thermodynamics,
we will also need an expression for the intrinsic entropy of the black hole.
In the leading semiclassical approximation (neglecting all gravitational back
reaction) it is easy to find the black hole entropy.  But in our analysis of
the RST model, back reaction effects of order $N\hbar$ are included, and we
will need to include a next-to-leading correction to the black hole entropy.
In this Section, we will derive this correction.

The leading semiclassical expression for the black hole entropy can be obtained
using thermodynamic reasoning, given the relation between the black hole mass
and the temperature of the Hawking radiation.  If we imagine that the black
hole is in equilibrium with a thermal radiation bath in a (small) cavity, we
may regard a process in which the black hole accretes or emits an infinitesimal
amount of radiation as a reversible thermodynamic process.  Integrating the
identity $dS=dM/T$ then determines the black hole entropy up to an additive
constant.

For the black hole in two-dimensional dilaton gravity (and the four-dimensional
magnetically-charged dilaton black hole to which it is intimately related), the
temperature $T_{\rm BH}=\lambda/2\pi$ is independent of its mass.  Because the
specific heat of the black hole is actually infinite, there are very large
fluctuations in thermal equilibrium; the black hole mass wanders randomly
[\cite{susskind2,rutgers}].  However, in the large $N$ limit, these
fluctuations are suppressed, and may be ignored.  (The characteristic {\it time
scale} of the fluctuations increases as $\sqrt{N}$ as $N$ increases.)  Thus,
the naive thermodynamic arguments are valid.  The leading expression for the
entropy becomes
$$
S_{\rm BH}=M_{\rm BH}/T_{\rm BH}=2e^{-2\phi_H}~,\eqno(bhleading)
$$
where $\phi_H$ denotes the value of the dilaton field $\phi$ at the apparent
horizon.

To go beyond this leading calculation, we wish to find the correction to the
relation between the $M_{\rm BH}$  and $\phi_H$, for a black hole in contact
with a radiation bath.  However, it is not even clear how to define $M_{\rm
BH}$ for a black hole surrounded by radiation---the ADM mass, for example,
includes both a contribution from the black hole and a contribution from the
bath.  We will therefore proceed in two steps.  For a black hole surrounded by
radiation in a (finite) cavity, we imagine adiabatically introducing a small
amount of additional left-moving matter, which eventually crosses the apparent
horizon and is accreted by the black hole.  The first step is to find how the
accretion process changes the value of $\phi_H$ (or equivalently $\Omega_H$).
Using thermodynamics, we can then find the relation between the change in
$\phi_H$ and the change in the total entropy contained in the cavity.

This first step is not quite the whole story, though, because the total entropy
is the sum of the entropy of the black hole and the entropy of the bath, both
of which change in this process.  The temperature of the bath is unchanged, but
when the black hole accretes the additional matter, the apparent horizon shifts
outward, concealing some of the radiation behind the apparent horizon, and thus
reducing the entropy of the bath.  The second step is to find how the horizon
shift changes the entropy of the radiation outside the apparent horizon.  Only
then can we infer the relation between the change in $\phi_H$ and the change in
$S_{\rm BH}$.

To carry out the first step of the calculation, we begin by noting that, for an
eternal black hole in equilibrium with a radiation bath, the quantum state of
the matter fields is the Kruskal vacuum, or ``Hartle-Hawking state''---there
are no quanta that are positive frequency with respect to the Kruskal
coordinates $x^{\pm}$ [\cite{hartlehawking}].  Now we recall that if we build
an arbitrary coherent state of left-moving matter on this vacuum, the general
solution to the field equations in Kruskal gauge has the form
$$
\Omega(x^+,x^-)=-\lambda^2x^+\left(x^-+{1\over
\lambda^2}P_+(x^+)\right)+{1\over \lambda} M(x^+)~,\eqno(kruskalomega)
$$
where $P_+$ is the total incoming Kruskal momentum up to retarded time $x^+$,
and $M$ is the total mass (at infinity) of the incoming matter.  (We have
chosen the origin of the Kruskal coordinate system to remove possible linear
terms in $x^+$ and $x^-$.)  If we assume that $P_+$ and $M$ are constants, then
the position of the apparent horizon, determined by the condition $\partial_+
\Omega=0$, is
$$
x^-_H(x^+)=-{1\over\lambda^2} P_+(x^+)~,
\eqno(kruskalapparent)
$$
and the value of $\Omega$ at the horizon is
$$
\Omega_H= {1\over\lambda}M(x^+)\eqno(kruskalomegahorizon)
$$
Therefore, if a pulse of left-moving matter that carries Kruskal momentum
$\Delta P_+$ and mass $\Delta M$ is accreted by the black hole, then the
horizon shifts outward according to
$$
\Delta x^-_H(x^+)=-{1\over\lambda^2} \Delta P_+(x^+)~,
\eqno(horizonshift)
$$
and $\Omega$ at the horizon changes according to
$$
\Delta \Omega_H= {1\over\lambda}\Delta M~.
\eqno(omegachange)
$$
We must recall, though, that the energy-momentum used in the field equations
has the unconventional normalization Eq.~\(tdef).  In thermodynamics, we should
use the conventionally normalized mass $M_{\rm conv}= (N/12\pi)M$, so that
$$
\Delta \Omega_H=  {12\pi\over N\lambda}\Delta M_{\rm conv}~.
\eqno(conv_omega_change)
$$
Now the identity $dS=dM/T$ becomes
$$
\Delta S_{\rm total}\equiv\Delta \left(S_{\rm BH}+S_{\rm matter}\right)
={1\over T}\Delta M_{\rm conv}={N\over 6}\Delta \Omega_H~.
\eqno(tot_entropy_change)
$$

We now proceed to the second step, which is to calculate $\Delta S_{\rm
matter}$, so that $\Delta S_{\rm BH}$ can be extracted from
Eq.~\(tot_entropy_change).  To carry out this step, we need a precise
definition of the entropy carried by the matter outside of the apparent
horizon.  Our proposal will be that $S_{\rm matter}$ is given by
Eq.~\(finegrained)---it is the fine-grained entropy of the matter fields
outside of the apparent horizon.\footnote{*}{The fine-grained entropy outside
the black hole horizon has also been discussed recently by Frolov and Novikov
[\cite{fronov}].}  It is not {\it a priori} obvious that this expression for
$S_{\rm matter}$ is correct or appropriate.  Ordinarily, the thermodynamic
entropy is a coarse-grained entropy [\cite{lebowitz}].  Surely, for a pure
state, $S_{\rm FG}=0$ would be a very poor estimate of the thermodynamic
entropy.  We are proposing that the quantum fields inside and outside the
horizon are so thoroughly entangled that it is reasonable to regard the
fine-grained entropy outside the horizon as the thermodynamic entropy.  In any
event, it is hard to think of another way to give the notion of the ``entropy
outside the apparent horizon'' any precise meaning.

For an eternal black hole, there is no reflecting boundary; the right-moving
modes and left-moving modes are uncorrelated.  The fine-grained entropy is
given by our curved-space generalization of the formula for the entropy on the
half line.  If the quantum state of $N$ scalar fields is a coherent state built
on the Kruskal vacuum, Eq.~\(curved_halfline_entropy) becomes
$$
S_{\rm FG}={N\over 6}\left( \rho_{H,K} + {1\over 2}~\ln\left({-x^+_{\rm max}
x^-_{\rm max}\over \delta^2}\right)\right)~,
\eqno(HHmatterentropy)
$$
where $x^{-}_{\rm max}$ and $x^+_{\rm max}$ are infrared cutoffs (in Kruskal
coordinates) for the right-movers and left-movers.
Of course, the conformal factor $\rho$ is gauge-dependent; the subscript
$K$ in
Eq.~\(HHmatterentropy) indicates that $\rho_{H,K}$ is evaluated in the Kruskal
gauge.

We can check that it is reasonable to interpret $S_{\rm FG}$ as the
thermodynamic entropy of the radiation bath by evaluating the infrared
divergent part of Eq.~\(HHmatterentropy).  The Kruskal coordinates $x^\pm$ are
related to the $\sigma^\pm$ coordinates (which become inertial in the
asymptotic region) by Eq.~\(sigma_define); thus the infrared divergent term in
$S_{\rm FG}$ is
$$
S_{\rm FG}\sim {N\over 12}\lambda\left(\sigma^+-\sigma^-\right)_{\rm max} =
{N\over 6}\lambda \sigma^1_{\rm max}~.
\eqno(large_sigma_plus)
$$
We may interpret $\sigma^1_{\rm max}$ as the size $L$ of the cavity that
contains the radiation.  Thus, Eq.~\(large_sigma_plus) agrees with the entropy
$$
S=2{\pi\over 6} T L\eqno(thermal_bath_entropy)
$$
of a thermal bath at temperature $T=\lambda/2\pi$, times a factor of $N$ for
the $N$ species. (The factor of two arises because both left-movers and
right-movers contribute to the entropy of the bath.)

When the black hole accretes some incoming matter, only the $\rho_H$ term in
Eq.~\(HHmatterentropy) is affected by the shift of the horizon.  Furthermore,
since in Kruskal gauge we have $\rho=\phi+{\rm constant}$, we conclude that
$$
\Delta S_{\rm matter}= {N\over 6}\Delta \phi_H~.
\eqno(matter_entropy_change)
$$
Combining with Eq.~\(tot_entropy_change), we find that
$$
\Delta S_{\rm BH}={N\over 6}\left(\Delta \Omega_H-\Delta\phi_H\right)~.
\eqno(bh_entropy_change)
$$
We can fix the arbitrary constant of integration by demanding that the black
hole entropy reaches zero when the apparent horizon meets the singularity, or
when $\phi_H=\phi_{\rm cr}=-{1\over 2}\ln(N/48)$; thus, from the expression
Eq.~\(odef) for $\Omega$ in terms of $\phi$, we obtain
$$
S_{\rm BH}= 2e^{-2\phi_H} -{N\over 12}\phi_H -{N\over 24}- {N\over
24}\ln\left({N\over 48}\right)~.
\eqno(corrected_bh_entropy)
$$
This is our corrected formula for the black hole entropy.

The formula Eq.~\(corrected_bh_entropy) for the black hole entropy has a
satisfying interpretation.  The action of two-dimensional dilaton gravity can
be obtained by spherical reduction of the four-dimensional action for a
near-extreme magnetically charged dilaton black hole.  When this reduction is
carried out, the area of the sphere of constant radius in four-dimensions
becomes the $\phi$-dependent prefactor of the Ricci scalar in the classical
two-dimensional action [\cite{CGHS}].  Now in the RST model, an extra term is
added to this prefactor.  The modified prefactor has just the form of the black
hole entropy in Eq.~\(corrected_bh_entropy).  Thus, loosely speaking, the
relation $S_{\rm BH}={1\over 4}A$ is satisfied by our corrected entropy
formula, but where $A$ is the {\it corrected} ``area'' of the RST model.

It may help to clarify the nature of the correction that we have found to
Eq.~\(BHentropy) if we restore the factors of $\hbar$ and  ``Newton's
constant'' $G$ that have been suppressed
until now. In the classical action Eq.~\(CGHS), there is a factor $G^{-1}$
multiplying the term $\left(R+4(\nabla \phi)^2+4\lambda^2\right)$, where $\hbar
G$ is dimensionless.  Thus the dilaton field $\phi$ is dimensionless, and
$\lambda^{-1}$ has the dimensions of length.  The leading term in the black
hole entropy is then
$$
S_{\rm BH,0}={2\pi M_{\rm BH}\over \hbar \lambda}~,
\eqno(entropy_hbar)
$$
and the correction is
$$
S_{\rm BH, 1}=-{N\over 12}\phi_H~.
\eqno(correction_hbar)
$$
Relative to the leading term, then, the correction is suppressed by
$$
{S_{\rm BH,1}\over S_{\rm BH,0}}=-{N\hbar\over 24\pi}~{\lambda\phi_H\over
M_{\rm BH}}={N\hbar\over 48\pi}~{\lambda\over M_{\rm BH}}~\ln\left({\pi
GM_{\rm BH}\over \lambda}\right)~.
\eqno(relative_correction_hbar)
$$
Thus, the correction is higher order in $\hbar$, but cannot be neglected in the
large-$N$ limit.  It is also suppressed, for a very massive black hole, by the
factor $\ln(M_{\rm BH})/M_{\rm BH}$.

In the RST model, it is possible to obtain a simple analytic expression for the
value $\Omega_H$ of $\Omega$ at the apparent horizon, on a general
time-dependent background.  There is no such simple expression for $\phi_H$, as
$\Omega$ and $\phi$ are related by the transcendental Eq.~\(odef).  Thus, we
cannot write down an analytic formula for $S_{\rm BH}$ or $S_{\rm FG}$ on a
general background.  However, when these quantities are added together, a
notable simplification occurs.
Combining Eq.~\(corrected_bh_entropy) and
Eq.~\(RSTFG), and comparing with Eq.~\(cdef), we see that
$$
S_{\rm BH}+S_{\rm FG}={N\over 6}\left(\chi_{H,\sigma}-{1\over 4} +\ln 2
+\ln{L\over \delta}\right)
\eqno(bhplusfg)
$$
can be expressed in terms of $\chi$, which obeys a simple field equation in the
RST model.
(Here $L=\sigma^+_H-\sigma_B^+$, as in Eq.~\(first_L_define).) Of
course, the value $\chi_H$ of $\chi$ at the apparent horizon is
gauge-dependent, while $S_{\rm BH}+S_{\rm FG}$ is not. This formula is
valid if $\chi_H$
is evaluated in the same coordinate system used to define the vacuum,
in other words, in ``sigma gauge''.  Recall that $\sigma^+$
is the null coordinate with respect to which the
incoming vacuum state is defined, and $\sigma^-$ must be chosen
so that $\sigma^+-\sigma^-$=constant at the boundary of the spacetime, as we
explained in Section III.  We should emphasize that Eq.~\(bhplusfg) is a
general formula that applies under the above conditions.  In particular, it
need not be assumed that the $\sigma^{\pm}$ coordinates are related to the
Kruskal coordinates by Eq.~\(sigma_define).

We will discuss the evolution of $S_{\rm BH}+S_{\rm FG}$ in the next Section.
For now, we remark that Eq.~\(bhplusfg), like Eq.~\(corrected_bh_entropy), has
an intriguing interpretation.  We observe that $\chi$ is proportional to the
coefficient of the scalar curvature $R$ in the {\it quantum-corrected}
effective
action of the (large-$N$) RST model.  Thus, if we neglect the logarithmic term
in Eq.~\(bhplusfg), we find that the sum $S_{\rm BH} + S_{\rm FG}$ is related
to the quantum-corrected Newton's constant just as $S_{\rm BH}$ is related to
the classical Newton's constant of the model.  This remark makes contact with
the observations in Ref.~[\cite{uglum}], where a connection
between entropy and the renormalization of Newton's constant is proposed.

(One is tempted to go further, and regard Eq.~\(bhplusfg) as a hint that the
proper way to define the fine-grained entropy is to use the ``$\chi$ metric''
$ds^2=-e^{2\chi_\sigma}d\sigma^+d\sigma^-$ when implementing the short-distance
cutoff.  Then Eq.~\(bhplusfg) could be interpreted as wholly due to the entropy
of entanglement between the regions outside and inside the black hole---there
would be no need to add in a separate Bekenstein-Hawking term.)

\head{VI. Evaporation and thermodynamics}

Equipped now with our formulas for the black hole entropy $S_{\rm BH}$ and the
fine-grained entropy $S_{\rm FG}$ outside the apparent horizon, we are prepared
to study the thermodynamics of a process in which a black hole forms from
infalling matter and then evaporates, as in Fig.~1.  We wish to find the
time-dependence of the total entropy in this process.  We will assume that the
incoming matter state is a coherent state built on the inertial $\sigma^+$
vacuum at ${\cal I}^-$.  For such states, we know how to evolve the geometry
using the RST equations, and we know how to calculate the fine-grained entropy.
 We will also make the further assumption that none of the infalling matter
reaches the reflecting boundary of the spacetime before the appearance of the
global event horizon.  This assumption simplifies the calculation of $S_{\rm
FG}$, as we explained in Section III.

In their analysis of the model, RST noted that the boundary condition
Eq.~\(rbc) can be re-imposed at the endpoint of black hole evaporation (when
the singularity meets the apparent horizon), and that the final quantum state
can be chosen to be the vacuum.  This prescription results in the emission of a
thunderpop.  Furthermore, the information about the quantum state of the
initial incoming matter is lost {\it by assumption}.  But we wish to emphasize
that the time-dependence of the entropy up until the apparent horizon meets the
singularity is insensitive to the RST prescription for continuing past this
point, and is not affected by the thunderpop.  It will be of interest to see
how the fine-grained entropy outside the horizon behaves as the black hole
approaches its demise.

We have seen that the fine-grained entropy depends on an arbitrary ultraviolet
cutoff.  However, the ultraviolet divergence is logarithmic, and the
cutoff-dependent term is a time-independent additive constant.  Thus, the
sensitivity to the cutoff does not prevent us from making definite statements
about how the entropy outside the black hole {\it changes}
during its evolution, or about the change in the intrinsic entropy of the black
hole itself.

\subhead{A. Boltzman entropy}

In the previous Section, we argued that, in the Hartle-Hawking vacuum state,
the fine-grained entropy $S_{\rm FG}$ could be regarded as the thermodynamic
entropy outside the event horizon of the black hole.  But for the black hole
formed from infalling matter, this assignment must be modified.  To see why,
cover the spacetime of Fig.~1a with a sequence of spacelike slices, as depicted
in Fig.~1b.  Slices I and II in the figure represent times prior to the
formation of the black hole.  Since there is no apparent horizon, the quantum
state ``outside'' the horizon on these slices is a pure coherent state, which
has $S_{\rm FG}=0$.

But even though the incoming matter is in a pure state, it surely carries
thermodynamic entropy.  We can assign a nonzero entropy to this state by
performing a coarse-graining procedure.  Our coherent state carries the {\it
left-moving} energy density
$$
{\cal E}(\sigma^+)\equiv  T^f_{++}(\sigma^+)~.
\eqno(etf)
$$
We may regard ${\cal E}$ as a measurable macroscopic quantity.  Given the
energy-density profile ${\cal E}$ of the incoming state, we assign an entropy
by counting the number of microscopic quantum states with this energy
profile---the entropy is the logarithm of the number of states.   We will refer
to the entropy defined by this procedure as $S_{\rm Boltz}$, the Boltzman
entropy of the incoming coherent state.

The spacetime is asymptotically flat, so we may use standard flat-space
thermodynamics on ${\cal I}^-$.  We may then appeal to the equivalence of the
microcanonical and canonical ensembles in the thermodynamic limit, and express
both the entropy density and the energy density in terms of a locally measured
temperature.  Fluctuations of the entropy and energy densities about these
values are suppressed in the large $N$ limit.  If the energy density is
conventionally normalized, we can express the energy density ${\cal E}_{\rm
conv}$ and entropy density ${\cal S}$ for $N$ left-moving massless free scalar
fields in terms of the temperature $T$ as
$$
{\cal E}_{\rm conv}=N{\pi\over 12}T^2~,\quad
{\cal S}= N{\pi\over 6}T~,
\eqno(convEandS)
$$
so that the entropy and energy densities are related by
$$
{\cal S}=\sqrt{{\pi\over3}N{\cal E}_{\rm conv}}~.
\eqno(convSfromE)
$$
The energy density in Eq.~\(etf) has the unconventional normalization
$$
{\cal E}={12\pi\over N}{\cal E}_{\rm conv}~,
\eqno(EfromconvE)
$$
so that the Boltzmann entropy can be written
$$
S_{\rm Boltz} = \frac{N}{6}\ \int_{{\cal I}^-} d\sigma^+ \sqrt{{\cal
E}(\sigma^+)}~,
\eqno(sblt)
$$

We can now evolve the incoming matter state from slice I of Fig.~1b to slice
II, which is still prior to the formation of the black hole.
In general, $S_{\rm Boltz}$ can change
under unitary evolution, but for a free field it is invariant as a
consequence of the curved space generalization of Liouville's theorem
[\cite{mtw}]. In the present context, this is simply the statement that
the energy profile ${\cal E}(\sigma^+)$ is unchanged.

The black hole is finally encountered on slice III.  Liouville's theorem
continues to apply here, so that $S_{\rm Boltz}$ is still unchanged.  However,
we are interested in the entropy of the matter outside the black hole.
Therefore, we divide slice III into two segments, $\Sigma_{\rm in}$ and
$\Sigma_{\rm out}$, inside and outside the apparent horizon.  The Boltzman
entropy
$S_{\rm BO}$ outside the apparent horizon is
$$
S_{\rm BO} =
\frac{N}{6} \int_{\Sigma_{\rm out}} d\sigma^+ \sqrt{{\cal E}(\sigma^+)}~.
\eqno(sbo)
$$

In defining $S_{\rm BO}$, we have chosen to divide the slice $\Sigma$ at the
{\it apparent} horizon.  We made the same choice when we defined the
fine-grained entropy $S_{\rm FG}$ outside the black hole in Section III.B.
Furthermore, our formula Eq.~\(corrected_bh_entropy) for the black hole entropy
$S_{\rm BH}$ has been expressed in terms of the value of the dilaton field at
the apparent horizon.  These choices deserve some explanation.  If we are
adopting the viewpoint of an observer who remains outside the black hole, it
may seem more logical to divide the slice at the {\it global} event horizon
instead.  After all, it is possible for the observer to cross the apparent
horizon (very carefully!) and return to tell about it.  However, we find it
more appropriate to define $S_{\rm BO}$, $S_{\rm FG}$ and $S_{\rm BH}$ using
the apparent horizon, for several reasons.  First of all, the position of the
apparent horizon can be determined locally in time, without any required
information about the global properties of the spacetime.  Our observer on a
time slice can readily identify the apparent horizon as the location where
$\partial_+\Omega$ vanishes.  Second, because the position of the apparent
horizon is determined by this local condition, it is easy to compute the
trajectory of the apparent horizon using the RST equations.  Third, if we use
the global horizon to define the entropy, the resulting expressions do not seem
to have a nice thermodynamic interpretation. In particular, the would-be second
law is
easily violated by sending in a very sharp pulse with large entropy and
energy density but small total entropy and energy.  The essential point is that
the value of the dilaton at the global horizon responds less sensitively to the
incoming pulse than does the dilaton at the apparent horizon.

\subhead{B. Total entropy}

Once the black hole forms, matter entropy can become concealed behind the
horizon, and the left-moving Boltzman entropy Eq.~\(sbo) can decrease.  If
physics perceived by an observer outside the black hole is to respect the
second law of thermodynamics, then (as Bekenstein argued [\cite{bek}]) we must
attribute entropy to the black hole.  Furthermore, we must not neglect the
entropy carried by the outgoing Hawking radiation.

We propose to adopt, as our definition of the total thermodynamic entropy
$$
S_{\rm tot}\equiv S_{\rm BH}+S_{\rm BO}+S_{\rm FG}~.
\eqno(totentropy)
$$
The fine-grained entropy $S_{\rm FG}$ outside the apparent horizon is dominated
by the entanglement of the right-moving modes outside the horizon with the
right-moving modes just inside the horizon.  It roughly corresponds to the
thermodynamic entropy of the of the outgoing Hawking radiation, while $S_{\rm
BO}$ is the entropy of the incoming matter.  We have seen that the fine-grained
entropy does not include the entropy of the incoming matter---an incoming
coherent state has the same $S_{\rm FG}$ as the vacuum state---so $S_{\rm BO}$
must be added on.

While the expression Eq.~\(totentropy) may appear (and indeed, is!)
somewhat strange, we
believe it to be a precise two-dimensional analog of the notion of
``total entropy'' used implicitly in discussions of four-dimensional
black hole thermodynamics.
This prescription might be interpreted as follows:  We
may consider, instead of a pure initial state, the mixed initial state $\rho$
that maximizes $-{\rm tr}\rho\ln\rho$, subject to the constraint that the
energy density is given by the specified function ${\cal E}(\sigma^+)$.  For
this mixed initial state, we have $S_{\rm Boltz}=-{\rm tr}\rho\ln\rho$.  What
we are adding to $S_{\rm BH}$ in Eq.~\(totentropy) is the fine-grained entropy
outside the horizon for this particular mixed initial state.\footnote{*}{Note
that we have not really established that this interpretation is correct.  In
particular, our expression for $S_{\rm FG}$ has been derived only for {\it
coherent} incoming states, and may not apply for arbitrary states.}  In any
event, we have not been able to find any other reasonable and precise
alternative to Eq.~\(totentropy) that obeys a generalized second
law.

As we noted at the end of Section V, the sum $S_{\rm BH}+S_{\rm FG}$ can be
expressed in terms of the field $\chi$ at the apparent horizon (in sigma
gauge), for which we can find an analytic expression.  Alternatively, we may
combine Eq.~\(corrected_bh_entropy) and Eq.~\(entropy_phi), to obtain directly
an expression in terms of the gauge invariant quantity $\Omega_H$.  We obtain
$$
S_{\rm BH}+S_{\rm FG}={N\over 6}\left(\Omega_H-{1\over 4}+{1\over 2}\lambda
L+\ln{L\over\delta}\right)~,
\eqno(bh_plus_fg_omega)
$$
where $L=\sigma^+_H-\sigma_B^+$, as in Eq.~\(first_L_define).
Now we may use the general solution Eq.~\(gsol) to the field
equations in Kruskal gauge, which applies if the state of the matter is a
coherent state built on the sigma vacuum.  Recalling that the apparent horizon
is defined by the condition $\partial_+\Omega=0$, we deduce from Eq.~\(gsol)
that
$$
\eqalign{
\Omega_H&={1\over 4}+{1\over\lambda}M(x_H^+) -{1\over
4}\ln\left(-4\lambda^2x_H^+
x_H^-)\right)~\cr
&={1\over 4}+{1\over\lambda}M(\sigma_H^+) -{1\over
4}\lambda\left(\sigma_H^+-\sigma_H^-\right) - {1\over 2}\ln 2~,\cr}
\eqno(omegahorizon)
$$
where
$$
M(\sigma_H^+)=\int^{\sigma_H^+}d\sigma^+{\cal E}(\sigma^+)~.
\eqno(masssigmaplus)
$$
is the total mass flowing in from ${\cal I}^-$ up until retarded time
$\sigma^+_H$.

Next, we express $\Omega_H$ in terms of the quantity $L=\sigma_H^+
-\sigma_B^+$.  Under the assumption that there is no infalling matter up
until retarded time $x^+_B$, the position of the boundary defined by
$\Omega=\Omega_{\rm cr}=1/4$ is given by Eq.~\(ldv_sigma_boundary).
For the point on the boundary with the same advanced time as the apparent
horizon (as in Fig.~6), we have $\sigma_B^-=\sigma_H^-$.  Combining
Eq.~\(ldv_sigma_boundary) with Eq.~\(omegahorizon), we find
$$
\Omega_H={1\over4}+{1\over\lambda} M -{1\over 4}\lambda L~,
\eqno(omegaHagain)
$$
Inserting into Eq.~\(bhplusfg) now yields
$$
S_{\rm BH}+ S_{\rm FG}={N\over 6}\left({1\over \lambda} M(\sigma^+_H) +{1\over
4}\lambda L +\ln{L\over \delta}\right)~.
\eqno(bhplusfgagain)
$$
Adding the Boltzman entropy Eq.~\(sbo) outside the black hole, we find
$$
S_{\rm tot}\equiv S_{\rm BH}+S_{\rm FG}+S_{\rm BO}={N\over
6}\left({1\over\lambda}M(\sigma^+_H) +{1\over 4} \lambda L +\ln {L\over
\delta}+\int_{\sigma_H^+}^{\infty} d\sigma^+\sqrt{{\cal E}(\sigma^+)}\right)~,
\eqno(final_total_entropy)
$$
our final expression for the total entropy.

It is instructive to compare $S_{\rm tot}$ and $S_{\rm Boltz}$ on the same time
slice, or equivalently, to compare $S_{\rm BH}+S_{\rm FG}$ with the Boltzman
entropy $S_{\rm BI}$ {\it inside} the apparent horizon.  Since we assume that
there is no incoming energy density before the retarded time
$\sigma^+=\sigma_B^+$, we can choose the lower limit of integration in
Eq.~\(masssigmaplus) to be $\sigma^+_B$, and we then have
$$
S_{\rm BH}+S_{\rm FG}-S_{\rm BI}={N\over
6}\left[\int_{\sigma^+_B}^{\sigma^+_H}d\sigma^+
{1\over\lambda}\left(\sqrt{{\cal E}(\sigma^+)}- {\lambda\over 2}\right)^2 +
\ln{L\over\delta}\right]~.
\eqno(entropysquare)
$$
This expression is always positive, so that $S_{\rm tot}$ is always greater
than $S_{\rm Boltz}$.  In particular, the total entropy  $S_{\rm tot}$
always jumps by a (cutoff dependent) positive amount when the apparent horizon
first appears.

The first term in Eq.~\(entropysquare) is minimized if we choose ${\cal
E}=\lambda^2/4$.  This incoming energy flux is the critical flux ${\cal E}_{\rm
cr}$ that matches the flux of the outgoing Hawking radiation.  (From
Eq.~\(EfromconvE), we see that ${\cal E}_{\rm cr}$ corresponds to the
conventionally normalized thermal flux ${\cal E}_{\rm conv}=N\pi T^2/12$, where
$T=\lambda/2\pi$.)  We see from Eq.~\(entropysquare)\footnote{*}{Since $S_{\rm
Boltz}=S_{\rm BO}+S_{\rm BI}$ is
conserved (by Liouville's theorem), the expression in Eq.~\(entropysquare)
differs from the total entropy by an additive constant.} that, even when the
black
hole is critically illuminated, the total entropy continues to grow like
$(N/6)\ln L$. This increasing term arises from the {\it long-distance}
correlations of the quantum fields outside the black hole with the fields in
the region behind the horizon.  The existence of this term is a bit of a
surprise, as one might have expected the critical illumination of the black
hole to be a thermodynamically reversible process.  Indeed, one might say that
the result Eq.~\(entropysquare) calls into question our proposal to identify
$S_{\rm tot}$ with the thermodynamic entropy---an expression without the $\ln
L$ term would look more plausible.  However, we will see in Section VII that
the
second law can be (mildly) violated for an appropriately chosen energy density
profile ${\cal E}(\sigma^+)$, if the $\ln L$ term is absent.

Note that for a very long-lived black hole, the $\ln L$ term becomes very
slowly varying, so that the total entropy of a critically illuminated black
hole does become very nearly constant.  This is how Eq.~\(entropysquare)
becomes reconciled with our calculation of the black hole entropy in Section
V, where we {\it did} assume that the emission of radiation by a black hole in
a thermal bath is thermodynamically reversible, so that the total entropy
remains unchanged.  In other words (and not so surprisingly), the process in
which a black hole immersed in a thermal bath accretes or emits a small net
amount of radiation becomes reversible only when it is carried out arbitrarily
slowly.

\subhead{C. Complete evaporation}

Let us now consider a process in which a black hole forms from infalling matter
and eventually evaporates completely.  Our semiclassical approximations
actually break down at the very end of this process, but we can still make
definite statements about how the total entropy behaves as the endpoint of the
process approaches.

The endpoint occurs when the apparent horizon and the singularity coincide, or
when $\Omega_H=\Omega_{\rm cr}=1/4$.  From Eq.~\(omegaHagain), we see that at
the endpoint
$$
M={1\over 4}\lambda^2 L={\cal E}_{\rm cr} L~.
\eqno(finalM)
$$
Eq.~\(finalM) simply says that, at the endpoint, the total energy $M$ that has
propagated in matches the total energy ${\cal E}_{\rm cr}L$ of the Hawking
radiation that has been emitted.\footnote{*}{Actually, this explanation does
not exclude a possible extra additive term on the right-hand side of
Eq.~\(finalM) that is subleading for large $L$, both because the Hawking flux
takes a short while to turn on, and because the emitted radiation
``overshoots'' (resulting in the emission of a negative energy thunderpop at
the endpoint).  But it turns out that this potential subleading term is
absent.}  The relation between $M$ and $L$ is
independent of the energy profile of the incoming matter, because the
temperature of the black hole is independent of its mass.

At the endpoint, the black hole entropy goes to zero, so we readily find the
fine-grained entropy to be
$$
S_{\rm FG}=S_{\rm BH}+S_{\rm FG}={N\over 6}\left[{2M\over
\lambda}+\ln\left({4M\over\lambda^2\delta}\right)\right]~.
\eqno(completelygone)
$$
We may regard Eq.~\(completelygone) as an expression for the amount of
information that is destroyed due to the formation and complete evaporation of
the black hole.  It is not entirely clear how to interpret the ultraviolet
divergence in this formula, since the amount of lost information should be
finite.  Presumably, in a complete description of the evaporation process,
there will be some quantum fuzziness in the endpoint, and hence in the position
of the global horizon.  It then seems plausible that $\delta$ would be replaced
by a (small) characteristic time scale for the final quantum-mechanical
transition that returns the quantum fields to the vacuum state.  Thus, we
expect that the first term in Eq.~\(completelygone) will actually dominate over
the cutoff-dependent term, in the evaporation of a sufficiently large black
hole.

It is easy to understand the origin of the two terms in Eq.~\(completelygone),
by referring to Eq.~\(RSTFG).  From Eq.~\(rho_phi_L), we see that the first
term is just $(N/6)\rho_{H,\sigma}$ evaluated
at the endpoint (where $\phi=\phi_{\rm cr}$).  As we have already discussed in
section IV, $e^{2\rho_{H,\sigma}}$ is the factor by which the modes emitted in
the late stages of the process have been redshifted, relative to frequencies
measured on ${\cal I}^-$. It is the {\it
very-short-distance} correlations between these modes just inside and just
outside the horizon that are responsible for the dominant contribution to the
entropy in Eq.~\(completelygone).  The subdominant second term in
Eq.~\(completelygone) arises from the {\it long-distance} correlations between
field modes inside and outside the horizon.

The first term in Eq.~\(completelygone) also has an interpretation in terms of
standard thermodynamics.  Recalling the relation Eq.~\(EfromconvE) between our
normalization of energy and the conventional normalization, we see that
Eq.~\(completelygone) can be reexpressed as
$$
S_{\rm FG}={2M_{\rm conv}\over T} + \cdots~,
\eqno(thermo_mass)
$$
in terms of the conventionally normalized mass that has been emitted by the
black hole during its lifetime.  The factor of 2 in Eq.~\(thermo_mass) arises
because the emission of thermal radiation into cold empty space is an
irreversible process [\cite{zurek}].  (This factor becomes $(D+1)/D$ in
$D$-dimensional space; to compute it we observe that the entropy $S$ of a
relativistic ideal gas is related to its energy $E$ by $S={D+1\over D}E/T$.  In
three dimensions, ${4\over 3}$ is modified by ``grey-body factors''
[\cite{page}], but there are no such factors in the RST model.)

While this factor of two agrees with thermodynamic expectations, that it
appears in the {\it fine-grained} entropy is nonetheless intriguing.  We have
found that if a black hole forms from collapse and then evaporates, the
fine-grained entropy of the emitted radiation is (approximately) twice as large
as the
Bekenstein-Hawking entropy of the black hole that initially formed.  We might
have expected, instead, that the amount of quantum-mechanical information that
is lost due to the collapse of a pure state is correctly quantified by $S_{\rm
BH}$, as it is often presumed [\cite{bek}] that the number of distinct quantum
states from which  the black hole could have formed is $\exp (S_{\rm BH})$.
Then the extra factor of two in the coarse-grained entropy of the emitted
radiation would not be due to an intrinsic loss of information; the
fine-grained entropy would be only half as large as the coarse-grained entropy,
because of subtle correlations among the quanta. Evidently, the radiation
outside the horizon is so thoroughly entangled with the degrees of freedom
behind the horizon that virtually {\it all} of its thermodynamic entropy can be
attributed to correlations with the fields behind the horizon, and hence to
``lost information.''  Indeed, we can attribute all of the thermodynamic
entropy to the exponential redshifting of the modes near the horizon, which, as
we noted above, allows shorter and shorter wavelength modes to make a
contribution to the fine-grained entropy as the black hole evolves.

Of course, we can make the mass $M$ in Eq.~\(completelygone) as large as we
please by maintaining the black hole for a long time; we just send in a
continuous flux of matter that compensates for the outgoing Hawking flux.  And
we can choose the infalling matter to be in a pure coherent state, with $S_{\rm
FG}=0$.  It is clear, then, that there is no limit to the amount of information
that can be destroyed by the black hole, or in other words, no limit to the
degree of entanglement of the fields outside the global horizon with those
inside, a conclusion that was already stated in Section IV.

The subdominant logarithmic term in Eq.~\(completelygone) arises from the
long-distance correlations of the quantum fields outside the horizon with those
inside.  This term indicates that the amount of missing information is even
greater than naive thermodynamic expectations can accommodate.  It would be
satisfying to find an interpretation of the logarithmic term in thermodynamic
language, but we know no such interpretation.

\head{VII. The Second Law}

Bekenstein conjectured that a generalized second law of thermodynamics applies
to processes involving black holes, so that the sum of the entropy outside the
black hole and the intrinsic black hole entropy is always non-decreasing
[\cite{bek}].  According to this conjecture, although entropy can disappear
behind the horizon, the increase in the area of the horizon always compensates
(and typically overcompensates) for the lost entropy.  Similarly, the emission
of Hawking radiation causes the  horizon to shrink, but the decrease in horizon
area is always compensated by the entropy of the emitted radiation.

We want to examine whether this conjecture holds in the RST model.
To show that Bekenstein's conjecture is correct, we need to attach a precise
meaning to the notion of the ``entropy outside the black hole.''  Our proposal
is that the entropy outside is $S_{\rm FG}+S_{\rm BO}$.  Bekenstein's
conjecture then becomes the statement that the quantity $S_{\rm tot}$ given by
Eq.~\(final_total_entropy) is non-decreasing.  This expression depends on the
short-distance cutoff $\delta$ that we introduced by smoothing the apparent
horizon.  But since the cutoff-dependent term is just an additive constant,
{\it changes} in the entropy are not sensitive to the cutoff, at times after
the formation of the black hole and before the endpoint of its evaporation.

Our task is to determine whether there is any energy density profile of the
incoming matter for which $S_{\rm tot}$ can decrease as the black hole evolves.
We continue to assume, as in Section VI, that the incoming matter is in a
coherent state built on the asymptotic vacuum state at ${\cal I}^-$, and that
no infalling matter reaches the boundary of the spacetime before the global
event horizon.  Under these assumptions, we will show that the second law is
valid.

To find the time evolution of $S_{\rm tot}$ in Eq.~\(final_total_entropy), we
will need to know how $L=\sigma_H^+-\sigma_B^+$ evolves, and hence how the
position $(\sigma_H^+,\sigma_H^-)$ of the apparent horizon evolves.  Since the
apparent horizon is defined by the condition $\partial_+\Omega|_H=0$, the
trajectory $x_H^-(x_H^+)$ of the apparent horizon in Kruskal coordinates
satisfies
$$
{dx_H^-\over dx_H^+}=-{\partial_+^2\Omega\over\partial_-\partial_+\Omega}
\Biggr|_H=-{1\over\lambda^2}\left(T_{++}^f (x_H^+)-{1\over 4
(x_H^+)^2}\right)~;
\eqno(kruskal_horizon_moves)
$$
in the second equality we have used the Eq.~\(cstr) satisfied by $\Omega$ in
the Kruskal gauge.  Recalling that $T_{++}^f$ transforms as a tensor, we may
re-express this condition in $\sigma$ coordinates as
$$
{d\sigma_H^-\over d\sigma_H^+}=-{1\over\lambda^2}e^{-\lambda(\sigma_H^+ -
\sigma_H^-)}\left({\cal E}(\sigma_H^+)-{1\over 4}\lambda^2\right) =e^{-\lambda
L}\left(1-{{\cal E}(\sigma_H^+)\over{\cal E}_{\rm cr}}\right)~,
\eqno(sigma_horizon_moves)
$$
where we have used Eq.~\(ldv_sigma_boundary), and have expressed the result in
terms of the critical (thermal) flux ${\cal E}_{\rm cr}={1\over 4}\lambda^2$.
We note that the trajectory of the apparent horizon is timelike if the incoming
flux is less than the outgoing flux due to Hawking radiation, and becomes null
when the incoming and outgoing flux match.

If we regard the total entropy as a function of the retarded time $\sigma_H^+$
at the apparent horizon, then we may use
$$
{dL\over d\sigma^+_H}\equiv{d\over
d\sigma_H^+}(\sigma_H^+-\sigma_B^+)=1+e^{-\lambda L}\left({{\cal E}\over{\cal
E}_{\rm cr}}-1\right)
\eqno(Lvaries)
$$
and Eq.~\(masssigmaplus) to see that the total entropy given by
Eq.~\(final_total_entropy) varies at the rate
$$
{d\over d\sigma_H^+}S_{\rm tot}={N\lambda\over 24}\left[\left(\sqrt{\tilde{\cal
E}(\sigma_H^+)}-1\right)^2 + e^{-\lambda L}\Biggr(\tilde{\cal
E}(\sigma_H^+)-1\Biggr)\left(1+{4\over \lambda  L}\right)+{4\over \lambda
L}\right]~,
\eqno(entropy_rate_change)
$$
which we have expressed in terms of
$$
\tilde{\cal E}(\sigma_H^+)={{\cal E}(\sigma_H^+)\over {\cal E}_{\rm cr}}~,
\eqno(tildeE)
$$
the ratio of the incoming flux to the thermal flux.  As expected, the rate of
change of the entropy does not depend on the short-distance cutoff $\delta$.

It is not hard to check that Eq.~\(entropy_rate_change) is {\it positive} for
any $\tilde{\cal E}\ge 0$ and any finite $L>0$.  For a fixed $L$, $dS_{\rm
tot}/d\sigma_H^+$ is minimized when the incoming flux is
$$
\tilde{\cal E}=\left(1+e^{-\lambda L} +{4\over \lambda L}e^{-\lambda
L}\right)^{-2}~,
\eqno(minimizing_flux)
$$
and the minimum value attained is
$$
{d\over d\sigma_H^+}S_{\rm tot}\Biggr|_{\rm min}={N\lambda\over 24}\left[
{4\over\lambda L}-{e^{-2\lambda L}\left(1+{4\over \lambda L}\right)^2\over
1 + e^{-\lambda L}\left(1+{4\over \lambda L}\right)}\right]
\eqno(minimum_flux)
$$
This expression is a monotonically decreasing function of $\lambda L$ that
approaches zero as $\lambda L\to \infty$.  Thus, we see that the total entropy
is always increasing, in accord with the generalized second law.

If the black hole is critically illuminated ($\tilde{\cal E}=1$), the mass
radiated away is matched exactly by the incoming matter flux.  We see from
Eq.~\(entropy_rate_change) that the total entropy nevertheless continues to
increase for $L<\infty$  (as we already noted in Section VI).  The entropy
increase is due to the $\ln(L/\delta)$ term in $S_{\rm tot}$, the term arising
from the long-distance correlations of the quantum fields outside the horizon
with those inside.  This term is consistent with the property that a black hole
can reach thermal equilibrium with a radiation bath, because the rate of change
of the entropy approaches zero as the age $L$ of the black hole gets
arbitrarily large.  Still, since the $\ln(L/\delta)$ term has no clear
thermodynamic interpretation, one is tempted to seek a reformulation of the
second law in which the long-distance contribution to the fine-grained entropy
is absent.

The obvious thing to try is to subtract the offending term away, and define a
new total entropy
$$
S_{\rm tot}^{(new)}=S_{\rm tot} - {N\over 6}\ln\left({L\over \delta}\right)
\eqno(new_entropy)
$$
The rate of change of this entropy is
$$
{d\over d\sigma_H^+}S_{\rm tot}^{(new)}={N\lambda\over 24}\left[(1+e^{-\lambda
L}) \left(\sqrt{\tilde{\cal E}} - {1\over 1+e^{-\lambda L}}\right)^2-
{e^{-2\lambda L}\over 1 + e^{-\lambda L}}\right]~.
\eqno(new_rate)
$$
We see that the new entropy does not strictly satisfy the second law.  The
entropy of a critically illuminated black hole is constant, but the entropy
decreases slowly if the incoming flux is slightly below critical.  On the other
hand, for $\lambda L>>1$
the violations of the new second law are extremely mild, and occur only
under very rare conditions.  The entropy is non-increasing unless the flux lies
in the narrow range
$$
1>\tilde{\cal E} >\tanh^2\left({\lambda L\over 2}\right)\approx 1-4e^{-\lambda
L}~.
\eqno(bad_range)
$$
Thus, for $\lambda L>>1$, the second law fails only when the flux is tuned to
be exponentially close to critical, and even then the rate of decrease of the
entropy is exponentially small.

We caution the reader again that our derivation of the second law applies only
under special conditions.  In particular, we have assumed that the incoming
matter is in a coherent state built on the inertial vacuum at ${\cal I}^-$.
When more general quantum states are considered, our proof breaks down.  We
will show in Appendix B that states can be constructed that carry, locally, a
large amount of fine-grained entropy and a small amount of energy, or carry
negative energy density without accompanying negative entropy
[\cite{holzhey,wilczek}].  (Neither of these pathologies occurs for the
coherent states built on the inertial vacuum.)  Thus, the second law, as we
have formulated it here, can be violated at least for a while by tossing matter
in such a state into the black hole.  Such examples show that if there is a
very general statement of the second law, our expression for the total entropy
cannot apply in all situations.

Boltzmann's derivation of the macroscopic second law
of thermodynamics
from the microscopic laws of statistical mechanics is one of the most
satisfying developments in the history of physics. We believe that there
should be an equally satisfying derivation  of Bekenstein's generalized
second law. In this paper, beginning from the microscopic laws of a specific
two-dimensional theory, we have given a derivation of Bekenstein's generalized
second law which is applicable to a wide range of processes. Yet we do not
feel that our derivation has provided complete insight into {\it why}
the generalized second law is (often) valid, because we
relied mainly on explicit calculation, rather than general reasoning.
Indeed, it is not evident from
our derivation that the  generalized second law will hold in variants of the
RST model. Thus, while we have made some progress, the true nature
of Bekenstein's generalized second law remains an outstanding enigma.

\head{Acknowledgments}

We have benefited from discussions with S. Das, L. Thorlacius, and especially
S. Mathur.
This work was supported in part by the U.~S.~Department of Energy under Grant
No. DOE-91ER40618 and Grant No. DE-FG03-92-ER40701.

\head{Appendix A}
\taghead{A.}
In our calculations of the fine-grained entropy in Section III, we considered
the quantum state of the scalar field to be either the inertial vacuum or a
``vacuum'' state that is conformally related to the inertial vacuum.  In this
appendix, we will generalize the results to include the case of a coherent
state built on such a ``vacuum.''  We will show that the fine-grained entropy
for the coherent state is the same as the fine-grained entropy of the vacuum
state.  Thus, if space is divided into two regions, building a coherent state
on the vacuum does not affect the degree of entanglement of the quantum fields
in the two regions.

To begin, we consider a toy problem that incorporates all of the essential
features of the general case.  Consider a system of two uncoupled harmonic
oscillators, with associated annihilation operators $a_1$ and $a_2$.  Perform a
Bogolubov transformation of the form
$$
\eqalign{
a_1^\gamma&={1\over \sqrt{1-\gamma^2}}\left( a_1-\gamma a_2^\dagger\right)
{}~,\cr
a_2^\gamma&={1\over \sqrt{1-\gamma^2}}\left(a_2 - \gamma a_1^\dagger\right)
{}~,\cr}
\eqno(osc_bogo)
$$
where $\gamma$ is real and $\gamma^2<1$.  This is the most general Bogolubov
transformation in
which $a_1^\gamma$ is a linear combination of an $a_1$ annihilation operator
and an $a_2^\dagger$ creation operator, up to phases that can be removed by
adjusting the phases of the $a_1$, $a_2$, $a_1^\gamma$, and $a_2^\gamma$.

We can now construct the ``$\gamma$-vacuum'' that is annihilated by
$a_1^\gamma$
and $a_2^\gamma$; it is
$$
|\gamma\rangle=\sqrt{1-\gamma^2}\exp\left({\gamma a_1^\dagger
a_2^\dagger}\right)|0,1\rangle\otimes
|0,2\rangle=\sqrt{1-\gamma^2}\sum_{n=0}^\infty \gamma^{n}|n,1\rangle\otimes
|n,2\rangle ~,
\eqno(gamma_vac)
$$
where $|n,1\rangle$ and $|n,2\rangle$ denote the $n$th excitation of oscillator
1 and 2 respectively.  The easiest way to verify the first equality in
Eq.~\(gamma_vac) is to use the representation of the commutation relations with
$$
a_1={\partial\over \partial a_1^\dagger}~,\quad a_2={\partial\over \partial
a_2^\dagger}~.
\eqno(a_rep)
$$
The conditions $a_1^\gamma|\gamma\rangle=a_2^\gamma|\gamma\rangle=0$ become two
coupled first order differential equations satisfied by the coefficient of
$|0,1\rangle\otimes |0,2\rangle$; the expression in Eq.~\(gamma_vac) is the
unique solution that yields a normalized state.

If we now trace over the state of the second oscillator to find a density
matrix for the first oscillator, we obtain
$$
\rho^\gamma_1\equiv {\rm
tr}_2\Biggr(|\gamma\rangle\langle\gamma|\Biggr)=\left(1-\gamma^2\right)
\sum_{n=0}^\infty\gamma^{2n}|n,1\rangle\langle n,1|~.
\eqno(rho_one)
$$
This has precise form of a thermal density matrix with inverse temperature
$\beta$ given by
$$
\gamma^2=e^{-\beta\omega}
\eqno(gamma_temp)
$$
where $\omega$ is the frequency of oscillator 1.  The calculation we have
performed is just what is needed to proceed from Eq.~\(minkmodeone) and
\(minkmodetwo) to Eq.~\(rhor).

A general coherent state built ``on top of'' the state $|\gamma\rangle$ has the
form
$$
|\gamma,\alpha_1,\alpha_2\rangle=N_{\alpha_1\alpha_2}
\exp\left(\alpha_1(a_1^\gamma)^{\dagger}\right)
\exp\left(\alpha_2(a_2^\gamma)^{\dagger}\right)|\gamma\rangle~,
\eqno(coherent_gamma)
$$
where $N_{\alpha_1\alpha_2}$ is a normalization constant.  This is the unique
normalized state that obeys the conditions
$$
\eqalign{
&\left(a_1^\gamma -
\alpha_1\right)|\gamma,\alpha_1,\alpha_2\rangle=0~,\cr
&\left(a_2^\gamma -
\alpha_2\right)|\gamma,\alpha_1,\alpha_2\rangle=0~.\cr}
\eqno(new_a_conditions)
$$
Thus, we may regard the coherent state as the ``vacuum'' state of the {\it
shifted} annihilation operators
$$
\eqalign{
&\hat a_1^\gamma=a_1^\gamma -\alpha_1~,\cr
&\hat a_2^\gamma=a_2^\gamma -\alpha_2~.\cr}
\eqno(shifted_a)
$$
If we also define shifted annihilation operators
$$
\eqalign{
&\hat a_1=a_1 -\left({\alpha_1+\gamma
\alpha_2^*\over\sqrt{1-\gamma^2}}\right)~,\cr
&\hat a_2=a_2 -\left({\alpha_2+\gamma
\alpha_1^*\over\sqrt{1-\gamma^2}}\right)~,\cr}
\eqno(old_shifted_a)
$$
then the Bogolubov transformation relating $\hat a^\gamma_{1,2}$ to $\hat
a_{1,2}$ is
$$
\eqalign{
\hat a_1^\gamma&={1\over \sqrt{1-\gamma^2}}\left( \hat a_1-\gamma \hat
a_2^\dagger\right)
{}~,\cr
\hat a_2^\gamma&={1\over \sqrt{1-\gamma^2}}\left(\hat a_2 - \gamma \hat
a_1^\dagger\right)
{}~,\cr}
\eqno(osc_bogo_hat)
$$
which has exactly the same form as Eq.~\(osc_bogo).
Since the shifted operators obey the standard commutation relations, the
same argument as before shows that the coherent state can be expressed as
$$
|\gamma,\alpha_1,\alpha_2\rangle=\sqrt{1-\gamma^2}\exp\left(\gamma\hat
a_1^\dagger\hat a_2^\dagger\right)|\hat 0,1\rangle\otimes|\hat 0,2\rangle~,
\eqno(hat_gamma_vac)
$$
where $|\hat 0,1\rangle$ and $|\hat 0,2\rangle$ are the ground states of the
{\it shifted} oscillators 1 and 2 (or, in other words, coherent states of the
unshifted oscillators).  We can trace over the second oscillator
just as before, and find
$$
\rho^{\gamma\alpha_1\alpha_2}_1\equiv {\rm tr}_2
\Biggr(|\gamma,\alpha_1,\alpha_2\rangle\langle \gamma,\alpha_1\alpha_2|\Biggr)
=\left(1-\gamma^2\right) \sum_{n=0}^\infty\gamma^{2 n }|\hat n,1\rangle\langle
\hat
n,1|~.
\eqno(hat_rho_one)
$$
This density matrix has exactly the same form as Eq.~\(rho_one), except that we
are now expanding in terms of the basis of states that have definite occupation
number with respect to the shifted oscillators.  The coherent state density
matrix, then, has exactly the same eigenvalues as the vacuum density matrix,
and it therefore also has exactly the same entropy.  Note that it is not quite
correct to describe Eq.~\(hat_rho_one) as a ``thermal density matrix,'' because
the eigenstates of the shifted number operator $\hat a_1^\dagger \hat a_1$ are
not eigenstates of the Hamiltonian $H=\omega a_1^\dagger a_1$.

Now we note that the case of two entangled oscillators described above is all
that we need to deal with when we compute the fine-grained entropy for a free
field.  It follows from Eq.~\(minkmodeone) and \(minkmodetwo) that
$$
\eqalign{
a_{1,\omega}&={1\over \sqrt{1-e^{-2\pi\omega}}}\left(
a_{R,\omega}-e^{-\pi\omega} a_{L,\omega}^\dagger\right)
{}~,\cr
a_{2,\omega}&={1\over \sqrt{1-e^{-2\pi\omega}}}\left(a_{L,\omega} -
e^{-\pi\omega}a_{R,\omega}^\dagger\right)
{}~,\cr}
\eqno(field_osc_bogo)
$$
are operators that annihilate modes that are positive frequency with respect to
Minkowski time; $a_{R,\omega}$ and $a_{L,\omega}$ denote the operators that
annihilate the modes of Rindler frequency $\omega$ that are localized in the
right and left wedge, respectively.  (The minus signs in Eq.~\(field_osc_bogo)
arise from the minus sign in the Klein-Gordon inner product of two negative
frequency modes.)
Thus, the expression Eq.~\(minvac) for the Minkowski vacuum is a tensor
product of states that have just the form Eq.~\(gamma_vac), with
$\gamma=e^{-\pi\omega}$.  Each field mode in
the right Rindler wedge is correlated with a particular mode in the left
Rindler wedge; for each such pair of modes, the entanglement of the state of
the right mode with the state of the left mode has exactly the same form as the
entanglement of oscillator 1 with oscillator 2 in the above discussion.

Furthermore, a general coherent state built on the Minkowski vacuum also has
the property that it can be factorized into a tensor product of correlated
states for pairs of modes.  The general coherent state can be expressed as
$$
|{\rm Minkowski~coherent}\rangle= N\prod_j \left(e^{\left(\alpha_{1,j}
a_{1,j}^\dagger\right)} e^{\left(\alpha_{2,j}
a_{2,j}^\dagger\right)}|0_M,j\rangle\right)
\eqno(field_coherent_minkowski)
$$
where $|0_M,j\rangle$ denotes the state that is annihilated by the Minkowski
annihilation operators $a_{1,j}$ and $a_{2,j}$.
Eq.~\(field_coherent_minkowski) is just a product of states of the form
$|\gamma_j=e^{-\pi\omega_j},\alpha_{R,j},\alpha_{L,j}\rangle$.  The evaluation
of the density matrix $\rho_R$ in the right Rindler wedge than proceeds as
above, and we find that it has the same eigenvalues for the coherent state as
for the Minkowski vacuum.

As our arguments in Section III show, the Minkowski vacuum still has the form
Eq.~\(minvac) when expressed in terms of the modes that are localized inside
and outside a finite region of space, and the general coherent state built on
the Minkowski vacuum still has the form Eq.~\(field_coherent_minkowski).  These
statements remain true if we consider, not the Minkowski vacuum, but a state
that is conformally related to it.  Also, the form
Eq.~\(field_coherent_minkowski) applies in curved space as well as in flat
space.

We conclude, finally, that our formula Eq.~\(RSTFG) for the fine-grained
entropy outside the apparent horizon of a black hole applies not just when the
incoming quantum state of the matter fields is the asymptotic inertial vacuum,
but also when the quantum state is an arbitrary coherent state built on the
inertial vacuum.

\head{Appendix B}
\taghead{B.}

In our derivation of the generalized second law in Section VII, we made some
restrictive assumptions about the incoming matter.  In particular, we assumed
that the quantum state of the matter is a coherent state built on the
asymptotic inertial vacuum state at ${\cal I}^-$.  In this Appendix, we will
examine what happens when this assumption is relaxed.  We will show that if
more general quantum states are allowed, the total entropy can decrease.  Thus,
the second law can be violated.

The crucial point is that quantum states can be constructed that pack a large
positive density of (fine-grained) entropy without carrying a large energy
density.  We can prepare matter in such a state, and allow the matter to fall
into a black hole.  Then the fine-grained entropy decreases sharply, but
without any compensating sharp increase in the black hole entropy.  Hence, the
total entropy decreases.

Alternately, we can make the total entropy decrease (momentarily)
by simply sending negative energy into the black hole.
It can be arranged that the black hole shrinks and loses entropy
without a compensating increase in the fine-grained entropy.

To demonstrate the existence of such states, consider an initial state of
left-moving matter than is in the ``vacuum'' state defined, not with respect to
the asymptotic inertial coordinate $\sigma^+$, but rather with respect to a
different coordinate $\hat x^+(\sigma^+)$.  In this quantum state, the incoming
energy flux, expressed in the $\sigma$ gauge, is [\cite{fulling}]
$$
{\cal E}(\sigma^+)\equiv\langle : \hat
T_{++}(\sigma^+):_\sigma\rangle=-\left({d\hat x^+\over
d\sigma^+}\right)^{3\over 2}{d^2\over d (\hat x^+)^2}\left({d\hat x^+\over
d\sigma^+}\right)^{1\over 2}
={3 (h')^2\over  4 h^2}- {h''\over 2h}~,
\eqno(hat_vacuum_energy)
$$
where
$$
h\equiv {d\hat x^+\over d\sigma^+}~,
\eqno(h_define)
$$
and the prime denotes differentiation with respect to $\sigma ^+$.  (Here we
have used the normalization convention of Eq.~\(tdef), and have assumed that
there are $N$ massless scalar matter fields.)  Note that the energy density is
not necessarily positive.  In this ``vacuum'' the equation for the trajectory
of the apparent horizon, in Kruskal coordinates, is,
$$
{dx^-_H\over dx^+_H}=- {\partial_+^2\Omega\over
\partial_-\partial_+\Omega}\Biggr|_H=-{1\over \lambda^2} t_+(x^+_H)
=-{1\over \lambda^4(x^+_H)^2}\left({\cal E}-{1\over 4}\lambda^2\right)~.
\eqno(weird_boundary)
$$
Thus, as in our previous analysis of coherent states built on the $\sigma$
vacuum, the condition for ``critical illumination'' is ${\cal
E}={\cal E}_{\rm cr} \equiv{1\over 4}\lambda^2$; when this condition is
satisfied, the incoming flux matches the flux of the outgoing Hawking
radiation, and the apparent horizon is null.

For this state, the expression Eq.~\(mirrorentropy) for the fine-grained
entropy outside of the apparent horizon becomes
$$
S_{\rm FG}={N\over 6}\left[\rho_{H,\sigma} -{1\over 2}\ln\left({d\hat
x_H^+\over d\sigma_H^+}{d\hat x_H^-\over d\sigma_H^-}\right) +\ln\left({\hat
x^+_H - \hat x^+_B\over \delta}\right)\right]~.
\eqno(weird_entropy)
$$
This formula differs from our old expression Eq.~\(RSTFG) in two respects.
First, the affine volume in the argument of the logarithm in the third term is
expressed in terms of the $\hat x^+$ coordinate that is used to define the
vacuum, rather than the inertial $\sigma^+$ coordinate.  Second, the term that
enters when we reexpress the cutoff in terms of the inertial coordinates at the
horizon is the conformal factor of the metric
in $\hat x$ coordinates.  This
differs from the conformal factor in $\sigma$ coordinates, which accounts for
the second term in Eq.~\(weird_entropy).

Now let us suppose that the function $\hat x^+(\sigma^+)$ is chosen so that the
black hole is critically illuminated at a particular retarded time
$\sigma_H^+$.  At that moment, $x^-_H$ is instantaneously constant, as is the
value $\Omega_H$ of $\Omega$ at the apparent horizon.  Thus, it is easy to
evaluate the rate at which
the fine-grained entropy is changing.  Using Eq.~\(horizon_rho_from_phi),
we find
$$
{dS_{\rm FG}\over d\sigma^+_H}={N\over 6}\left( {1\over 2}\lambda -{h'\over
2h}+{h\over \hat x_H^+-\hat x_B^+}\right)~.
\eqno(weird_entropy_rate)
$$
It is clear from Eq.~\(weird_entropy_rate) that we can make the rate of change
of $S_{\rm FG}$ large and negative by choosing $\hat x^+(\sigma^+)$ so that
$h'$ is large and positive.  Furthermore, we may simultaneously
arrange that $h''$ is
large, so that ${\cal E}(\sigma^+)$ in Eq.~\(hat_vacuum_energy) obeys
the critical illumination condition.  Finally, under critical illumination, the
black hole entropy is constant, so that no increase in the black hole entropy
compensates for the decrease in the fine grained entropy, and the Boltzman
entropy outside the black hole is also decreasing.  Hence, the total entropy
decreases.

Another way to make the total entropy momentarily
decrease is to throw negative energy
into the black hole. Evidently this can be achieved by choosing $h'=0$
and $h''>0$ in Eq.~\(hat_vacuum_energy). The black hole will then shrink and
decrease its entropy, but there will not in general be any compensating
increase in $S_{FG}$.  It is not clear, however, how an analog of the Boltzman
entropy should be defined for these states that carry negative energy density.

A preliminary investigation of the properties of states with the above
properties indicates that such an imbalance between entropy and energy cannot
be sustained indefinitely [\cite{prtr}].  We expect that there are fundamental
limitations on the severity and duration of these violations of the generalized
second law.

\head{References}

\refis{CGHS} C.~G.~Callan, S.~B.~Giddings, J.~A.~Harvey and A.~Strominger,
 {\sl Phys.\ Rev. D  } {\bf 45}
(1992) R1005; For reviews see J.~A.
Harvey and A.~Strominger,
{\it Quantum Aspects of Black Holes}, in the proceedings of the 1992 TASI
Summer School in Boulder, Colorado (World Scientific, 1993),
and S.~B.~Giddings, {\it Toy Models for Black Hole Evaporation}, in the
proceedings of the
International Workshop of Theoretical Physics, 6th Session, June 1992, Erice,
Italy (World Scientific, 1993).

\refis{bos} T.~Banks, M.~O'Loughlin and A.~Strominger, {\sl Phys.\ Rev. D  }
{\bf 47}
(1993) 4476.

\refis{emod} A.~Bilal and C.~G.~Callan, {\sl Nucl. Phys. B} {\bf 394}
(1993) 73; S.~P.~de Alwis,
 {\sl Phys. Lett. B} {\bf 289\/} (1992) 278;
 {\sl Phys. Lett. B} {\bf 300\/} (1993) 330;
 S. B. Giddings and A. Strominger, {\sl Phys. Rev. D} {\bf 46\/} (1993) 2454.

\refis{AS} A.~Strominger, unpublished.

\refis{rst} J.~G.~Russo, L.~Susskind, and L.~Thorlacius, {\sl Phys. Rev.
D\/}{\bf 46\/} (1992) 3444; {\sl Phys. Rev. D\/} {\bf 47\/} (1993) 533.

\refis{prtr} J.~Preskill and S.~Trivedi, unpublished.

\refis{gine} S.~B.~Giddings and W.~M.~Nelson,
{\sl Phys. Rev. D} {\bf 46\/} (1992) 2486.

\refis{lowe} D.~A.~Lowe,
{\sl Phys. Rev. D\/} {\bf 47} (1993) 2446.

\refis{hawk} S.~W.~Hawking,
 {\sl Comm. Math. Phys.} {\bf 43} (1975) 199.

\refis{hawk2} S.~W.~Hawking,
 {\sl Phys. Rev. D} {\bf 14} (1976) 2460; {\sl Comm. Math. Phys.} {\bf 87}
(1982) 395.

\refis{verl}E.~Verlinde and H.~Verlinde,
{\it A Unitary S-matrix for 2D Black Hole Formation and Evaporation,}
Princeton Preprint, PUPT-1380, IASSNS-HEP-93/8, hep-th/9302022 (1993);
 K. Schoutens, E.~Verlinde, and H.~Verlinde, {\sl Phys. Rev. D\/}{\bf 48}
(1993) 2670.

\refis{pira}T.~Piran and A.~Strominger, {\sl Phys. Rev. D\/}{\bf 48} (1993)
4729.

\refis{bps} T.~Banks, M.~E.~Peskin and L.~Susskind,
{\sl Nucl. Phys. B\/} {\bf 244} (1984) 125.

\refis{jacobson} T.~Jacobson, {\sl Phys.\ Rev. D\ }{\bf 44} (1991) 173; {\sl
Phys.\ Rev.  D\ }{\bf 48} (1993) 728.

\refis{bek} J.~D.~Bekenstein, {\sl Phys. Rev. D\ }{\bf 7} (1973) 2333; {\sl
Phys. Rev. D\ }{\bf 9} (1974) 3292.

\refis{mtw} C.~Misner, K.~Thorne, and J.~Wheeler, {\it Gravitation},
(W.~H.~Freeman, New York, 1973)).

\refis{zurekthorne} W. G. Unruh and R. M. Wald, {\sl Phys.~Rev.\ }
 {\bf D25}, 942
 (1982); W.~H.~Zurek and K.~S.~Thorne, {\sl Phys. Rev. Lett.\ }{\bf
54} (1985) 2171; K.~S.~Thorne, W.~H.~Zurek, and R.~H.~Price, in {\it Black
Holes: The Membrane Paradigm}, edited by K.~S.~Thorne, R.~H.~Price, and
D.~A.~MacDonald (Yale University Press, New Haven, 1986), p. 280;
 R. M. Wald, in {\it Black Hole Physics}, edited by V.~
 De Sabbata and Zhenjiu Zhang (Kluwer, Dordrecht, 1992), p.~55;
 I. D. Novikov and V. P. Frolov, {\it Physics of Black
 Holes} (Kluwer, Dordrecht, 1989) and references therein.

\refis{frolov} V.~P.~Frolov and D.~N.~Page, {\sl Phys. Rev. Lett.\ }{\bf 71}
(1993) 3902.

\refis{remnant} Y. Aharonov, A. Casher, and S. Nussinov, {\sl Phys. Lett.\ }
{\bf 191B} (1987) 51; T. Banks, A. Dabholkar, M. R. Douglas, amd M.
O'Loughlin, {\sl Phys. Rev. D\ }{\bf 45} (1992) 3607; T. Banks and M.
O'Loughlin, {\sl Phys. Rev. D\ }{\bf 47} (1993) 540; T. Banks, M. O'Loughlin,
and A. Strominger, {\sl Phys. Rev. D\ }{\bf 47} (1993) 4476;
S. B. Giddings  {\sl Phys. Rev. D\ }{\bf 49} (1994) 947.

\refis{gibbons} G.~W.~Gibbons and K, Maeda, {\sl Nucl. Phys. B\ }{\bf 298}
(1988) 741; D. Garfinkle, G. Horowitz, and A. Strominger,
{\sl Phys. Rev. D\ }
{\bf 43} (1991) 3140.

\refis{unruh} W. G. Unruh, {\sl Phys. Rev. D\ }{\bf 14} (1976) 870.

\refis{holzhey} C. Holzhey, Princeton University Ph.D. Thesis (1993),
unpublished.

\refis{hooft} G. 't Hooft, {\sl Nucl. Phys. B\ }{\bf 335} (1990) 138, and
references therein.

\refis{hooft2} G. 't Hooft, {\sl Nucl. Phys. B\ }{\bf 256} (1985) 727.

\refis{bombelli} L. Bombelli, R. K. Koul, J. Lee, and R. Sorkin, {\sl Phys.
Rev. D\ }{\bf 34} (1986) 373.

\refis{wilczek} F. Wilczek, in {\it Black Holes, Membranes, Wormholes,and
Superstrings}, edited by S. Kalara and D.~V.~Nanopoulos (World Scientific,
Singapore, 1993), p. 1.

\refis{stromtriv} A.~Strominger and S.~Trivedi, {\sl Phys. Rev. D\ }{\bf 48}
(1993) 5778.

\refis{cpt} A.~Strominger, {\sl Phys. Rev. D\ }{\bf 48}
(1993) 5769.

\refis{zurek} W.~H.~Zurek, {\sl Phys. Rev. Lett.\ }{\bf 49} (1982) 1683.

\refis{page} D.~N.~Page, {\sl Phys. Rev. D\ }{\bf 14} (1976) 3260; {\sl Phys.
Rev. Lett.\ }{\bf 50} (1983) 1013.

\refis{susskind} L. Susskind, {\sl Phys. Rev. Lett.\ }{\bf 71} (1993) 2367; L.
Susskind and L. Thorlacius, {\sl Phys. Rev. D\ }{\bf 49} (1994) 966; L.
Susskind, {\it Strings, Black Holes and Lorentz Contraction}, Stanford Preprint
SU-ITP-93-21, hep-th/9308139 (1993); L. Susskind, {\it Some Speculations about
Black Hole Entropy in String Theory}, Rutgers Preprint RU-93-44, hep-th/9309145
(1993).

\refis{mathur} E.~Keski-Vakkuri and S.~D.~Mathur, {\it Evaporating Black Holes
and Entropy}, MIT preprint, hep-th/9312194 (1993).

\refis{uglum} L. Susskind and J. Uglum, {\it Black Hole Entropy in Canonical
Quantum Gravity and Superstring Theory}, Stanford University Preprint
SU-ITP-94-1, hep-th/9401070 (1994).

\refis{callan} C. Callan and F. Wilczek, {\it On Geometric Entropy}, Institute
for Advanced Study Preprint IASSNS-HEP-93/87, hep-th/9401072 (1994).

\refis{strassler} D. Kabat and M.~J.~Strassler, {\it A Comment on Entropy and
Area}, Rutgers University Preprint RU-94-10, hep-th/9401125 (1994).

\refis{dowker} J.~S.~Dowker, {\it Remarks on Geometric Entropy}, University of
Manchester Preprint MUTP/94/2, hep-th/9401159 (1994).

\refis{polyakov} A.~M.~Polyakov, {\sl Phys. Lett.\ }{\bf 103B} (1981) 207.

\refis{numerical} S.~W.~Hawking, {\sl Phys. Rev. Lett.\ }{\bf 69} (1992) 406.,
B. Birnir, S.~B.~Giddings, J.~A.~Harvey, and A. Strominger,
{\sl Phys. Rev. D\ }
{\bf 46} (1992) 638; S.~W.~Hawking and J.~M.~Stewart, {\sl Nucl Phys. B\ }{\bf
400} (1993) 393.

\refis{fulling} S.~M.~Christensen and S.~A.~Fulling, {\sl Phys. Rev. D\ }{\bf
15} (1977) 2088.

\refis{andyandlarus} A. Strominger and L. Thorlacius, {\it Conformally
Invariant
Boundary Conditions for Dilaton Gravity}, to appear;  A. Strominger,
L. Thorlacius and S. Trivedi, in progress.

\refis{srednicki} M. Srednicki, {\sl Phys. Rev. Lett.\ }{\bf 71} (1993) 666,
and private communication.

\refis{rutgers} N. Seiberg, S. Shenker, L. Susskind, L. Thorlacius, and J.
Tuttle (1993) unpublished.

\refis{susskind2} L. Susskind, L Thorlacius, and J. Uglum,
{\sl Phys. Rev. D\ }
{\bf 48} (1993) 3743.

\refis{hartlehawking} J.~B.~Hartle and S.~W.~Hawking, {\sl Phys. Rev. D\ }{\bf
13} (1976) 2188.

\refis{lebowitz} For a cogent recent review, see J.~L.~Lebowitz, {\sl Physica
A\ }{\bf 194} (1993) 1.

\refis{page2} D.~N.~Page, {\sl Phys. Rev. Lett.\ }{\bf 71} (1993) 3743.

\refis{fronov} V. Frolov and I. Novikov, {\sl Phys. Rev. D\ }{\bf 48} (1993)
4545.

\refis{wald} R. M. Wald, {\sl Comm. Math. Phys.\ }{\bf 45} (1975) 9.

\refis{carlitz} R.~D.~Carlitz and R.~S.~Willey, {\sl Phys. Rev. D\ }{\bf 36}
(1987) 2336.

\refis{jp} J. Preskill, in {\it Black Holes, Membranes, Wormholes, and
Superstrings}, edited by S. Kalara and D.~V.~Nanopoulos (World Scientific,
Singapore, 1993), p. 22.

\refis{holzhey_wilczek} C. Holzhey, F. Larsen, and F. Wilczek, {\it Geometric
and Renormalized Entropy in Conformal Field Theory}, Princeton Preprint PUPT
1454, hep-th/9403108 (1994).

\endreferences
\endit
\head{Figures}

\item{\bf Fig.~1}{\bf (a).} The two-dimensional spacetime of a black hole that
forms due to the collapse of a shock wave, and then evaporates completely.
After the black hole forms, the apparent horizon recedes along a timelike
trajectory, eventually meeting the singularity at the ``endpoint.''  The
timelike boundary and the spacelike singularity are in the strongly-coupled
region.  RST boundary conditions are imposed where the boundary is timelike.
{\bf (b).} Five spacelike slices through the spacetime, referred to in the
text.

\item{\bf Fig. 2} Rindler spacetime.  The ``right wedge,'' with $U<0$ and
$V>0$, is accessible to an observer that accelerates uniformly to the right.
The ``left wedge,'' with $U>0$ and $V<0$, is accessible to an observer that
accelerates uniformly to the left.

\item{\bf Fig. 3} A spacelike slice through flat spacetime.  By tracing over
the field degrees of freedom on the portion of the slice in the region between
the points $P_1=(\hat U_1, \hat V_1)$ and $P_2=(\hat U_2,\hat V_2)$, we obtain
a density matrix $\rho_{\rm out}$ for the fields on the portion of the slice
outside that region.

\item{\bf Fig. 4} A spacelike slice through the moving mirror spacetime.
Coordinates have been chosen so that the trajectory of the mirror is $\hat
V(\hat U)=\hat U$. By tracing over the field degrees of freedom on the portion
of the slice in the region between the points $P_1=(\hat U_1, \hat V_1)$ and
$P_2=(\hat U_2,\hat V_2)$, we obtain a density matrix $\rho_{\rm out}$ for the
fields on the portion of the slice outside that region.

\item{\bf Fig. 5}{\bf (a).} A spacelike slice through the moving mirror
spacetime.  Coordinates have been chosen so that the trajectory of the mirror
is $\hat V(\hat U)=\hat U$.  By tracing over the field degrees of freedom on
the portion of the slice between the point $P=(\hat U_P,\hat V_P)$ and the
mirror, we obtain a density matrix $\rho_{\rm out}$ for the fields on the
portion of the slice to the right of the point $P$.  {\bf (b).}  If coordinates
are not chosen so that $\hat V=\hat U$ at the mirror, we define $V_B$ as the
retarded time of an incoming null ray that reflects off the mirror and then
passes through $P$.

\item{\bf Fig. 6} A spacelike slice $\Sigma$ through the black hole spacetime.
The slice crosses the apparent horizon at the point
$P=(\sigma_H^-,\sigma_H^+)$.  We define $\sigma_B^+$ as the retarded time of an
incoming null ray that reflects off the boundary and then passes through $P$.
Incoming null rays with retarded time between $\sigma_B^+$ and $\sigma_H^+$
cross $\Sigma$ inside the apparent horizon.

\endit

Another feature of Eq.~\(halflineentropy) deserves comment.  In deriving it, it
was not necessary to be specific about the boundary conditions that are imposed
to implement the infrared cutoff.  However, if the field obeys periodic
boundary conditions, then there is a zero-frequency mode that can vary over an
infinite range.  Thus, since the zero mode surely entangles the left and right
wedges, it should by itself make an {\it infinite} contribution to the entropy.
 Fixed end or antiperiodic boundary conditions would remove the zero mode, so
that Eq.~\(halflineentropy) is the correct expression for the entropy in these
cases.  Still, the question arises why this sensitivity to boundary conditions
was not apparent in our derivation.

The answer is that we have implicitly chosen as our basis a set of modes that
{\it vanish} at $U=0$.  Though this might not be apparent in Eq.~\(landrmodes),
the point is that the normalizable wavepackets that are constructed as
superpositions of modes that are positive frequency with respect to Rindler
time necessarily vanish as, say, $u_R\to -\infty$.  Physically, this is because
it takes an infinite amount of Rindler time for a wavepacket to reach the
Rindler horizon.  Thus, in our calculation of the density matrix $\rho_R$, the
contribution of the zero mode was not included.  This is not a serious
omission.  In the infinite volume limit, different values of the constant mode
of the scalar field correspond to distinct superselection sectors of the
quantum theory.  It is appropriate to project out a particular value of the
zero mode before computing the fine-grained entropy.